\documentclass[11.5pt]{article}
\usepackage{amsmath,amssymb}
\usepackage{amsthm,subcaption}
\usepackage{multirow,booktabs}
\usepackage{hyperref}
\usepackage{natbib}
\usepackage{graphicx}
\usepackage{color}
\usepackage{caption}
\newtheorem{theorem}{Theorem}
\newtheorem{definition}{Definition}
\newtheorem{assumption}{Assumption}
\newtheorem{lemma}{Lemma}
\newtheorem{lem}{Lemma}[section]
\newtheorem*{assumption*}{Assumption}

\usepackage[plain,noend]{algorithm2e}
\newcommand{\beq}{\begin{equation}}
\newcommand{\eeq}{\end{equation}}

\newcommand{\bSigma}{\mbox{\boldmath $\Sigma$}}

\newcommand{\bLambda}{\mbox{\boldmath $\Lambda$}}
\newcommand{\bxi}{\mbox{\boldmath $\xi$}}
\newcommand{\boldeta}{\mbox{\boldmath $\eta$}}
\newcommand{\bXi}{\mbox{\boldmath $\Xi$}}
\newcommand{\bD}{\mbox{\boldmath $D$}}
\newcommand{\bX}{\mbox{\boldmath $X$}}
\newcommand{\bI}{\mbox{\boldmath $I$}}
\newcommand{\bB}{\mbox{\boldmath $B$}}
\newcommand{\bY}{\mbox{\boldmath $Y$}}
\newcommand{\bS}{\mbox{\boldmath $S$}}

\begin{document}

\title{Estimation of the number of spiked eigenvalues in a covariance matrix by bulk eigenvalue matching analysis}
\author{Zheng Tracy Ke, Yucong Ma  and Xihong Lin 
\thanks{Zheng Tracy Ke is Assistant Professor and Yucong Ma is Graduate Student, both in the Department of Statistics of Harvard University. Xihong Lin is Professor of Biostatistics  at Harvard T.H. Chan School of Public  Health and Professor of Statistics at Faculty of Arts and Sciences, Harvard University. This work was supported by the National Science Foundation grant DMS-1712958, and the National Institutes of Health grants R35-CA197449,  U01-HG009088, U19-CA203654. The authors thank Rounak Dey and Derek Shyr for their help on downloading and pruning the 1000 Genomes dataset,  Zhigang Bao and Xiucai Ding for helpful pointers on the random matrix theory,  and the reviewers for helpful comments that improved the paper.
} }
\date{}

\maketitle
\thispagestyle{empty}
\baselineskip=18pt


\begin{abstract} 
The spiked covariance model has gained increasing popularity in high-dimensional data analysis. A fundamental problem is determination of the number of spiked eigenvalues, $K$. For estimation of $K$, most attention has focused on the use of {\it top} eigenvalues of sample covariance matrix, and there is little investigation into proper ways of utilizing {\it bulk} eigenvalues to estimate $K$. We propose a principled approach to incorporating bulk eigenvalues in the estimation of $K$. Our method imposes a working model on the residual covariance matrix, which is assumed to be a diagonal matrix whose entries are drawn from a 
gamma distribution.  Under this model,  the bulk eigenvalues are asymptotically close to the quantiles of a fixed parametric distribution. 
This motivates us to propose a two-step method: the first step uses bulk eigenvalues to estimate parameters of this distribution, and the second step leverages these parameters to assist the estimation of $K$. The resulting estimator $\hat{K}$ aggregates information in a large number of bulk eigenvalues. We show the consistency of $\hat{K}$ under a standard spiked covariance model. 
 We also propose a confidence interval estimate for $K$.   Our extensive simulation studies show that the proposed method is robust and outperforms the existing methods in a range of scenarios.  We apply the proposed method to analysis of a lung cancer microarray data set and the 1000 Genomes data set. 

\end{abstract}
\medskip


\newpage
\pagestyle{plain}
\setcounter{page}{1}

\section{Introduction}
The spiked covariance model \citep{Johnstone2001} has been widely used to model the covariance structure of high-dimensional data. In this model, the population covariance matrix has $K$ large eigenvalues, called {\it spiked eigenvalues}, where $K$ is presumably much smaller than the dimension. Estimation of $K$ is of great interest in practice, as it helps determination of the latent dimension of data. For example, in a clustering model with $K_0$ clusters \citep{JKW}, the pooled covariance matrix has $(K_0-1)$ spiked eigenvalues; therefore, an estimate of $K$ tells the number of clusters. Similarly, in Genome-Wide Association Studies (GWAS), the number of spiked eigenvalues of a genetic covariance matrix   reveals the number of ancestry groups in the study \citep{patterson2006population}. In high-dimensional covariance matrix estimation, $K$ is often required as input for factor-based covariance estimation \citep{fan2013large}.  

In this paper, we assume the data vectors $\bX_1, \bX_2, \ldots, \bX_n\in\mathbb{R}^p$ are independently generated from a multivariate distribution with covariance matrix $\bSigma\in\mathbb{R}^{p\times p}$, which has positive values $\mu_1\geq\mu_2\geq \ldots\geq\mu_K$ and mutually orthogonal unit-norm vectors $\bxi_1,\bxi_2,\ldots, \bxi_K\in\mathbb{R}^p$ such that 
\beq \label{mod-Sigma0}
\bSigma=\sum_{k=1}^K\mu_k\bxi_k\bxi_k^{\top} + \bD, \qquad \mbox{where}\quad \bD=\mathrm{diag}(\sigma^2_1,\sigma_2^2, \ldots,\sigma^2_p). 
\eeq
Here, $\bD$ is called the residual covariance matrix. The goal is to estimate $K$ from $\bX_1,\bX_2, \ldots,\bX_n$. We are primarily interested in the settings where $K$ is finite and $p/n\to\gamma$, for a constant $\gamma >0$. 
Throughout the paper, we denote by $\lambda_1\geq\lambda_2\geq\ldots\geq\lambda_p$ the eigenvalues of $\bSigma$, and denote by $\hat{\lambda}_1\geq\hat{\lambda}_2\geq\ldots\geq\hat{\lambda}_{n\wedge p}$ the nonzero eigenvalues of the sample covariance matrix.

In the literature, there are several approaches for estimating $K$. The first is the information criterion approach, which finds $\hat{K}$ that minimizes an objective of the form $L_n(K)+P_n(K)$, where $L_n(K)$ is a measure of goodness-of-fit and $P_n(K)$ is a penalty on $K$. An influential work is \cite{bai2002determining}, who let $L_n(K)$ be the sum of squared residuals after fitting a $K$-factor model and studied a few choices of the penalty function $L_n(K)$. Other examples include \cite{wax1985detection}, where $L_n(K)$ is a function of the arithmetic and geometric means of $(n-K)$ smallest eigenvalues. 
However, the information criterion approach requires the spiked eigenvalues to be sufficiently large. In \cite{bai2002determining}, the spiked eigenvalues are at the order of $p$, which is much larger than the necessary order. It has been recognized that correct estimation of $K$ is possible even when the spiked eigenvalues are at the constant order \citep{BAP}.

The second approach finds a big ``gap" between eigenvalues of the sample covariance matrix. Recall that $\hat{\lambda}_k$ is the $k$th eigenvalue of the sample covariance matrix.  \cite{onatski2009testing} introduced a test statistic, $\max_{K_0< k\leq K_{\max}}(\hat{\lambda}_i-\hat{\lambda}_{i+1})/(\hat{\lambda}_{i+1}-\hat{\lambda}_{i+2})$, for testing against the null hypothesis $K=K_0$ and then applied it sequentially to estimate $K$. \cite{cai2017limiting} proposed an iterative algorithm for estimating $K$ that searches for a gap of $\gtrsim O(n^{-2/3})$ between eigenvalues. \cite{passemier2014estimation} suggested estimating $K$ by finding two consecutive gaps in eigenvalues.  Such methods rely on sharp limiting distributions of the first $K$ empirical eigenvalues, which theoretically requires a large magnitude of the spiked eigenvalues. Additionally, while utilizing eigengap is a neat idea in theory, its practical use faces challenges, since the actual eigengaps in many real data sets are slowly varying, without a clear cut.

The last approach estimates $K$ by thresholding the empirical eigenvalues. For this approach, the key is to calculate a proper data-driven threshold. The threshold should reflect the ``scaling" of the residual matrix $\bD$. One idea is to first standardize the data matrix so that each variable has a unit variance and then use a scale-free threshold. Examples include the empirical Kaiser's criterion \citep{braeken2017empirical} and parallel analysis \citep{horn1965rationale}, where the scale-free threshold is determined by asymptotic behavior of the largest eigenvalue of sample covariance matrix associated with $\bX_i\overset{iid}{\sim}N(0, \bI_p)$.  Another idea is to estimate $\bD$ by the diagonal of the sample covariance matrix and then calculate the threshold via a deterministic algorithm \citep{dobriban2015efficient}. The success of both ideas rely on regularity conditions to ensure that the low-rank part in Model \eqref{mod-Sigma0} has a negligible effect on the diagonal of $\bSigma$; for example, the population eigenvalues cannot be enormously large and the population eigenvectors have to satisfy ``delocalization" conditions. \cite{dobriban2019deterministic} improved the algorithm in \cite{dobriban2015efficient} by a recursive procedure to remove leading eigenvalues and eigenvectors, but their method still requires some ``delocalization" conditions on eigenvectors. Other related work  includes \cite{onatski2010determining}, which used a convex combination of $\hat{\lambda}_{K_{\max}+1}$ and $\hat{\lambda}_{2K_{\max}+1}$ as the threshold, where $K_{\max}$ is a pre-specified upper bound of $K$, and \cite{fan2019estimating}, which introduced an unbiased estimator for each of the first few eigenvalues of the population correlation matrix, and estimated $K$ by thresholding these unbiased estimators at $1+\sqrt{p/n}$.

To address the limitations of these methods, we propose a new estimator of $K$. Different from the existing work, our attention is largely focused on how to better utilize the {\it bulk} empirical eigenvalues in the estimation of $K$, especially those eigenvalues in the middle range: 
\[
\bigl\{\hat{\lambda}_k: \alpha (n\wedge p)\leq k\leq (1-\alpha)(n\wedge p)\bigr\}, \qquad \mbox{for some constant }\alpha\in (0,1/2). 
\] 
It is well-known in random matrix theory that these bulk eigenvalues are almost not affected by the low-rank part in Model \eqref{mod-Sigma0} (e.g., see \cite{bloemendal2016principal}). We can use these eigenvalues to gauge the ``scaling" of $\bD$ and determine an appropriate threshold for top eigenvalues. To this end, we impose a working model on the diagonal matrix $\bD$. Let $\mathrm{Gamma}(a,b)$ denote the gamma distribution with shape parameter $a$ and rate parameter $b$. Fixing $\sigma>0$ and $\theta>0$, we assume  
\beq \label{mod-D}
 \sigma^2_j\, \overset{iid}{\sim}\, \mathrm{Gamma}(\theta,\; \theta/\sigma^2), \qquad 1\leq j\leq p.
\eeq
The mean and variance of $\mathrm{Gamma}(\theta,\theta/\sigma^2)$ is $\sigma^2$ and $\sigma^4/\theta$, respectively. As a result, the diagonal entries of $\bD$ are centered around $\sigma^2$, where the level of dispersion is controlled by $\theta$. As $\theta\to\infty$, $\mathrm{Gamma}(\theta, \theta/\sigma^2)$ converges to a point mass at $\sigma^2$, and it yields $\bD=\sigma^2 \bI_p$. This case corresponds to the standard spiked covariance model which is frequently studied in the literature \citep{Johnstone2001,donoho2018optimal}. Combining Model~\eqref{mod-D} with Model~\eqref{mod-Sigma0}, we now have a flexible spiked covariance model that includes the standard spiked covariance model as a special case.

Under Models~\eqref{mod-Sigma0}-\eqref{mod-D}, the empirical spectral distribution (ESD) converges to a limit, which is a fixed distribution with two parameters $(\sigma^2,\theta)$ \citep{silverstein2009stieltjes}. Since the empirical eigenvalues are nothing but quantiles of the ESD, we expect that all the bulk eigenvalues are asymptotically close to the corresponding quantiles of the limit of ESD. 
We thus estimate $(\sigma^2,\theta)$ by minimizing the sum of squared differences between bulk eigenvalues and quantiles of the limiting distribution. Once $(\hat{\sigma}^2,\hat{\theta})$ are available, we borrow the idea of parallel analysis \citep{horn1965rationale} to decide a threshold for the top eigenvalues by Monte Carlo sampling. This gives rise to a new method for estimating $K$, which we call  {\it bulk eigenvalue matching analysis (BEMA)}. Analogous to the orators' bema in Athens, our BEMA is a platform for gathering a large number of bulk eigenvalues and utilizing them efficiently in the estimation of $K$.  Additional to the point estimator, we also propose a confidence interval for $K$.

\begin{figure}[t]
\centering
\includegraphics[width=0.55\textwidth]{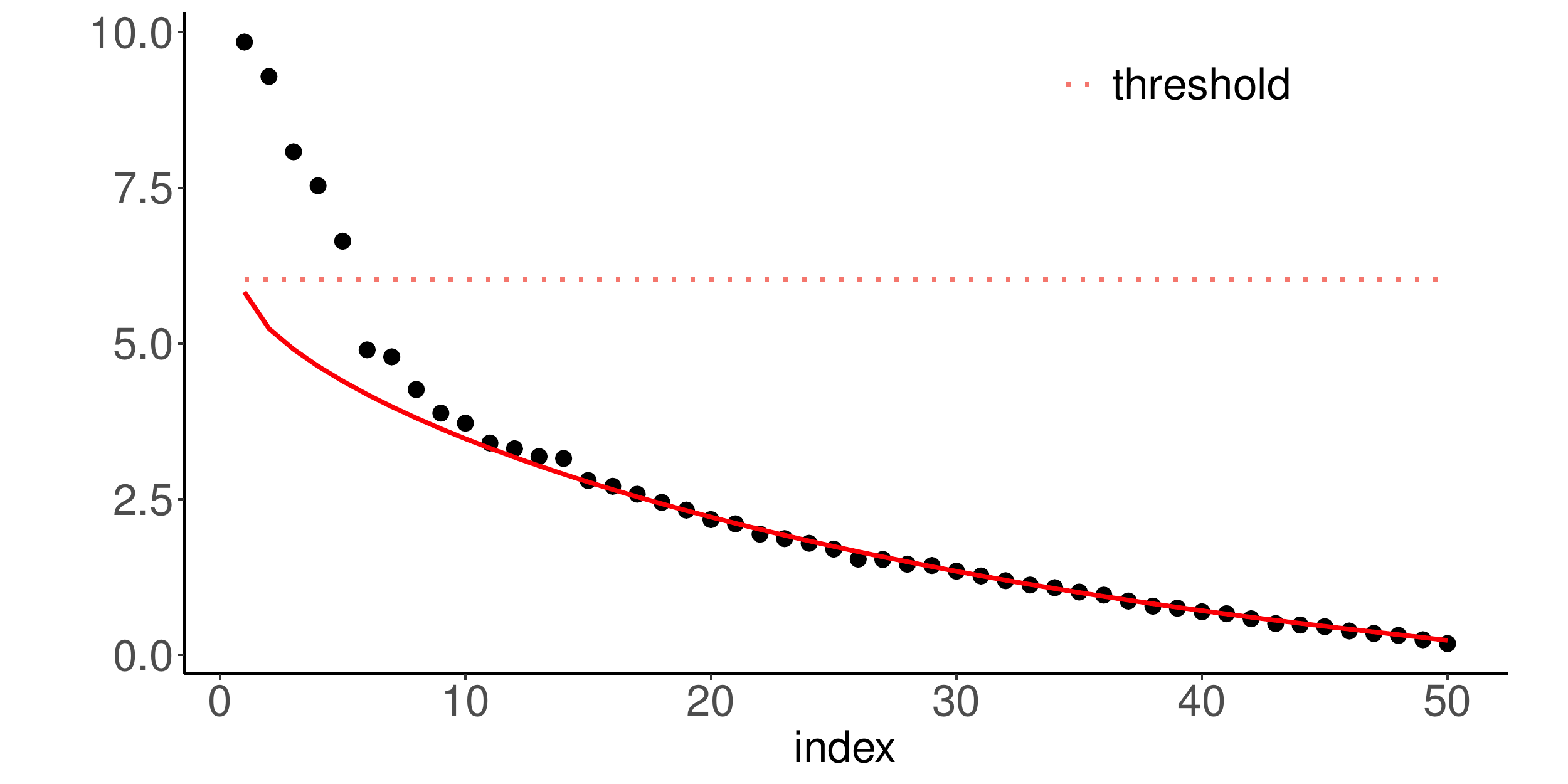} 
\caption{Illustration of BEMA via a scree plot. The red solid curve shows the quantiles of the theoretical limit of Empirical Spectral Distribution (ESD) under Models~\eqref{mod-Sigma0}-\eqref{mod-D}. It is a parametric curve with two parameters $(\sigma^2,\theta)$, and by random matrix theory, it should fit the bulk eigenvalues well. BEMA first uses bulk eigenvalues to estimate $(\sigma^2,\theta)$ and then extends the estimated curve to the left boundary to get a threshold for top eigenvalues.} \label{fig:intro}
\end{figure}

Our method has an intuitive explanation in terms of a scree plot. Figure~\ref{fig:intro} shows the scree plot of a simulated example. There are multiple elbow points, and it is hard to decide where the true $K$ is. The core idea of our method is to explore the ``shape" of the scree plot in the middle range and fit it with a parametric curve; this curve is determined by the theoretical quantiles of the limit of ESD, governed by two parameters $\sigma^2$ and $\theta$. Then, this curve can be extended to the left boundary of the scree plot to produce a threshold for top eigenvalues.

The goodness-of-fit check of Model~\eqref{mod-D} on real datasets can also be done via the scree plot. If the middle range of the scree plot can be well approximated by the estimated parametric curve, then it suggests that the model indeed fits the real data. In Section~\ref{sec:realdata}, we shall see that Model~\eqref{mod-D} is well suited to gene microarray data and GWAS data. We remark that assuming the diagonal entries of $\bD$ are generated from a fixed distribution is only a mild assumption. Similar conditions appear in the literature (often implicitly as regularity conditions in the theory); e.g., \cite{dobriban2019deterministic} and \cite{fan2019estimating} assume that the histogram of population eigenvalues of $\bD$ converges to a fixed limit. We make one step ahead by assuming that this fixed distribution is a gamma distribution. At the first glance, restricting to the gamma family seems restrictive, but Model~\eqref{mod-D} is in fact much more flexible than expected. With only two parameters $(\sigma^2,\theta)$, it can accommodate various kinds of real data and even misspecified models (see Section~\ref{sec:simulation}).

The special case of $\theta=\infty$ is of independent interest. It corresponds to the standard spiked covariance model \citep{Johnstone2001}, where $\bD=\sigma^2 \bI_p$.  This model has attracted a lot of attention \citep{BAP,Paul2007,donoho2018optimal}. In this special case, BEMA reduces to a simpler algorithm. We conduct theoretical analysis under this model. First, we give an explicit error bound for estimating $\sigma^2$. This is connected to the robust estimation of $\sigma^2$ in the literature of reconstruction of spiked covariance matrices \citep{donoho2018optimal,shabalin2013reconstruction}. 
In our method, we obtain a new robust estimator of $\sigma^2$ as a byproduct, and we study it theoretically. Second, we prove the consistency of estimating $K$ under minimal conditions. Our results impose no assumptions on the population eigenvectors $\bxi_1,\ldots,\bxi_K$ and only require the spiked eigenvalues $\lambda_1,\ldots,\lambda_K$ to be larger than a constant. In comparison, literature works often either require some regularity conditions on eigenvectors or need much larger spiked eigenvalues. 
We also provide theory for the general case of $\theta<\infty$, which has never been studied. 



The remaining of this paper is organized as follows: In Section~\ref{sec:BEMA0}, we describe BEMA for the standard spiked covariance model (i.e., $\theta=\infty$); in this case, the idea is easier to understand and the algorithm is simpler. In Section~\ref{sec:BEMA}, we describe BEMA for the general case. Section~\ref{sec:theory} states the theoretical properties. Section~\ref{sec:simulation} and Section~\ref{sec:realdata} provide simulation study results and real data analysis, respectively. Section~\ref{sec:discussion} concludes the paper. Proofs are relegated to the Appendix.

\section{BEMA for the standard spiked covariance model}  \label{sec:BEMA0}
In this section, we consider the standard spiked covariance model \citep{Johnstone2001}, a special case of Models \eqref{mod-Sigma0}-\eqref{mod-D} with $\theta=\infty$. Since each $\sigma^2_j$ is equal to $\sigma^2$, the model is re-written as  
\beq \label{mod-Sigma}
\bSigma=\sum_{k=1}^K\mu_k\bxi_k\bxi_k^{\top} + \sigma^2 \bI_p.
\eeq
The first $K$ eigenvalues of $\bSigma$ are $\lambda_k=\mu_k+\sigma^2$, and the remaining eigenvalues are $\sigma^2$. The sample covariance matrix is
 $\bS = \frac{1}{n}\sum_{i=1}^n (\bX_i-\bar{\bX})(\bX_i-\bar{\bX})^{\top}$, where $\bar{\bX}=\frac{1}{n}\sum_{i=1}^n \bX_i$.
With probability 1, $\bS$ has $n\wedge p$ distinct nonzero eigenvalues \citep{Uhlig94}, denoted as 
$\hat{\lambda}_1>\hat{\lambda}_2> \ldots>\hat{\lambda}_{n\wedge p}$. 


We first review some existing results about the asymptotic behavior of empirical eigenvalues. 

\begin{definition} \label{def:MP}
Given a parameter $\gamma>0$, the zero-excluded Machenko-Pastur (MP) distribution is defined by the density
\beq  \label{MPdensity}
f_{\gamma}(x; \sigma^2) = \frac{1}{2\pi\sigma^2}\frac{1}{ x(\gamma \wedge 1)}\sqrt{(x-\sigma^2 h_-)(\sigma^2 h_+-x)}\cdot 1\bigl\{ \sigma^2 h_- < x < \sigma^2  h_+\bigr\},
\eeq
where $h_{\pm}=(1\pm\sqrt{\gamma})^2$. We let $F_{\gamma}(x; \sigma)$ denote its cumulative distribution function. 
\end{definition}  

\noindent
When $\gamma\leq 1$, this definition is the same as the classical MP law; when $\gamma>1$,  it excludes the point mass at zero in the classical MP law.
The zero-excluded empirical spectral distribution (ESD) is given by $F_n(x) = \frac{1}{n\wedge p}\sum_{i=1}^{n\wedge p}1\{\hat{\lambda}_i\leq x\}$.  
For convenience, we shall omit the word `zero-excluded' and still call them MP and ESD.

\begin{figure}[htb!]
\centering
\includegraphics[width=0.5\textwidth, trim=30 80 30 50, clip=true]{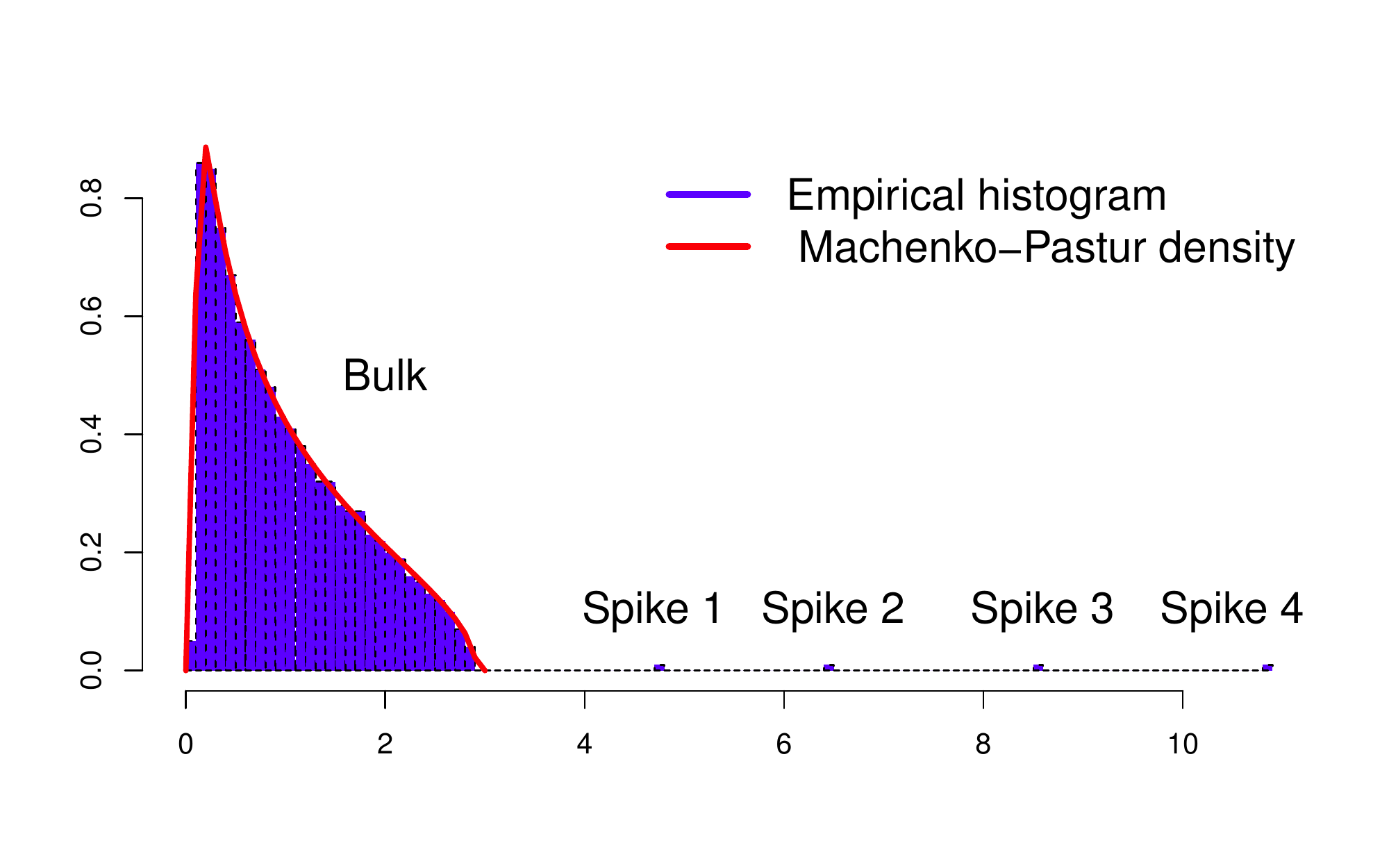}
\caption{The asymptotic behavior of empirical eigenvalues. The histogram of bulk eigenvalues converges to an MP distribution, and $K$ top eigenvalues are outside the support.} \label{fig:ESD}
\end{figure}

When $\bSigma$ satisfies \eqref{mod-Sigma}, $K$ is fixed and $p/n\to\gamma $ for a constant $\gamma\in (0,\infty)$, 
under mild regularity conditions, the following statements are true \citep{bloemendal2016principal}:
\begin{itemize}
\item The ESD converges to the MP distribution with parameter $\gamma$;
more precisely, it holds that $\mathbb{E}[\sup_{x}|F_n(x)-F_{\gamma}(x)|]=O(n^{-1/2})$ \citep{gotze2004rate}.
\item If $\mu_K\geq \sigma^2 \sqrt{\gamma}+n^{-1/3}$, the first $K$ empirical eigenvalues are located outside the support of the MP distribution with high probability. 
\end{itemize}
See Figure~\ref{fig:ESD} for an illustration via simulated data ($n=1000$, $p=500$). 

Inspired by the asymptotic behavior of empirical eigenvalues, we propose a two-step approach to estimating $K$. In the first step, 
we use bulk eigenvalues to fit an MP distribution. The density $f_{\gamma}(x; \sigma^2)$ in \eqref{MPdensity} has two parameters $(\gamma, \sigma^2)$, where $\gamma$ can be approximated by $\gamma_n=p/n$. It reduces to considering $f_{\gamma_n}(x; \sigma^2)$, for all possible $\sigma^2$. 
We aim to find $\hat{\sigma}^2$ such that $f_{\gamma_n}(x;\hat{\sigma}^2)$ is the best fit to the histogram of empirical eigenvalues. In the second step, we determine $K$ by comparing top eigenvalues with the right boundary of the support of the estimated MP density, namely, $\hat{\sigma}^2(1+\sqrt{\gamma_n})^2$.

Now, we describe the method in detail. 
First, consider the estimation of $\sigma^2$.  
Fixing a constant $\alpha\in (0,1/2)$, we take only a faction of nonzero eigenvalues: 
\[
\{\hat{\lambda}_k: \alpha (n\wedge p)\leq k\leq (1-\alpha) (n\wedge p)\}.
\] 
Since $K$ is fixed and $n\wedge p\to\infty$, any $\alpha$ guarantees that the first $K$ eigenvalues are excluded. The choice of $\alpha$ does not matter. We usually set $\alpha=0.2$, so that $60\%$ of the nonzero eigenvalues in the middle range are used.
Write for short $\tilde{p}=n\wedge p$. By definition, $\hat{\lambda}_k$ is the $(k/\tilde{p})$-upper-quantile of the ESD. Let $q_k=q_k(\gamma_n)$ denote the $(k/\tilde{p})$-upper-quantile of the MP distribution associated with $\gamma=\gamma_n$ and $\sigma^2=1$, that is,  
\beq \label{MPquantile}
\mbox{$q_k$ is the unique value such that } \int_{q_k}^{(1+\sqrt{\gamma_n})^2}f_{\gamma_n}(x; 1)dx = k/\tilde{p}. 
\eeq
These $q_k$'s can be easily computed (e.g., via the R package {\it RMTstat}). For an MP distribution with a general $\sigma^2$, its $(k/\tilde{p})$-upper-quantile equals to $\sigma^2 q_k$. Since the ESD is asymptotically close to the MP distribution, we expect that 
\[
\hat{\lambda}_k \approx \sigma^2\cdot q_k. 
\]
It motivates us to use $\{(q_k, \hat{\lambda}_k)\}_{\alpha\tilde{p}\leq k\leq (1-\alpha)\tilde{p}}$ to fit a line without intercept, and this can be done by a simple least-squares. The slope of this line is an estimator of $\sigma^2$.

\RestyleAlgo{boxed}
\begin{algorithm}[t!]
{\bf Algorithm 1}. BEMA for the standard spiked covariance model.
\vspace*{2pt}

{\it Input}: Nonzero eigenvalues $\hat{\lambda}_1,\ldots,\hat{\lambda}_{n\wedge p}$, $\alpha\in (0,1/2)$ and $\beta\in (0,1)$.\\
{\it Output}: An estimate of $K$.
\vspace*{5pt}

\noindent
Step 1: Write $\tilde{p}=n\wedge p$. For each $\alpha \tilde{p}\leq k\leq (1-\alpha)\tilde{p}$, obtain $q_k$, the $(k/\tilde{p})$-upper- quantile of the MP distribution associated with $\sigma^2=1$ and $\gamma_n=p/n$. Compute
 \[
   \hat{\sigma}^2 = \frac{\sum_{\alpha\tilde{p}\leq k\leq (1-\alpha)\tilde{p}}q_k\hat{\lambda}_k}{\sum_{\alpha\tilde{p}\leq k\leq (1-\alpha)\tilde{p}}q_k^2}. 
\]

\noindent
Step 2: Obtain $t_{1-\beta}$, the $(1-\beta)$-quantile of Tracy-Widom distribution. Estimate $K$ by
\[
\hat{K} =\#\bigl\{ 1\leq k\leq \tilde{p}:\;  \hat{\lambda}_k > \hat{\sigma}^2\bigl[ (1+\sqrt{\gamma}_n)^2 + t_{1-\beta} \cdot   n^{-\frac{2}{3}}\gamma_n^{-\frac{1}{6}}\bigl(1+\sqrt{\gamma_n}\bigr)^{\frac{4}{3}}  \bigr] \bigr\}. 
\] 
\end{algorithm}

Next, we use $\hat{\sigma}^2$ to determine a threshold for the top eigenvalues. A natural choice of threshold is $\hat{\sigma}^2(1+\sqrt{\gamma_n})^2$, but it has a considerable  probability of over-estimating $K$. We slightly increase this threshold by taking an advantage of another result in random matrix theory. When $\mu_K>\sigma^2 \sqrt{\gamma}$, it is known that \citep{Johnstone2001,bloemendal2016principal} 
\beq \label{TWlimit}
\frac{\hat{\lambda}_{K+1}- \sigma^2(1+\sqrt{\gamma_n})^2}{\sigma^2 n^{-\frac{2}{3}}\gamma_n^{-\frac{1}{6}}\bigl(1+\sqrt{\gamma_n}\bigr)^{\frac{4}{3}}}\quad \overset{d}{\to}\quad \mbox{type-I Tracy-Widom distribution}. 
\eeq
We propose thresholding the top eigenvalues at
\[
\hat{T}=\hat{\sigma}^2\Bigl[ (1+\sqrt{\gamma}_n)^2 + t_{1-\beta} \cdot   n^{-\frac{2}{3}}\gamma_n^{-\frac{1}{6}}\bigl(1+\sqrt{\gamma_n}\bigr)^{\frac{4}{3}}  \Bigr],
\]
where $t_{1-\beta}$ denotes the $(1-\beta)$-quantile of the Tracy-Widom distribution. Then, the probability of over-estimating $K$ is controlled by $\beta$.

Algorithm 1 has two tuning parameters $(\alpha, \beta)$. 
The output of the algorithm is insensitive to $\alpha$ if $\alpha$ is not too small, and we set $\alpha=0.2$ by default. $\beta$ controls the probability of over-estimating $K$ and is specified by the user. In theory, the ideal choice of $\beta$ should satisfy that $\beta\to 0$ at a properly slow rate (see Section~\ref{sec:theory}). In practice, choosing a moderate $\beta$ often yields the best finite-sample performance. Our numerical experiments suggest that $\beta=0.1$ is a good choice for most settings.

\begin{figure}[!htb]
\centering
\includegraphics[width=0.326\textwidth, trim=10 0 10 0, clip=true]{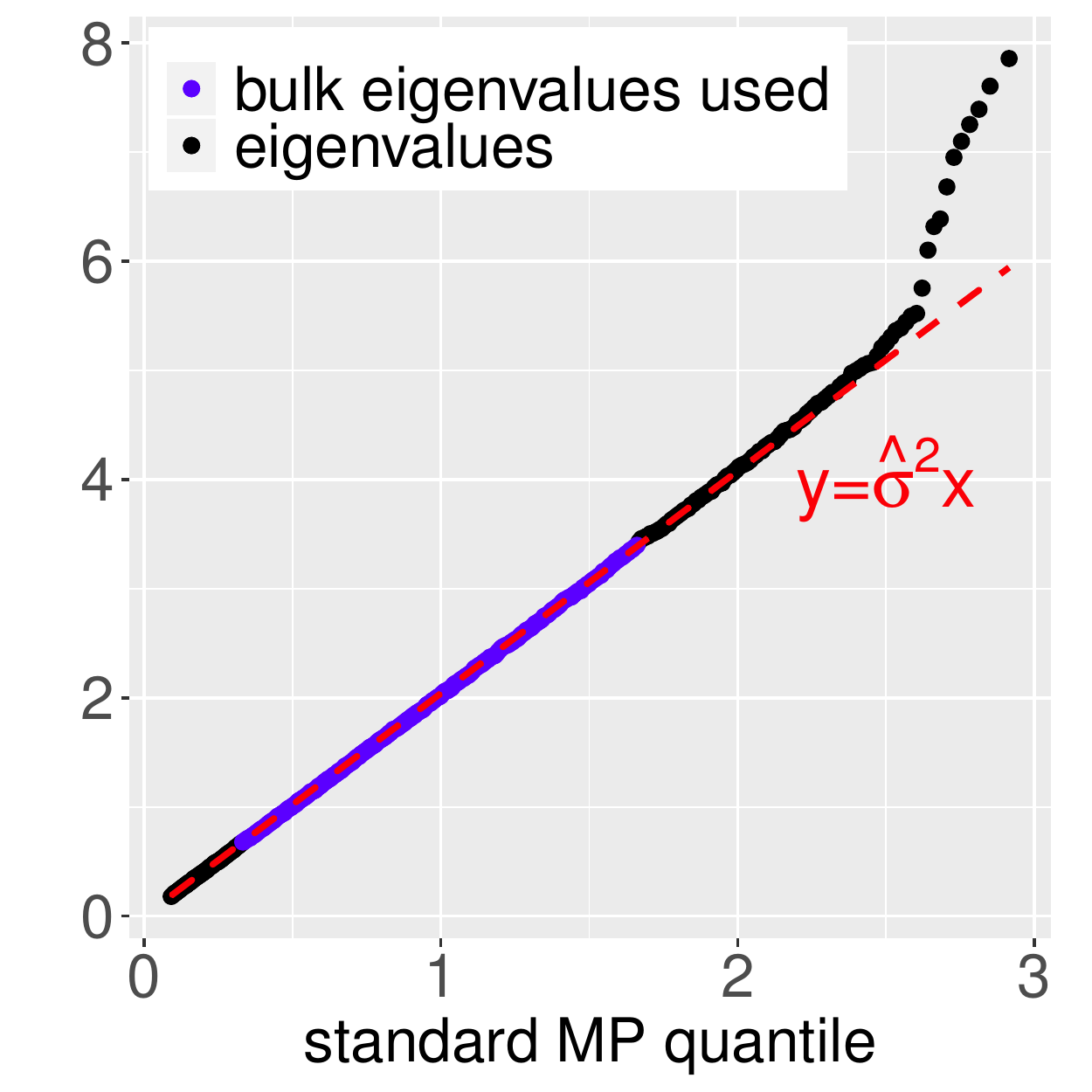}  
\includegraphics[width=0.326\textwidth, trim=10 0 10 0, clip=true]{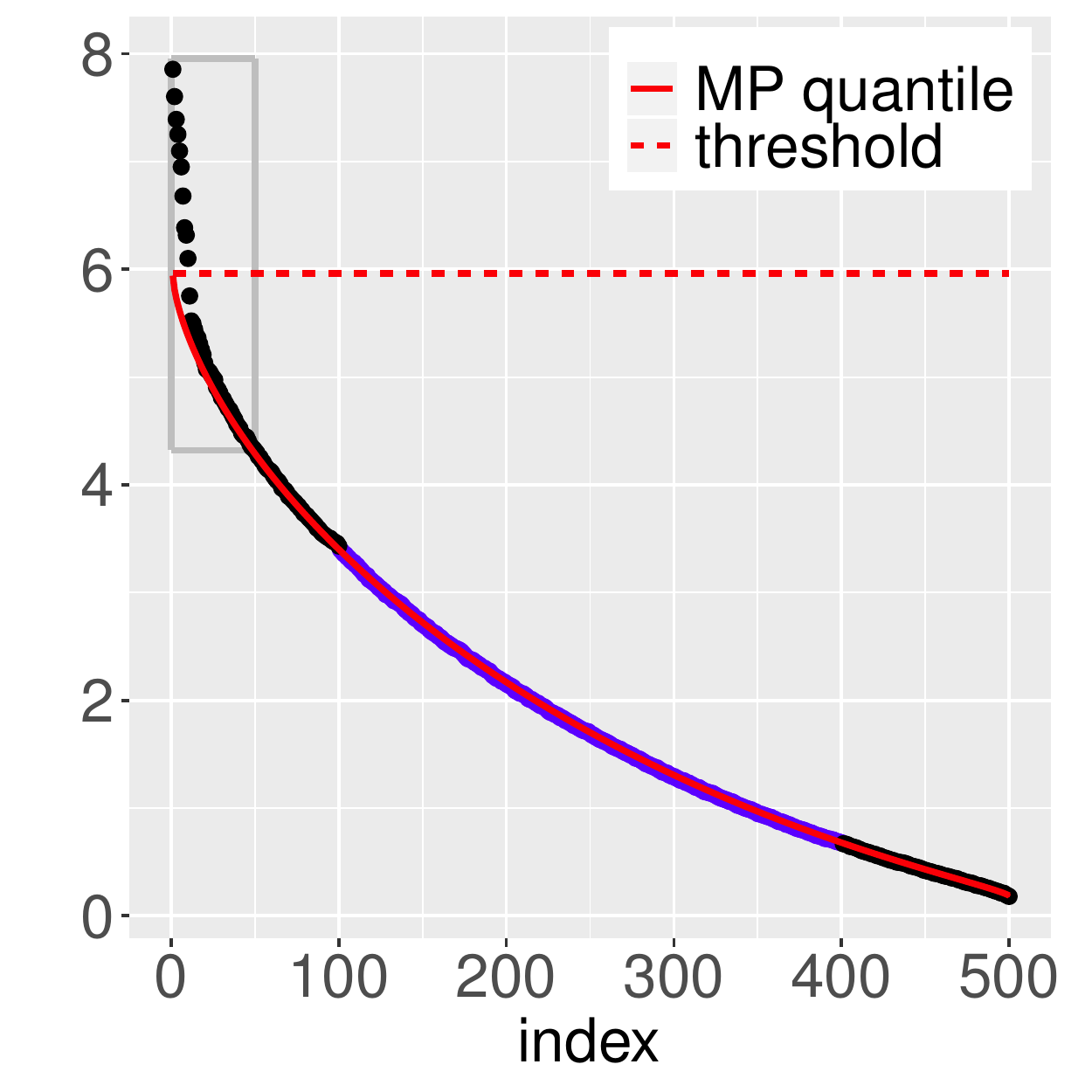} 
\includegraphics[width=0.323\textwidth, trim=10 0 10 0, clip=true]{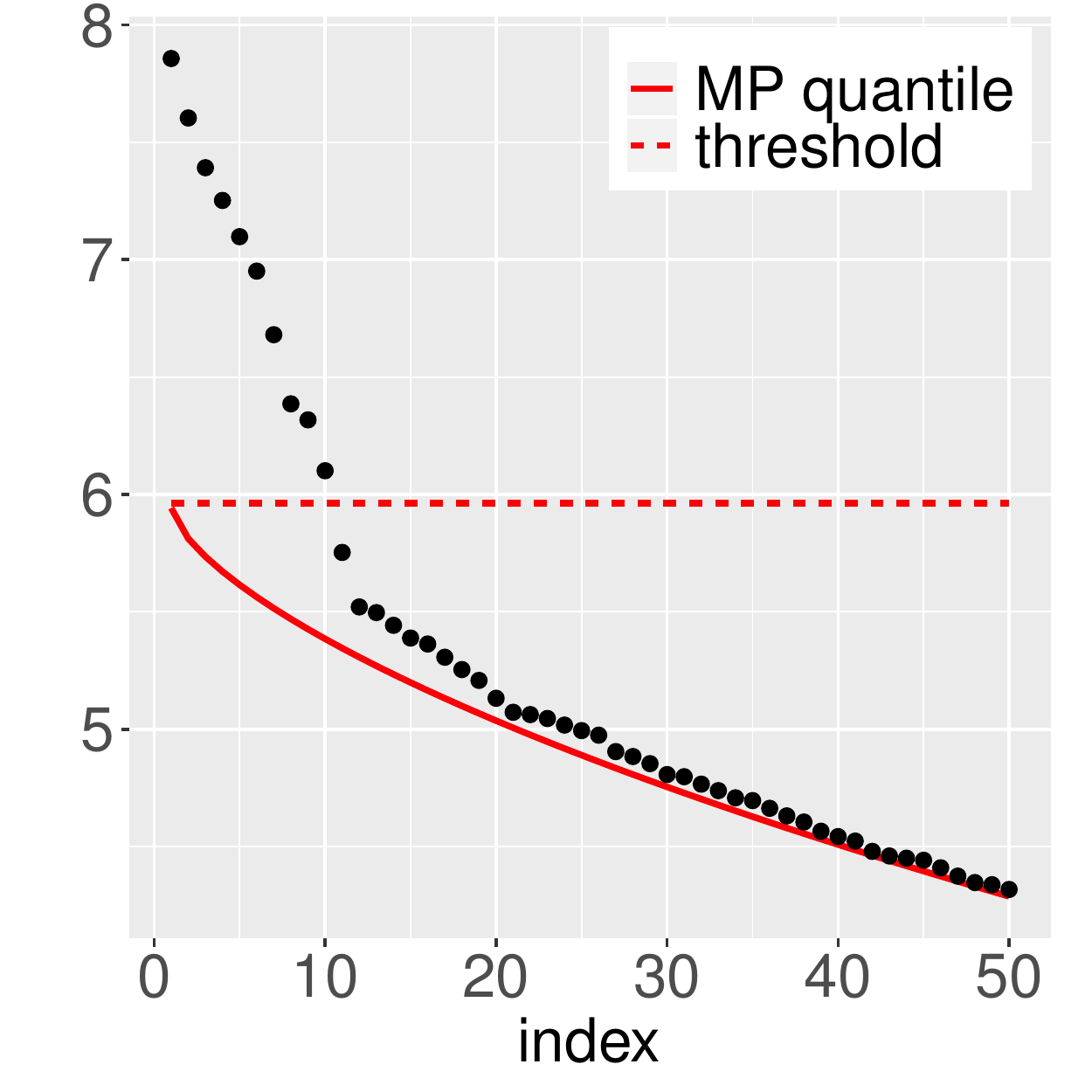}  
\caption{Illustration of BEMA for the standard spiked covariance model (simulated data, $n=1000$, $p=500$, $K=10$). The left panel plots $\hat{\lambda}_k$ versus $q_k$, where $q_k$ is the $(k/\tilde{p})$-upper-quantile of the standard MP distribution. The dashed red line is the fitted regression line on bulk eigenvalues (blue dots), whose slope is an estimate of $\sigma^2$. The middle panel plots $\hat{\lambda}_k$ versus $k$, which is the scree plot. The red solid curve is $\hat{\sigma}^2 q_k$ versus $k$. It fits the bulk eigenvalues (blue dots) very well. When this curve is extended to the left boundary, it hits $\hat{\sigma}^2(1+\sqrt{\gamma_n})^2$. Our threshold for the top eigenvalues, which is the $(1-\beta)$-quantile of the Tracy-Widom distribution, is slightly larger than this value and  shown by the dotted red line. The right panel zooms into the grey square area of the middle panel. It shows that 10 empirical eigenvalues exceeds the threshold, resulting in $\hat{K}=10$.} \label{fig:illustration}
\end{figure}

{\bf A simulation example}. We illustrate Algorithm 1 on a simulation example. Fix $(n, p, K)=(1000, 500, 10)$. We generate $\bX_i\overset{iid}{\sim}N(0, \bSigma)$, where $\bSigma$ is a diagonal matrix whose first $K$ diagonals equal to $5.4$ and the remaining diagonals equal to $\sigma^2=2$. In the left panel of Figure~\ref{fig:illustration}, we plot $\hat{\lambda}_k$ versus $q_k$. Except for a few top eigenvalues, it fits well to a straight line crossing the origin. We use 300 bulk eigenvalues $\{\hat{\lambda}_k\}_{100< k\leq 400}$ (the blue dots) to fit a regression line (the red dotted line). The slope of this line gives the estimate $\hat{\sigma}^2 = 2.04$. In the middle panel of Figure~\ref{fig:illustration}, we plot $\hat{\lambda}_k$ versus $k$. The red solid line is the curve of $\hat{\sigma}^2 q_k$ versus $k$. Although it is estimated using the blue dots only, we can extend this curve to the left boundary, which gives rise to the value $\hat{\sigma}^2(1+\sqrt{\gamma_n})^2$. We then use this value and the Tracy-Widom distribution to calculate a threshold for the top eigenvalues. The estimator $\hat{K}$ equals to the number of top eigenvalues that exceed this threshold. The right panel of Figure~\ref{fig:illustration} is a zoom-in of the middle panel. As $k$ gets smaller (e.g., $k<50$), the eigenvalues stay above the fitted MP quantile curve. This is because these $\hat{\lambda}_k$ are influenced by the spiked eigenvalues of $\bSigma$. Such eigenvalues are already excluded in the estimation of $\sigma^2$. The right panel can also be viewed as a scree plot. Finding the elbow point of the scree plot is a common ad-hoc method for estimating $K$. In this plot, the elbow points are $\{6, 7, 10, 11\}$, hard to decide the true $K$. In contrast, our method correctly picks $\hat{K}=10$.

{\bf Remark {\it (Connection to the robust estimation of $\sigma^2$)}}. As a byproduct, the BEMA algorithm yields a new estimator for $\sigma^2$ in the standard spiked covariance model, which can be useful for other problems such as  reconstruction of spiked covariance matrices. \cite{gavish2014optimal} proposed a robust estimator of $\sigma^2$, which is the ratio between the median of eigenvalues and the median of a standard MP distribution. Viewed in the Q-Q plot (left panel of Figure~\ref{fig:illustration}), their method is equivalent to using {\it a single point} to decide the slope. In comparison, our method uses a number of bulk eigenvalues to decide the slope and is thus more robust. \cite{kritchman2009non} proposed an estimator of $\sigma^2$ by solving a non-linear system of equations, and \cite{shabalin2013reconstruction} estimated $\sigma^2$ by minimizing the Kolmogorov-Smirnov distance between the ESD and its theoretical limit. In comparison, our estimator of $\sigma^2$ is from a simple least-squares and is much easier to compute. 
In Section~\ref{sec:theory}, we also give an explicit error bound for our estimator.

\section{BEMA for the general spiked covariance model} \label{sec:BEMA}

We now consider the general case where the residual covariance matrix can have unequal diagonal entries. We shall modify Algorithm 1 to accommodate this setting. 
Re-write Models~\eqref{mod-Sigma0}-\eqref{mod-D} as 
\beq \label{mod-Sigma2}
\bSigma=\sum_{k=1}^K \mu_k \bxi_k\bxi_k^{\top} + \mathrm{diag}(\sigma^2_1,\sigma_2^2, \ldots,\sigma^2_p), \qquad \mbox{where}\;\;  \sigma^2_k\overset{iid}{\sim}\mathrm{Gamma}(\theta,\, \theta/\sigma^2).  
\eeq
Same as before, let $\hat{\lambda}_1>\hat{\lambda}_2> \ldots>\hat{\lambda}_{n\wedge p}$ denote the nonzero eigenvalues of the sample covariance matrix. Below, in Section~\ref{subsec:main-RMT}, we first state some well-known results from random matrix theory and motivate our methodology idea. In Section~\ref{subsec:main-alg}, we formally introduce the BEMA algorithm. In Section~\ref{subsec:main-CI}, we provide an asymptotic confidence interval for $K$.

\subsection{The asymptotic behavior of empirical eigenvalues} \label{subsec:main-RMT}
Under Model~\eqref{mod-Sigma2}, the asymptotic behavior of bulk eigenvalues and top eigenvalues exhibit some similarity to the case of standard spiked covariance model: 
\begin{itemize}
\item The empirical spectral distribution (ESD) converges to a fixed limit. 
\item The first $K$ empirical eigenvalues stand out of the bulk. 
\end{itemize}
However, the precise statement is more sophisticated. 

We first consider the ESD. 
When $K$ is finite and $p/n\to\gamma$, the ESD converges to a distribution $F_{\gamma}(x;\sigma^2, \theta)$. This distribution is parametrized by $(\sigma^2,\theta)$, but it does not have an explicit form. It is defined implicitly by an equation of its Stieltjes transform \citep{marchenko1967distribution}. Let $H_{\sigma^2,\theta}(t)$ be the CDF of $\mathrm{Gamma}(\theta,\, \theta/\sigma^2)$.
For each $z\in\mathbb{C}^+$, there is a unique $m=m(z; \gamma,\sigma^2, \theta)\in \mathbb{C}^+$ such that
\begin{equation}\label{density1}
z=-\frac{1}{m}+\gamma\int \frac{t}{1+tm}dH_{\sigma^2,\theta}(t). 
\end{equation}
The density of $F_{\gamma}(x; \sigma^2, \theta)$, denoted by $f_{\gamma}(x; \sigma^2, \theta)$, satisfies that
\beq \label{density2}
 f_{\gamma}(x; \sigma^2, \theta) =   \lim_{y\to 0+} \biggl\{ \frac{1}{\pi(\gamma\wedge 1)}\, \Im\bigl( m(x+\mathrm{i}y; \gamma,\sigma^2, \theta)\bigr)\biggr\}, \footnote{The factor $1/(\gamma\wedge 1)$ is due to considering the zero-excluded ESD. If we consider the classical ESD, this factor should be $1/\gamma$.} 
\eeq
where $\Im(\cdot)$ denotes the imaginary part of a complex number. 

We aim to estimate $(\sigma^2,\theta)$ by comparing the bulk eigenvalues with the corresponding quantiles of $F_{\gamma}(x;\sigma^2,\theta)$. In the special case of $\theta= \infty$, $F_{\gamma}(x;\sigma^2, \theta)$ reduces to the MP distribution. Therefore, we can compute its quantiles explicitly and estimate $\sigma^2$ by a simple least-squares. For the general case, we have to compute the quantiles of $F_{\gamma}(x;\sigma^2,\theta)$ numerically. There are two approaches, one is solving the density from equations \eqref{density1}-\eqref{density2} and then computing the quantiles, and the other is using Monte Carlo simulations. We will describe them in Section~\ref{subsec:main-alg}. 

Next, we consider the top eigenvalues. It requires a precise definition of ``standing out" of the bulk. We use the distribution of $\hat{\lambda}_{K+1}$ under Model~\eqref{mod-Sigma2} as a benchmark, i.e., $\hat{\lambda}_k$ needs to be much larger than a high-probability upper bound of $\hat{\lambda}_{K+1}$ in order to be called ``standing out." Fortunately, the behavior of $\hat{\lambda}_{K+1}$ has been studied in the literature of random matrix theory. We define the following null model, which is a special case of Model~\eqref{mod-Sigma2} with $K=0$:
\beq \label{mod-null}
\bSigma= \mathrm{diag}(\sigma^2_1,\sigma_2^2, \ldots,\sigma^2_p), \qquad \mbox{where}\;\;  \sigma^2_k\overset{iid}{\sim}\mathrm{Gamma}(\theta,\, \theta/\sigma^2). 
\eeq 
Let $\hat{\lambda}_1^*$ denote the largest eigenvalue of the sample covariance matrix under this null model. By eigenvalue sticking result (see \cite{bloemendal2016principal}, \cite{ knowles2017anisotropic}  and a detailed discussion in Section~\ref{subsec:theory-remarks}), the distribution of $\hat{\lambda}_{K+1}$ is asymptotically close to the distribution of $\hat{\lambda}_1^*$. We now re-frame the statement that ``the first $K$ empirical eigenvalues stand out" as follows: 
Under some regularity conditions, 
each of $\hat{\lambda}_1,\ldots,\hat{\lambda}_K$ is significantly larger than $\hat{\lambda}_1^*$ associated with Model~\eqref{mod-null}.

We aim to threshold the top eigenvalues by the $(1-\beta)$-quantile of the distribution of $\hat{\lambda}_1^*$, where $\beta$ controls the probability of over-estimating $K$. In the special case of $\theta=\infty$, the distribution of $\hat{\lambda}_1^*$ converges to a Tracy-Widom distribution, so that we have a closed-form expression for the threshold. In the general case, we calculate this threshold by Monte Carlo simulation, where we simulate data from the null model to approximate the distribution of $\hat{\lambda}_1^*$. We relegate the details to Section~\ref{subsec:main-alg}.

\subsection{The algorithm of estimating $K$} \label{subsec:main-alg}
Same as before, the BEMA algorithm has two steps: Step 1 estimates $(\sigma^2,\theta)$ from bulk eigenvalues, and Step 2 calculates a threshold for the top eigenvalues. 

Consider Step 1. Write $\tilde{p}=p\wedge n$ and $\gamma_n=p/n$. For a constant $\alpha\in (0,1/2)$, we take the $(1-2\alpha)$-fraction of bulk eigenvalues in the middle range, i.e.,
$\{\hat{\lambda}_k: \alpha \tilde{p}\leq k\leq (1-\alpha)\tilde{p}\}$. Each empirical eigenvalue $\hat{\lambda}_k$ is also the $(k/\tilde{p})$-upper-quantile of the ESD. We recall that $F_{\gamma_n}(x;\sigma^2,\theta)$ is the theoretical limit of ESD as defined in \eqref{density1}-\eqref{density2}. Let $\bar{F}^{-1}_{{\gamma}_n}(k/\tilde{p}; \sigma^2,\theta)$ denote the $(k/\tilde{p})$-upper-quantile of this distribution. We expect to see
\[
\hat{\lambda}_k\approx \bar{F}^{-1}_{{\gamma}_n}(k/\tilde{p}; \sigma^2,\theta). 
\]
It motivates the following estimator of $(\sigma^2,\theta)$:
\beq \label{SQMgamma-estimate}
(\hat{\sigma}^2, \hat{\theta})=\mathrm{argmin}_{(\sigma^2, \theta)}\biggl\{ \sum_{\alpha\tilde{p}\leq k\leq (1-\alpha)\tilde{p}}\bigl[ \hat{\lambda}_k - \bar{F}^{-1}_{{\gamma}_n}(k/\tilde{p}; \sigma^2, \theta)\bigr]^2  \biggr\}. 
\eeq

We now describe how to solve \eqref{SQMgamma-estimate}. This is a two-dimensional optimization. As long as we can evaluate the objective function for arbitrary $(\sigma^2,\theta)$, we can solve it via a simple grid search. To further simplify the objective, we first get rid of $\sigma^2$ and reduce it to an optimization on $\theta$ only. 
Note that $\mathrm{Gamma}(\theta,\theta/\sigma^2)$ is equivalent to $\sigma^2\cdot \mathrm{Gamma}(\theta,\theta)$. We can deduce from \eqref{density1}-\eqref{density2} that a similar connection holds between $F_{\gamma_n}(x; \sigma^2, \theta)$ and $ F_{\gamma_n}(x;1, \theta)$. Then, their quantiles satisfy
\[
\bar{F}^{-1}_{\gamma_n}(k/\tilde{p}; \sigma^2,\theta)=\sigma^2\cdot \bar{F}^{-1}_{\gamma_n}(k/\tilde{p}; 1,\theta). 
\] 
We re-write \eqref{SQMgamma-estimate} as
\[
\min_{\theta} H(\theta), \qquad \mbox{where}\quad H(\theta)\equiv \min_{\sigma^2}\biggl\{ \sum_{\alpha\tilde{p}\leq k\leq (1-\alpha)\tilde{p}}\bigl[ \hat{\lambda}_k - \sigma^2 \bar{F}^{-1}_{{\gamma}_n}(k/\tilde{p}; 1, \theta)\bigr]^2  \biggr\}. 
\]
As long as we can compute $\bar{F}^{-1}_{{\gamma}_n}(y; 1,\theta)$ for any $\theta>0$ and $y\in [0,1]$, we can obtain $H(\theta)$ for each $\theta$ by least squares regression of the $\hat{\lambda}_k$'s on the $ \bar{F}^{-1}_{{\gamma}_n}(k/\tilde{p}; 1, \theta)$'s. Given $H(\theta)$, we can solve the optimization by a grid search on $\theta$.

This is described in Step 1 of Algorithm 2. Suppose there is an available algorithm \texttt{GetQT} that computes $\bar{F}^{-1}_{{\gamma}_n}(y; 1,\theta)$ for any $\theta>0$ and $y\in [0,1]$. Fix a set of grid points $\{\theta_j\}_{j=1}^N$. For each $\theta_j$, we first compute $\bar{F}^{-1}_{{\gamma}_n}(k/\tilde{p}; 1,\theta_j)$ for all $\alpha\tilde{p}\leq k\leq (1-\alpha)\tilde{p}$. Given $\theta_j$, the value of $\sigma^2$ that minimizes \eqref{SQMgamma-estimate} is obtained by regressing $\{\hat{\lambda}_k\}_{\alpha\tilde{p}\leq k\leq (1-\alpha)\tilde{p}}$ on $\{\bar{F}^{-1}_{{\gamma}_n}(k/\tilde{p}; 1,\theta_j)\}_{\alpha\tilde{p}\leq k\leq (1-\alpha)\tilde{p}}$ with a least-squares. Let $\hat{\sigma}^2(\theta_j)$ denote this optimal value of $\sigma^2$, and let $v_j$ denote the objective in \eqref{SQMgamma-estimate} associated with $\{\theta_j, \hat{\sigma}^2(\theta_j)\}$. We select $j^*$ so that $v_j$ is minimized and set $\hat{\theta}=\theta_{j^*}$ and $\hat{\sigma}^2=\hat{\sigma}^2(\theta_{j^*})$. 

What remains is the design of an algorithm \texttt{GetQT}($y,\gamma_n,\theta$) to compute the $y$-upper-quantile of the distribution $F_{\gamma_n}(\cdot; 1, \theta)$ for arbitrary $(\theta, y)$.
We note that $F_{\gamma_n}(x; 1,\theta)$ only has an implicit definition through equations \eqref{density1}-\eqref{density2}. In the appendix, we propose two algorithms that serve for this purpose: 
\begin{itemize}
\item \texttt{GetQT1} first utilizes the definition \eqref{density1}-\eqref{density2} to solve the density $f_{\gamma_n}(x;1,\theta)$ and then uses the density to compute  quantiles. \item \texttt{GetQT2} takes advantage of the fact that $F_{\gamma_n}(x; \sigma^2,\theta)$ is also the theoretical limit of the ESD of the null model \eqref{mod-null}. This algorithm simulates data from Model~\eqref{mod-null} with $\sigma^2=1$ to get the Monte Carlo approximation of the target quantile. 
\end{itemize} 
The two \texttt{GetQT} algorithms have comparable numerical performance, but each has an advantage on running time in some cases; see the appendix for more discussions.

\RestyleAlgo{boxed}
\begin{algorithm}[!tb]
{\bf Algorithm 2}. BEMA for the general spiked covariance model. 
\vspace*{2pt}

{\it Input}: Nonzero eigenvalues $\hat{\lambda}_1,\ldots,\hat{\lambda}_{n\wedge p}$, $\alpha\in (0,1/2)$, $\beta\in (0,1)$, a grid of values $0<\theta_1<\theta_2<\ldots<\theta_N$, 
an algorithm \texttt{GetQT}, and an integer $M\geq 1$.\\
\vspace*{2pt}
{\it Output}: An estimate of $K$.
\vspace*{5pt}

\noindent
Step 1: Write $\tilde{p}=n\wedge p$ and $\gamma_n=p/n$. For each $1\leq j\leq N$:
\vspace*{5pt}
\begin{itemize}
\item For each $\alpha \tilde{p}\leq k\leq (1-\alpha)\tilde{p}$, run the algorithm \texttt{GetQT}($k/\tilde{p}$, $\gamma_n, \theta_j$) to obtain $q_{kj}$.   
\item Compute $\hat{\sigma}^2(\theta_j)=(\sum_{\alpha\tilde{p}\leq k\leq (1-\alpha)\tilde{p}}q_{kj}\hat{\lambda}_k)/(\sum_{\alpha\tilde{p}\leq k\leq (1-\alpha)\tilde{p}}q_{kj}^2)$. 
\item Let $v_j=\sum_{\alpha\tilde{p}\leq k\leq (1-\alpha)\tilde{p}}[\hat{\lambda}_k-\hat{\sigma}^2(\theta_j)\cdot q_{kj}]^2$.
\vspace*{-5pt}
\end{itemize}
Find $j^*=\mathrm{argmin}_{1\leq j\leq N}v_j$. Let $\hat{\theta}=\theta_{j^*}$ and $\hat{\sigma}^2=\hat{\sigma}^2(\theta_{j^*})$. 

\vspace*{10pt}

\noindent
Step 2: For $1\leq m\leq M$:
\vspace*{-5pt}
\begin{itemize}
\item Sample $d^*_j\sim\mathrm{Gamma}(\hat{\theta},\hat{\theta})$, independently for $1\leq j\leq p$. 
Sample $X^*_i(j)\sim N(0, \hat{\sigma}^2d^*_j)$, independently for $1\leq i\leq n$ and $1\leq j\leq p$.  
\item 
Compute the largest singular value of $n^{-1/2}X^*$, where $X^*=[X_1^*, X_2^*, \ldots, X_n^*]^{\top}$. Let $\hat{\lambda}^{*}_{1(m)}$ be the square of this singular value.
\vspace*{-5pt}
\end{itemize}
Let $\hat{T}$ be the $(1-\beta)$-quantile of $\{\hat{\lambda}_{1(m)}^*\}_{1\leq m\leq M}$. Output $\hat{K} =\#\{ 1\leq k\leq \tilde{p}:\;  \hat{\lambda}_k > \hat{T}\}$. 

\end{algorithm}

Consider Step 2. We estimate $K$ by comparing each top eigenvalue with the $(1-\beta)$-quantile of the distribution of $\hat{\lambda}_1^*$ under the null model \eqref{mod-null}, with $(\hat{\sigma}^2,\hat{\theta})$ plugged in. The threshold is 
\beq \label{SQMgamma-threshold}
\hat{T} = \left\{\begin{array}{l}
\mbox{$(1-\beta)$-quantle of the distribution of $\hat{\lambda}_1^*$ under the null model}\\
\mbox{$\bSigma=\mathrm{diag}(\sigma^2_1,\sigma^2_2, \ldots,\sigma^2_p)$, where $\sigma^2_j\overset{iid}{\sim}\mathrm{Gamma}(\hat{\theta},\hat{\theta}/\hat{\sigma}^2)$}\\
\end{array}\right\}. 
\eeq
The $\hat{T}$ here generalizes the threshold in Algorithm 1. The threshold in Algorithm 1 is a special case of $\hat{T}$ at $\hat{\theta}=\infty$, which happens to have an explicit formula.

We compute $\hat{T}$ via Monte Carlo simulations. We first draw $\bSigma$ from the null model in \eqref{SQMgamma-threshold}, and then draw the data matrix from multivariate normal distributions and compute the largest eigenvalue of the sample covariance matrix. By repeating these steps multiple times, we obtain the sampling distribution of $\hat{\lambda}^*_1$ in \eqref{SQMgamma-threshold}. This is described in Step 2 of Algorithm 2.

The BEMA algorithm has three tuning parameters $(\alpha, \beta, M)$, where $\alpha$ controls the percentage of bulk eigenvalues used for estimating $(\sigma^2,\theta)$ and $M$ is the number of Monte Carlo repetitions for approximating $\hat{T}$. The performance of the algorithm is insensitive to $(\alpha, M)$ (see Section~\ref{sec:simulation}). We set $\alpha=0.2$ and $M=500$ by default. The parameter
$\beta$ controls the probability of over-estimating $K$. Theoretically, if the spiked eigenvalues are large enough, we should use a diminishing $\beta$ (i.e., $\beta\to 0$ as $n\to\infty$) so that the probability of over-estimating $K$ tends to zero. In practice, it often happens that the spiked eigenvalues are only moderately large. We thus need a moderate $\beta$ to strike a balance between the probability of over-estimating $K$ and the probability of under-estimating $K$. We leave it to the users to decide. It is analogous to the situation in false discovery rate control, where the users select the target false discovery rate. In our numerical experiments, we find that $\beta=0.1$ is a good choice.

\begin{figure}[!htb]
\centering
\includegraphics[width=0.325\textwidth, trim=20 0 10 0, clip=true]{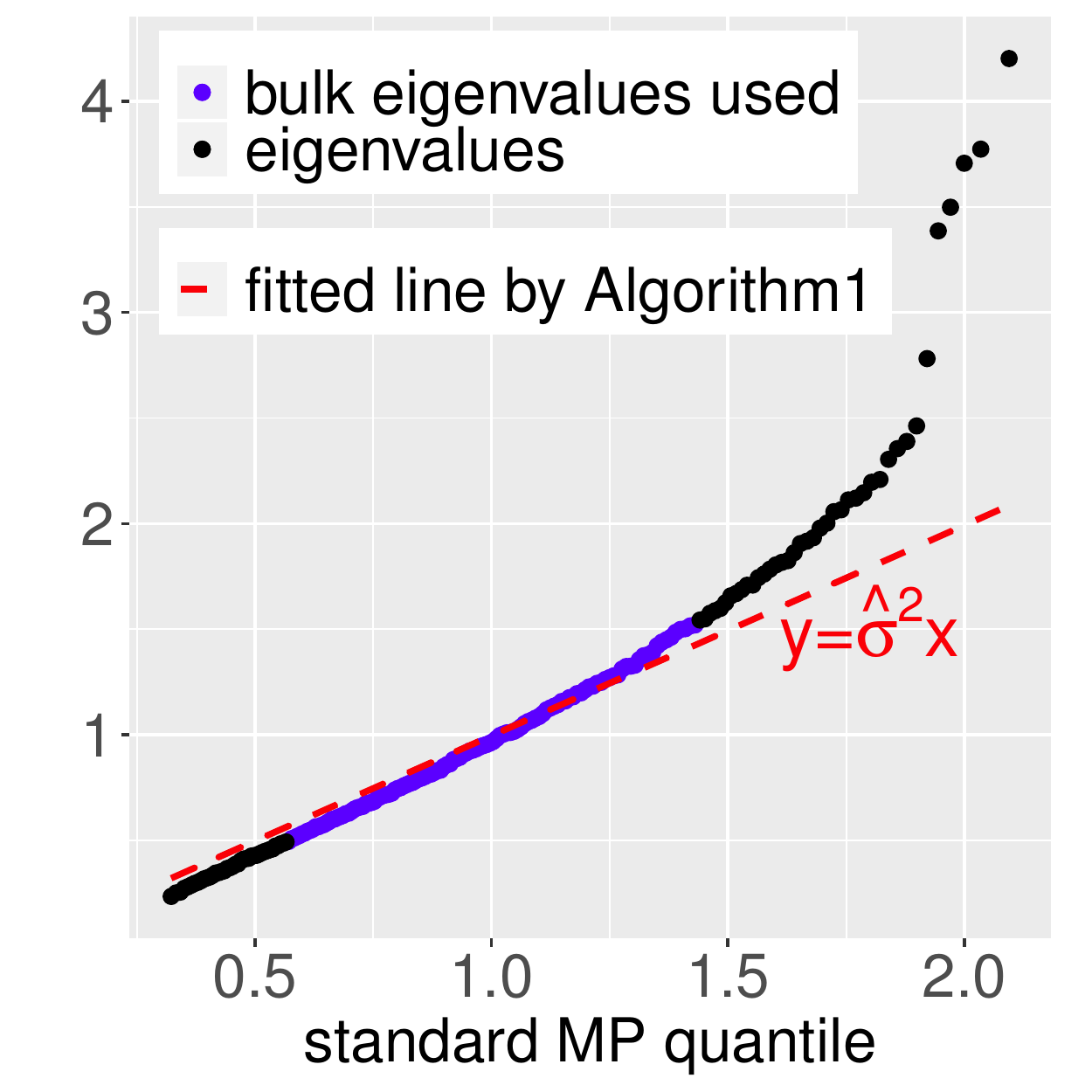}  
\includegraphics[width=0.332\textwidth, , trim=20 12 15 0, clip=true]{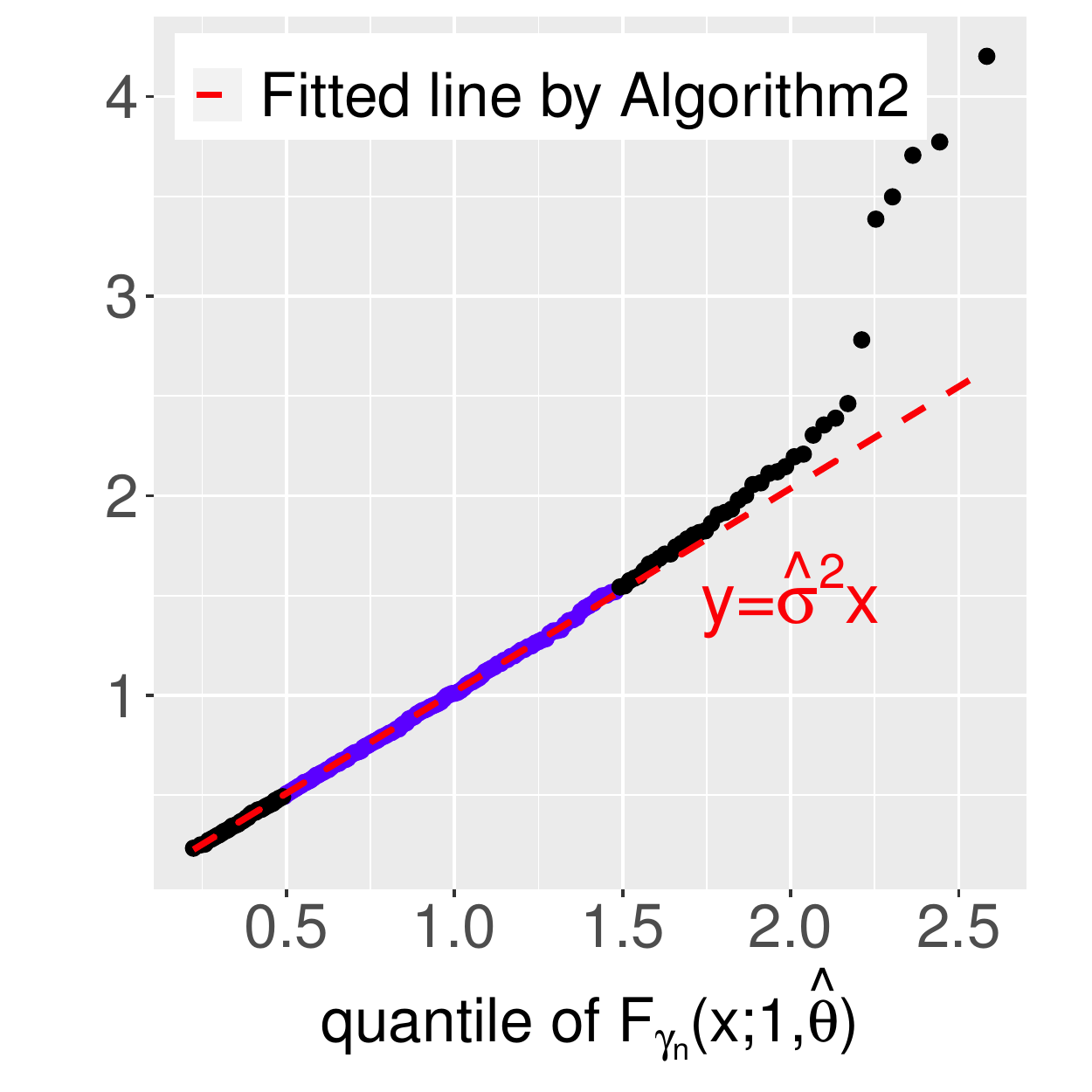} 
\includegraphics[width=0.323\textwidth, trim=15 0 15 0, clip=true]{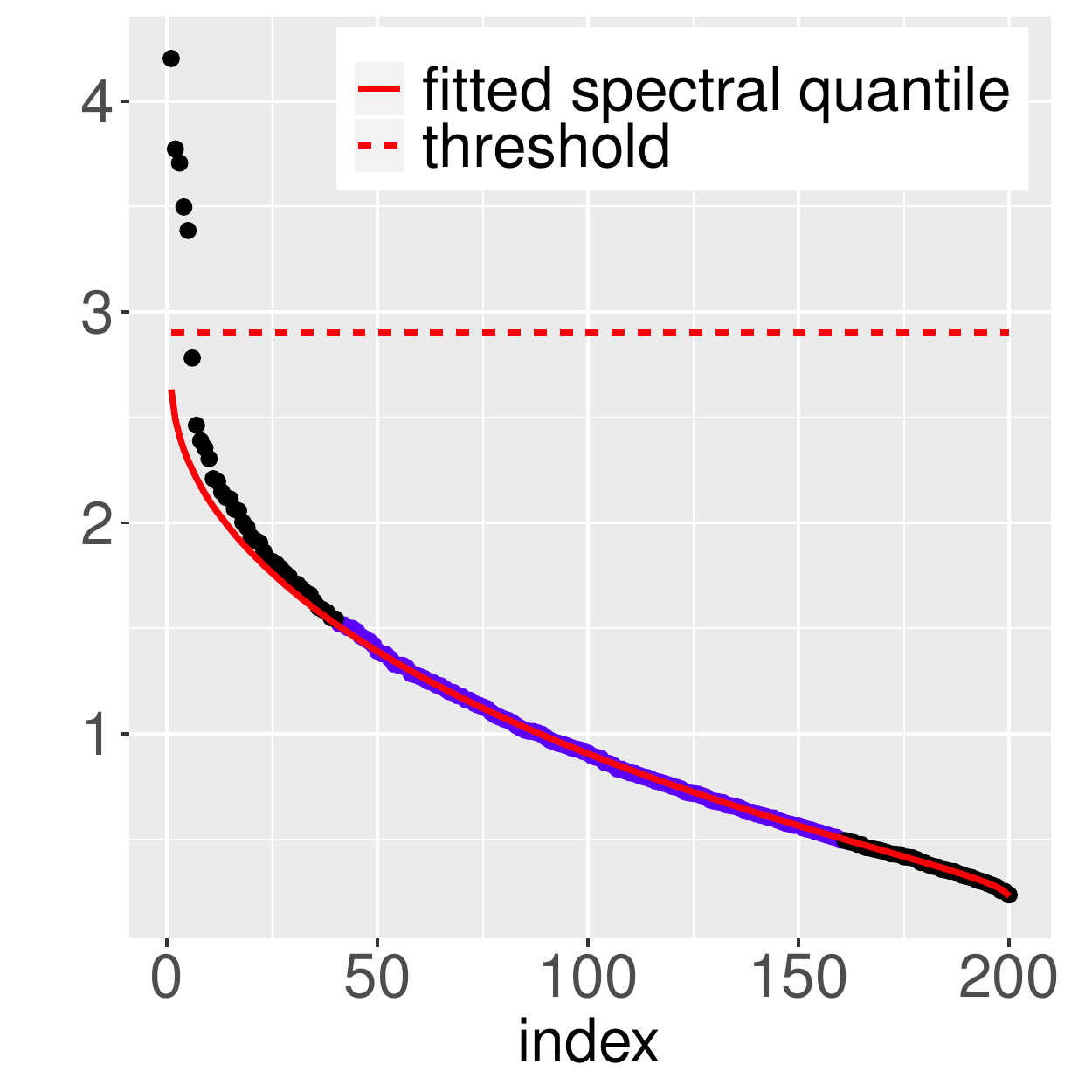}  
\caption{Illustration of BEMA for the general spiked covariance model. The left panel plots $\hat{\lambda}_k$ versus $q_k$, where the $q_k$'s are quantiles of the standard MP distribution. It fits the regression line poorly, suggesting that Algorithm 1 is no longer working for this general model. The middle panel plots $\hat{\lambda}_k$ versus $\bar{F}^{-1}_{\gamma_n}(x;1,\hat{\theta})$, where $\hat{\theta}$ is an estimate of $\theta$ by Algorithm 2. The bulk eigenvalues (blue dots) fit the regression line very well. The right panel is the scree plot, where the red solid curve is $\bar{F}^{-1}_{\gamma_n}(x;\hat{\sigma},\hat{\theta})$ versus $k$. A threshold (red dotted line) is given by the $90\%$-quantile of the distribution of $\hat{\lambda}_1^*$ from a null model; see \eqref{SQMgamma-threshold}. There are $5$ empirical eigenvalues exceeding this threshold, which gives $\hat{K}=5$.} \label{fig:illustration-gamma}
\end{figure}

{\bf A Simulation Example}. We illustrate Algorithm 2 using a simulation example. Fix $(n, p, K)=(1000, 200, 5)$ and $(\sigma^2, \theta)=(1, 10)$. We generate $\bX_i$ $iid$ from $N(0, \bSigma)$, where $\bSigma$ satisfies model \eqref{mod-Sigma2} with $\mu_k=2.3$ for $1\leq k\leq K$. The left panel of Figure~\ref{fig:illustration-gamma} shows the plot of $\hat{\lambda}_k$ versus the MP quantiles $q_k$. It does not fit a line crossing the origin, suggesting that Algorithm 1 does not work for this general covariance model. The middle panel contains the plot of $\hat{\lambda}_k$ versus $\bar{F}_{\gamma_n}^{-1}(k/\tilde{p}; 1,\hat{\theta})$, where $\hat{\theta}$ is from Algorithm 2. Except for a few top eigenvalues, it fits very well a line crossing the origin, suggesting that Algorithm 2 is successful in this setting. The estimated parameters are $(\hat{\sigma}^2, \hat{\theta})=(1.02, 10.39)$, which is close to the true values. The right panel contains the plot of $\hat{\lambda}_k$ versus $k$, and the fitted curve of $\hat{\sigma}^2\cdot \bar{F}_{\gamma_n}^{-1}(k/\tilde{p}; 1,\hat{\theta})$ versus $k$ (solid red line). The threshold $\hat{T}$ is also shown by the dashed line. It yields $\hat{K}=5$, which is the same as the ground truth. 

{\bf Remark {\it(Connection to parallel analysis)}}. Parallel analysis \citep{horn1965rationale} is a popular method for estimating the number of spiked eigenvalues in real applications. It samples data from a {\it null covariance model} that has no spiked eigenvalues, and estimates $K$ by comparing the distribution of top empirical eigenvalues on simulated data to those actually observed from the original data. The most common version of parallel analysis first standardizes the data matrix so that each variable has a unit variance and then uses $\bSigma=\bI_p$ as the null model. Our algorithm has a similar spirit as parallel analysis, but we adopt a more sophisticated null covariance model, Model~\eqref{mod-null}, and estimate parameters of this null model carefully from bulk eigenvalues.   

{\bf Remark {\it (Memory use of BEMA)}}. The input of BEMA includes nonzero eigenvalues of the sample covariance matrix. These eigenvalues can be computed by eigen-decomposition on either the $p\times p$ matrix $\bX^{\top}\bX$ or the $n\times n$ matrix $\bX\bX^{\top}$. 
Therefore, the memory use depends on the minimum of $n$ and $p$. 
In many real applications, $p$ is very large but $n$ is relatively small, and BEMA is still implementable under even strict memory constraints. 

\subsection{A confidence interval of $K$} \label{subsec:main-CI}
By varying $\beta$ in Algorithm 2, we get different estimators of $K$, where the over-shooting probability is controlled at different levels. We use these estimators to construct a confidence interval for $K$.  
\begin{definition}[Confidence interval of $K$]
Denote the output of Algorithm 2 by $\hat{K}_\beta$ to indicate its dependence on $\beta$.   Given any $\omega_0\in (0,1)$, we introduce the following $(1-\omega_0)$-confidence interval of $K$ as $[\hat{K}_{\omega_0/2},\; \hat{K}_{1-\omega_0/2}]$. 
\end{definition}

We explain why the confidence interval is asymptotically valid. Let $\hat{T}=\hat{T}_\beta$ be the threshold in \eqref{SQMgamma-threshold}, and let $\hat{\lambda}_1^*$ be the largest eigenvalue of the sample covariance matrix when data are from the null model \eqref{mod-null}. We use $\mathbb{P}_0$ to denote the probability measure associated with Model~\eqref{mod-null}. By definition of $\hat{T}_\beta$, $
\mathbb{P}_0\bigl\{\hat{\lambda}_1^* \leq t \bigr\}\Big|_{t=\hat{T}_\beta} = 1-\beta$. 
At the same time, the eigenvalue sticking result \citep{bloemendal2016principal, knowles2017anisotropic} states that, under some regularity conditions, the distribution of $\hat{\lambda}_{K+1}$ is asymptotically close to the distribution $\hat{\lambda}_1^*$. It follows that
\begin{align*}
\mathbb{P}\bigl\{\hat{K}_{\omega_0/2}>K\bigr\}&\leq \mathbb{P}\bigl\{\hat{\lambda}_{K+1} > \hat{T}_{\omega_0/2}\bigr\}\approx\mathbb{P}_0\bigl\{\hat{\lambda}^*_1> t\bigr\}\Big|_{t=\hat{T}_{\omega_0/2}}= \omega_0/2, \cr
\mathbb{P}\bigl\{\hat{K}_{1-\omega_0/2}<K\bigr\}&\leq \mathbb{P}\bigl\{\hat{\lambda}_{K} \leq \hat{T}_{1-\omega_0/2}\bigr\}\leq  \mathbb{P}\bigl\{\hat{\lambda}_{K+1} \leq \hat{T}_{1-\omega_0/2}\bigr\}\approx\mathbb{P}_0\bigl\{\hat{\lambda}^*_1\leq t\bigr\}\Big|_{t=\hat{T}_{1-\omega_0/2}}= \omega_0/2. 
\end{align*}

\section{Theoretical properties} \label{sec:theory}


We study in this section the theoretical properties of the proposed BEMA method. In Section~\ref{subsec:theory-standard}, we focus on the standard spiked covariance model ($\theta=\infty$), where we derive the error rate of $\hat{\sigma}^2$ and the consistency of $\hat{K}$. In Section~\ref{subsec:theory-general}, we study the general spiked covariance model ($\theta<\infty$). This setting is much more complicated.
It connects to an unsolved problem in random matrix theory, that is, how to get sharp asymptotic theory for eigenvalues when the limiting spectrum of $\bSigma$ is unbounded and has convex decay in the tail. Only partial results are known \citep{kwak2019extremal}.  To overcome the technical difficulty, in our theoretical investigation,  we approximate Model~\eqref{mod-Sigma2} by a proxy model where $\sigma^2_j$ are {\it iid} generated from a truncated Gamma distribution. Under this proxy model, we derive the rate of convergence for $(\hat{\sigma}^2, \hat{\theta})$ and the consistency of $\hat{K}$. In Section~\ref{subsec:theory-remarks}, we connect Model~\eqref{mod-Sigma2} to the proxy model and discuss the theory for Model~\eqref{mod-Sigma2}.


Through this section, we assume $\bX_1,\bX_2,\ldots,\bX_n$ are generated as follows: 
\begin{assumption} \label{cond:factorM}
Let $\bY=[\bY_1,\bY_2,\ldots,\bY_n]^{\top}\in\mathbb{R}^{n\times p}$ be a random matrix with independent but not necessarily identically distributed entries, where $\mathbb{E}[\bY_i(j)] = 0$ and $\mathrm{Var}(\bY_i(j))=1$, for $1\leq i\leq n$, $1\leq j\leq p$.
Given $\sigma_1,\sigma_2,\ldots,\sigma_p>0$, $\mu_1\geq \mu_2\geq\ldots\geq\mu_K>0$, and orthonormal vectors $\bxi_1,\bxi_2,\ldots,\bxi_p\in\mathbb{R}^p$, let $\bSigma=\sum_{k=1}^r(\sigma_k^2+\mu_k)\bxi_k\bxi_k^{\top}+\sum_{j=r+1}^p \sigma_j^2\bxi_j\bxi_j^{\top}$. We assume $\bX_i =\bSigma^{1/2}\bY_i$, for $1\leq i\leq n$. 
\end{assumption}



\noindent
Under this assumption, each $\bX_i$ is a linear transform of a random vector $\bY_i$ that has independent entries. This is stronger than assuming $\mathrm{Cov}(\bX_i)=\bSigma$ but is conventional in the literature.

\begin{assumption} \label{cond:moments}
For each integer $m\geq 1$, there exists a universal constant $C_m>0$ such that $\sup_{1\leq i\leq n,1\leq j\leq p}\mathbb{E}[|\bY_i(j)|^m]\leq C_m$.
\end{assumption}

\noindent
This assumption can be further relaxed. For example, we do not actually need the inequality to hold for every $m\geq 1$ but only for $1\leq m\leq M$, where $M$ is a properly large integer \citep{bloemendal2016principal, knowles2017anisotropic}. We use the current assumption for convenience. 
 
We will use the following notation frequently, which is conventional in random matrix theory: 

\begin{definition} \label{def:stochastic-dominance}
Let $U_n$ and $V_n$ be two sequences of random variables indexed by $n$. We say that  $U_n$ is stochastically dominated by $V_n$, if for any $\epsilon>0$ and $s>0$ there exists $N=N(\epsilon, s)$ such that $\mathbb{P}(U_n>n^{\epsilon}V_n)\leq n^{-s}$ for all $n\geq N$. We write $U_n\prec V_n$. 
\end{definition}

\subsection{The standard spiked covariance model} \label{subsec:theory-standard}
The standard spiked covariance model \citep{Johnstone2001} assumes $\bD=\sigma^2 \bI_p$. In this case, BEMA simplifies to Algorithm 1. It outputs $\hat{\sigma}^2$ and $\hat{K}$. We first give an error bound on estimating $\sigma^2$.
\begin{theorem}[Estimation error of $\hat{\sigma}^2$] \label{thm:sigma}
Suppose $\bX_1,\bX_2,\ldots,\bX_n$ satisfy Assumptions~\ref{cond:factorM}-\ref{cond:moments} with $\sigma_j^2\equiv \sigma^2$. Suppose $K\geq 1$ is fixed and $p/n\to \gamma$ for a constant $\gamma>0$. Let $\hat{\sigma}^2$ be the estimator of $\sigma^2$ by Algorithm 1, where the tuning parameter $\alpha$ is a constant in $(0,1/2)$. 
Then,  $|\hat{\sigma}^2-\sigma^2|\prec n^{-1}$. 
\end{theorem}
This result is connected to the robust estimation of $\sigma^2$ in a standard spiked covariance model  \citep{gavish2014optimal,kritchman2009non,shabalin2013reconstruction}. In these work, there are only consistency results available \citep{donoho2018optimal} which say that $\hat{\sigma}^2\to \sigma^2$ almost surely, but there are no explicit error rates. 
Using the recent advancement in random matrix theory on sharp large-deviation bounds for individual empirical eigenvalues (see \cite{ke2016detecting} for a survey), we can leverage those results to obtain an explicit bound for $|\hat{\sigma}^2-\sigma^2|$. 

We then establish the consistency on estimating $K$.
\begin{theorem}[Consistency of $\hat{K}$] \label{thm:K}
Suppose $\bX_1,\bX_2,\ldots,\bX_n$ satisfy Assumptions~\ref{cond:factorM}-\ref{cond:moments} with $\sigma_j^2\equiv \sigma^2$. Suppose $K\geq 1$ is fixed, $p/n\to \gamma\in (0,\infty)$, and $\mu_K\geq \sigma^2(\sqrt{\gamma}+\tau_n)$, where $\tau_n\gg n^{-1/3}$. Let $\hat{K}$ be the estimator of $K$ by Algorithm 1, where the tuning parameters are such that $\alpha\in (0,1/2)$ is a constant and that $\beta\to 0$ at a properly slow rate. As $n\to\infty$,  $\mathbb{P}\bigl\{  \hat{K}=K \bigr\}= 1- o(1)$.
\end{theorem}
We compare the conditions required for consistent estimation of $K$ with those in other work. 
Let $\lambda_1\geq\lambda_2\geq \ldots\geq \lambda_p$ denote the eigenvalues of $\bSigma$. In our model, $\lambda_k=\mu_k+\sigma^2$ for $1\leq k\leq K$. The condition in Theorem~\ref{thm:K}  translates to 
\[
\lambda_K>\sigma^2(1+\sqrt{\gamma}+\tau_n), \qquad \tau_n\gg n^{-1/3}.
\]
It is weaker than the conditions in \cite{bai2002determining} and \cite{cai2017limiting}, where the former requires $\lambda_K\asymp p$ 
and the latter needs $\lambda_K\to \infty$.
Our condition on $\lambda_K$ matches with the critical phase transition threshold in \cite{BAP} and is hardly improvable. In fact, \cite{fan2019estimating} showed that if $\lambda_K\leq \sigma^2 (1+\sqrt{\gamma})$ then there exists no consistent estimator of $K$. \cite{dobriban2019deterministic} impose the same condition on $\lambda_K$,  but they need stronger conditions on population eigenvectors. Their ``delocalization" condition states as $\|\bXi\bLambda^{1/2}\|_{\infty}\to 0$, where $\bXi=[\bxi_1,\ldots,\bxi_K]$, $\bLambda=\mathrm{diag}(\lambda_1,\ldots,\lambda_K)$, and $\|\cdot\|_\infty$ is the maximum absolute row sum. It requires the eigenvectors to be incoherent (i.e., $\max_{1\leq k\leq K}\|\bxi_k\|_\infty$ is sufficiently small) and that the eigenvalues cannot be too large. Examples such as equal-correlation matrices (i.e., $\bSigma(i,j)=a$, for all $i\neq j$, where $a\in (0,1)$ is a constant) are excluded. We do not need such a de-localization condition.\footnote{We remark that the comparison is for the standard spiked covariance model only. For this model, our method has the weakest conditions for consistent estimation of $K$. On the other hand, other methods apply to some other settings, which are not considered in the comparison.}



The proof of Theorem~\ref{thm:K} is an application of the {\it eigenvalue sticking} theory  \citep{bloemendal2016principal}. It compares the distribution of empirical eigenvalues $\{\hat{\lambda}_k\}$ under the spiked covariance model with the distribution of empirical eigenvalues $\{\hat{\lambda}^*_k\}$ under the null model $\bSigma=\sigma^2 \bI_p$. The claim is that the distribution of $\hat{\lambda}_{K+s}$ is asymptotically close to the distribution of $\hat{\lambda}^*_s$, for a wide range of $s$. We use this result to study the thresholding step in Algorithm 1.

\subsection{The truncated Gamma-based general spiked covariance model} \label{subsec:theory-general}
The general spiked covariance model ~\eqref{mod-Sigma2} assumes $\sigma^2_j$ are {\it iid} drawn from $\mathrm{Gamma}(\theta,\theta/\sigma^2)$. It differs from the conventional settings in random matrix theory because $\bSigma$ is not a deterministic matrix and because the limiting spectral density of $\bSigma$ does not have a compact support. Unfortunately, there is no existed random matrix theory that deals with this setting directly \citep{Bao-communication}. We thus approximate Model~\eqref{mod-D} by 
\beq \label{mod-D-truncated}
\sigma_j^2\; \, \overset{iid}{\sim}\;\, \mathrm{TruncGamma}(\theta,\, \theta/\sigma^2,\, \sigma^2 T_1, \, \sigma^2\, T_2), \qquad 1\leq j\leq p, 
\eeq
where $\mathrm{TruncGamma}(\alpha,\beta, l, u)$ denotes the truncated Gamma distribution with rate and shape parameters $\alpha$ and $\beta$ and truncations at $l$ and $u$. 
When $(T_1, T_2)=(0, \infty)$, it reduces to Model~\eqref{mod-D}. Given fixed $0<T_1<T_2<\infty$, the limiting spectral density of $\bSigma$ has a compact support, so that we can take advantage of the existing random matrix theory \citep{knowles2017anisotropic,ding2020spiked}. 
We first present the theory for Model~\eqref{mod-D-truncated} and then discuss how to extend it to $(T_1, T_2)=(0,\infty)$. 

Fixing $0<T_1<T_2<\infty$ and two intervals ${\cal J}_{\sigma^2}=[a, b]\subset(0,\infty)$ and ${\cal J}_\theta=[c, d]\in (0,\infty)$, let ${\cal Q}(T_1, T_2, {\cal J}_{\sigma^2}, {\cal J}_\theta)$ be the family of distributions $\mathrm{TruncGamma}(\theta,\theta/\sigma^2,\sigma^2T_1,\sigma^2T_2)$ satisfying that $\sigma^2\in {\cal J}_{\sigma^2}$ and $\theta\in {\cal J}_{\theta}$. The following Lemma is a result of Theorem 3.12 and Example 2.9 in \cite{knowles2017anisotropic}, and its proof is omitted.  

\begin{lemma} \label{lem:eigenvalue-rigidity}
Suppose $\bX_1,\bX_2,\ldots,\bX_n$ satisfy Assumptions~\ref{cond:factorM}-\ref{cond:moments} with $\sigma_j^2$ generated from Model~\eqref{mod-D-truncated}. 
Suppose $K\geq 1$ is fixed and $p/n\to\gamma$ for a constant $\gamma \neq 1$. 
Suppose the truncated Gamma distribution in \eqref{mod-D-truncated} is from the family ${\cal Q}(T_1, T_2, {\cal J}_{\sigma^2}, {\cal J}_\theta)$, for fixed $(T_1, T_2, {\cal J}_{\sigma^2}, {\cal J}_\theta)$. Let $H_{\sigma^2, \theta, T_1, T_2}(t)$ be the CDF of $\mathrm{TruncGamma}(\theta,\theta/\sigma^2,\sigma^2T_1,\sigma^2T_2)$. Define a distribution $F_{\gamma_n}(\cdot; \sigma^2,\theta,T_1,T_2)$ in the same way as in  \eqref{density1}-\eqref{density2}, with $H_{\sigma^2,\theta}(t)$ replaced by  $H_{\sigma^2,\theta,T_1,T_2}(t)$ and $\gamma$ replaced by $\gamma_n=p/n$. Let $q_i\equiv \bar{F}^{-1}_{\gamma_n}(i/\tilde{p};\sigma^2, \theta,T_1,T_2)$ be the $(i/\tilde{p})$-upper-quantile of this distribution, where $\tilde{p}=n\wedge p$. As $n\to\infty$, for every $K< i\leq \tilde{p}$, we have $|\hat{\lambda}_i-q_i|\prec [i\wedge (\tilde{p}+1-i)]^{-1/3} n^{-2/3}$. 
\end{lemma}

Given $(T_1, T_2)$, we estimate $\sigma^2$ and $\theta$ by 
\beq \label{generalM-estimate}
(\hat{\sigma}^2, \hat{\theta})=\mathrm{argmin}_{(\sigma^2, \theta)\in {\cal J}_{\sigma^2}\times {\cal J}_{\theta}}\biggl\{ \sum_{\alpha\tilde{p}\leq i\leq (1-\alpha)\tilde{p}}\bigl[ \hat{\lambda}_i - \bar{F}^{-1}_{{\gamma}_n}(i/\tilde{p}; \sigma^2, \theta, T_1, T_2)\bigr]^2  \biggr\}. 
\eeq
It can be solved by a slight modification of Step 1 of Algorithm 2. We note that \eqref{mod-D-truncated} is equivalent to $\sigma_j^2/\sigma^2\overset{iid}{\sim}\mathrm{TruncGamma}(\theta,\theta,T_1, T_2)$. Hence, the quantiles satisfy that $\bar{F}^{-1}_{{\gamma}_n}(i/\tilde{p}; \sigma^2, \theta, T_1, T_2)=\sigma^2\cdot \bar{F}^{-1}_{{\gamma}_n}(i/\tilde{p}; 1, \theta, T_1, T_2)$. We first modify \texttt{GetQT} so that it outputs the quantiles of $F_{\gamma_n}(\cdot;1,\theta,T_1,T_2)$ for any given $\theta$. Next, we mimic Step 1 of Algorithm 2 to solve \eqref{generalM-estimate}, where we run a least-squares for every $\theta$ and then optimize over $\theta$ via a grid search. The details are relegated to the Appendix. 

\begin{theorem}[Estimation error of $\hat{\sigma}^2$ and $\hat{\theta}$] \label{thm:generalM-error}
Suppose the conditions of Lemma~\ref{lem:eigenvalue-rigidity} hold, where $K$, $\gamma$, $T_1$, $T_2$, ${\cal J}_{\sigma^2}$, and ${\cal J}_{\theta}$ are fixed. Let 
\[
\Phi(\theta)=\Phi(\theta;T_1,T_2)=\frac{\bigl[\int_{T_1}^{T_2}x^{\theta+1}exp(-\theta x)dx\bigr]\bigl[\int_{T_1}^{T_2}x^{\theta-1}exp(-\theta x)dx\bigr]}{\bigl[\int_{T_1}^{T_2}x^{\theta}exp(-\theta x)dx\bigr]^2}.
\]
Suppose there exists a constant $\omega=\omega(T_1, T_2, {\cal J}_{\theta})$ such that $\sup_{\theta\in {\cal J}_\theta}\Phi'(\theta)\leq - \omega$. Let $\hat{\sigma}^2$ and $\hat{\theta}$ be the estimators from \eqref{generalM-estimate}, where the tuning parameter $\alpha$ satisfies $\alpha\tilde{p}>K$ and $\alpha\tilde{p}=O(n/\log(n))$. As $n\to\infty$, we have $|\hat{\sigma}^2-\sigma^2|\prec n^{-1}$ and $|\hat{\theta}-\theta|\prec n^{-1}$. 
\end{theorem}

Theorem~\ref{thm:generalM-error} assumes $\sup_{\theta\in {\cal J}_\theta}\Phi'(\theta)\leq - \omega$ for some constant $\omega>0$. It is a regularity condition on $({\cal J}_\theta, T_1, T_2)$. The next lemma shows that this condition is mild.

\begin{lemma} \label{lem:sensitivity-to-theta}
For any fixed ${\cal J}_{\theta}=[c,d]$ and $\omega<d^{-2}$, there exist constants $0<T_1^*<T_2^*<\infty$ such that $\sup_{\theta\in {\cal J}_\theta}\Phi'(\theta; T_1, T_2)\leq -\omega$ holds for all $T_1\leq T_1^*$ and $T_2\geq T_2^*$. 
\end{lemma}

With the estimates $\hat{\sigma}^2$ and $\hat{\theta}$, we then slightly modify Step 2 of Algorithm 2 by thresholding all the empirical eigenvalues at 
\beq \label{generalM-threshold}
\hat{T}_\beta = \left\{\begin{array}{l}
\mbox{$(1-\beta)$-quantle of the distribution of $\hat{\lambda}_1^*$ under the null model}\\
\mbox{$\bSigma=\mathrm{diag}(\sigma^2_1,\ldots,\sigma^2_p)$, where $\sigma^2_j\overset{iid}{\sim}\mathrm{TruncGamma}(\hat{\theta},\hat{\theta}/\hat{\sigma}^2, \hat{\sigma}^2T_1,\hat{\sigma}^2T_2)$}\\
\end{array}\right\}. 
\eeq
This threshold can be computed via Monte Carlo simulations, similarly as in Step 2 of Algorithm 2. We estimate $K$ by the number of empirical eigenvalues exceeding $\hat{T}$.

To establish the consistency of $\hat{K}$, we introduce the function 
\beq \label{define-right-edge}
G(x) =-\frac{1}{x} + \gamma\int\frac{1}{t^{-1}+x}dH_{\sigma^2,\theta,T_1,T_2}(t).
\eeq
By Example 2.9 of \cite{knowles2017anisotropic}, $G(x)$ has 2 critical points $0>x_1^*>x_2^*$ (the definition of critical points can be found in \cite{knowles2017anisotropic}), and the distribution $F_{\gamma}(\cdot; \sigma^2,\theta,T_1,T_2)$ defined in Lemma~\ref{lem:eigenvalue-rigidity} has the support $[G(x^*_2), G(x^*_1)]$. 
The next theorem is proved in the appendix. It uses a result in \cite{ding2020spiked} about the top empirical eigenvalues. 

\begin{theorem}[Consistency of $\hat{K}$] \label{thm:generalM-K}
Suppose the conditions of Lemma~\ref{lem:eigenvalue-rigidity} and Theorem~\ref{thm:generalM-error} hold. Let $x_1^*$ be the largest critical point of the function $G(x)$ in \eqref{define-right-edge}. We assume $-1/(T_1+\mu_K)\geq x_1^*+\tau$,\footnote{In our model (see Assumption~\ref{cond:factorM}), the spiked eigenvalues of $\bSigma$ are $\{\mu_k+\sigma_k\}_{1\leq k\leq K}$. Therefore, $\mu_K+T_1$ is a lower bound of these spiked eigenvalues.} where $\tau>0$ is a constant and $T_1$ is a truncation point in \eqref{mod-D-truncated}. Let $\hat{K}=\#\{1\leq i\leq (n\wedge p): \hat{\lambda}_i>\hat{T}_\beta\}$, where $\hat{T}_\beta$ is as in \eqref{generalM-threshold} with $\beta\to 0$ at a properly slow rate. As $n\to\infty$,  $\mathbb{P}\bigl\{  \hat{K}=K \bigr\}= 1- o(1)$.
\end{theorem}


\subsection{Remarks on extension to the  Gamma-based general spiked covariance matrix}  \label{subsec:theory-remarks}
We now discuss extension of the theoretical results to the Gamma-based  general spiked covariance model~\eqref{mod-Sigma2}, which is an extreme case of Model~\eqref{mod-D-truncated} at $T_1=0$ and $T_2= \infty$. As mentioned earlier, this setting is unconventional because the eigenvalues of $\bSigma$ are stochastic and the support of the limiting spectral density of $\bSigma$ is unbounded. 

First, we discuss the estimation of $(\sigma^2,\theta)$. The accuracy of $(\hat{\sigma}^2, \hat{\theta})$ depends on whether we have similar large deviation bounds to those in Lemma~\ref{lem:eigenvalue-rigidity}. Our conjecture is that the stochasticity and unboundedness of the spectrum of $\bSigma$ has a negligible effect on the eigenvalues deep into the bulk. To see why, we note that the classical result about weak convergence of ESD \citep{marchenko1967distribution} does not need the limiting spectrum of $\bSigma$ to have a compact support; therefore, the unboundedness is not an issue. The stochasticity is not an issue, either, because  
almost surely, the spectral distribution of $\bSigma$ converges weakly to $\mathrm{Gamma}(\theta,\theta/\sigma^2)$. We conclude that the weak convergence of ESD still holds. This further implies that the bulk eigenvalues still converge to the corresponding quantiles of the theoretical limit of ESD. 

The open question is whether we have the rates of convergence as in Lemma~\ref{lem:eigenvalue-rigidity}. The stochasticity and unboundedness of the spectrum of $\bSigma$ affect the rates of convergence of large eigenvalues. We thus do not expect Lemma~\ref{lem:eigenvalue-rigidity} to hold for all $i$. Fortunately, the estimation of $(\sigma^2, \theta)$ in BEMA only involves bulk eigenvalues in the middle range, i.e.,  $\alpha \tilde{p}\leq i\leq (1-\alpha)\tilde{p}$, where $\alpha\in (0,1/2)$ is a constant. We conjecture that Lemma~\ref{lem:eigenvalue-rigidity} continues to hold for these eigenvalues. If our conjecture is correct, then we can show similar results for $\hat{\sigma}^2$ and $\hat{\theta}$ as those in Theorem~\ref{thm:generalM-error}.

Next, we discuss the consistency of $\hat{K}$. The stochasticity and unboundedness of the spectrum of $\bSigma$ together yields a significant change of the behavior of edge eigenvalues. This can be seen from a relevant setting in \cite{kwak2019extremal}--- $\bSigma$ is a diagonal matrix whose diagonal entries are {\it iid} drawn from a density $\rho(t)\propto (1-t)^bf(t)\cdot 1\{l\leq t\leq 1\}$, where $b>1$ and $l\in (0,1)$ are constants and $f\in C^1([l,1])$. This setting has no spike. They showed that the limiting distribution of the largest eigenvalue, $\hat{\lambda}_1^*$, is not a Tracy-Widom distribution; it is a Weibull distribution if $\gamma<\gamma_0$ and a Gaussian distribution if $\gamma>\gamma_0$, where $\gamma_0$ is a positive constant. Our model is even more complicated, where the Gamma density exhibits a similar convex decay on the right tail but has an unbounded support.  We do not expect $\hat{\lambda}_1^*$ to follow a Tracy-Widom distribution any more. 

However, this does not eliminate the consistency of $\hat{K}$. To prove consistency, we first need that the stochastic threshold \eqref{SQMgamma-threshold}  in BEMA well approximates the $(1-\alpha)$-upper-quantile of $\hat{\lambda}_1^*$, where $\hat{\lambda}_1^*$  is the largest eigenvalue of the null model with no spike. This follows from the nature of Monte Carlo simulations, no matter whether $\hat{\lambda}_1^*$ converges to a Tracy-Widom distribution. Furthermore, the implementation of \eqref{SQMgamma-threshold} does not need any knowledge of the limiting distribution of $\hat{\lambda}_1^*$. 

To prove consistency, we also need to show that, under Model~\eqref{mod-Sigma2}, when $\mu_K$ is appropriately large, (i) the distribution of $\hat{\lambda}_{K+1}$ is asymptotically close to the distribution of $\hat{\lambda}_1^*$ in the null model (this is the ``eigenvalue sticking" argument), and (ii) each of $\hat{\lambda}_1, \hat{\lambda}_2, \ldots, \hat{\lambda}_K$ is significantly larger than the $(1-\alpha)$-upper-quantile of $\hat{\lambda}_1^*$. We conjecture that both (i)-(ii) are correct, provided that $\mu_K\gg\log(n)$. If our conjectures are correct, then we can obtain the consistency of $\hat{K}$ as in Theorem~\ref{thm:generalM-K}, under the slightly stronger condition that $\mu_K\gg \log(n)$.


The rigorous proofs of our conjectures require re-development of several fundamental results in random matrix theory for Model~\eqref{mod-Sigma2}, such as the local law on bulk eigenvalues and the limiting behavior of edge eigenvalues (including the spiked and non-spiked ones). It is beyond the scope of this paper, and we leave for future work.  

\section{Simulation studies} \label{sec:simulation}
We examine the performance of our methods in simulations. To differentiate between Algorithm~1 and Algorithm~2, we call the former BEMA0 and the latter BEMA. BEMA0 is a simplified version of BEMA, specifically designed for the standard spiked covariance model. The tuning parameters are fixed as $(\alpha, \beta)=(0.2, 0.1)$ for BEMA0 and $(\alpha, \beta, M)=(0.2, 0.1, 500)$ for BEMA when not particularly specified.

In Section~\ref{subsec:theory-general}, we also introduced a modification of BEMA using the truncated Gamma-based spiked mode for technical needs in our theoretical studies. We showed that this algorithm has desirable theoretical properties. It however requires two additional tuning parameters $(T_1, T_2)$. Our simulation studies  (not reported here) show that  the performance of the modified BEMA is similar to that of BEMA, when $T_1$ is appropriately small and $T_2$ is appropriately large. For this reason, we use BEMA, instead of the modified BEMA, in the following simulation studies.

We compare our methods with a few methods in the literature, including the deterministic parallel analysis (DDPA) from  \cite{dobriban2019deterministic}, the empirical Kaiser's criterion (EKC) from \cite{braeken2017empirical}, the information criteria $IC_{p1}$ (Bai$\&$Ng) from \cite{bai2002determining} and the eigen-gap detection (Pass\&Yao) from \cite{passemier2014estimation}.

\paragraph{Simulation 1.} 
This experiment is for the standard spiked covariance model, where we investigate the performance of BEMA0 and the confidence interval for $K$ as described in Section~\ref{subsec:main-CI}. We generate data from $\bX_i\overset{iid}{\sim}N(0, \bSigma)$, $1\leq i\leq n$, where $\bSigma$ satisfies Model~\eqref{mod-Sigma} with 
\[
\mu_1=\mu_2=\cdots=\mu_K= \rho\cdot \sigma^2\sqrt{p/n}, \qquad \mbox{for some }\rho>0. 
\] 
The value of $\rho$ controls the magnitude of spiked eigenvalues. $\rho\leq 1$ is the region where consistent estimation of $K$ is impossible \citep{BAP, fan2019estimating}. 
We examine the performance of BEMA0 in the region of $\rho>1$. 


\begin{figure}[!tb]
\centering
\hspace*{5pt} 
\includegraphics[width=0.296\textwidth]{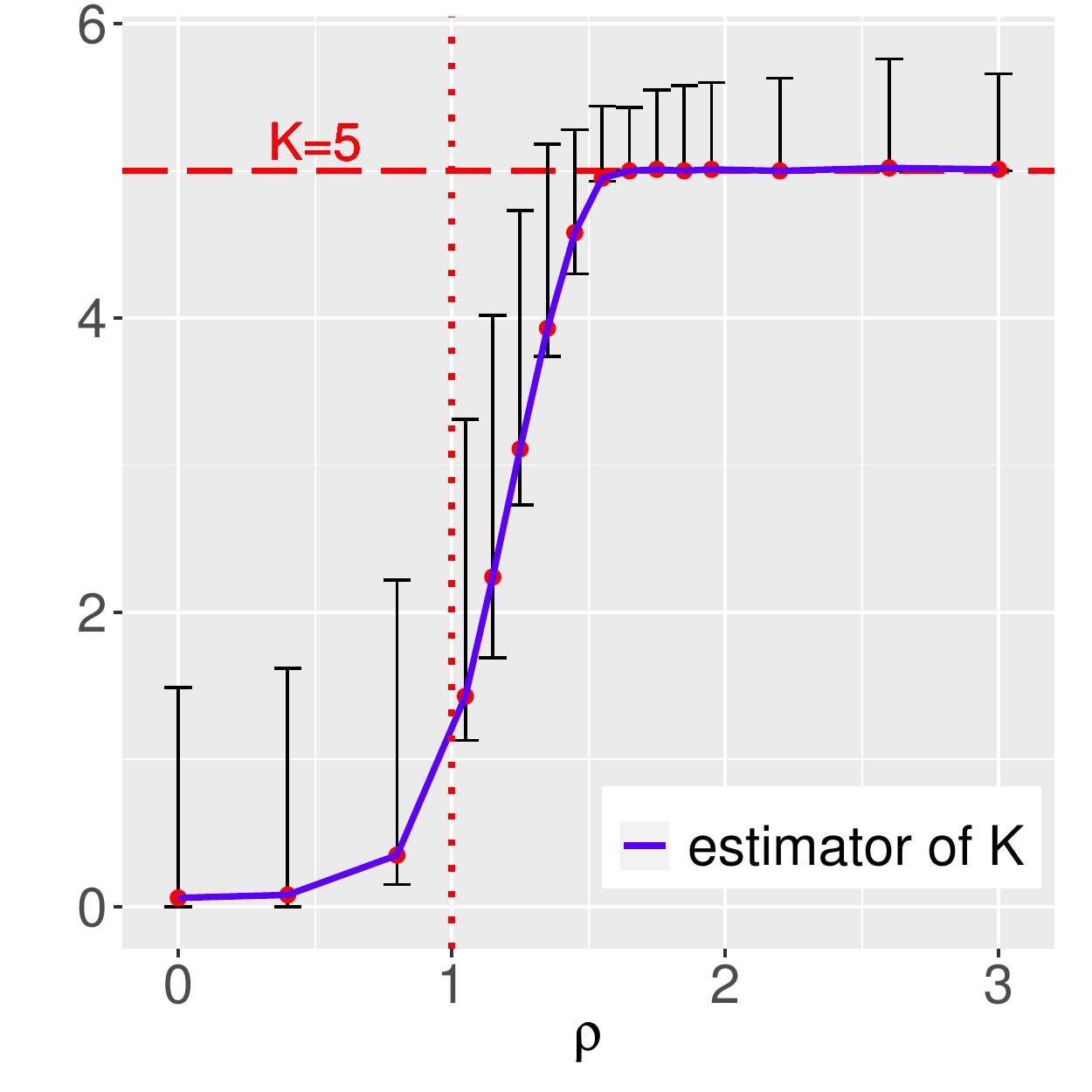} 
\includegraphics[width=0.296\textwidth]{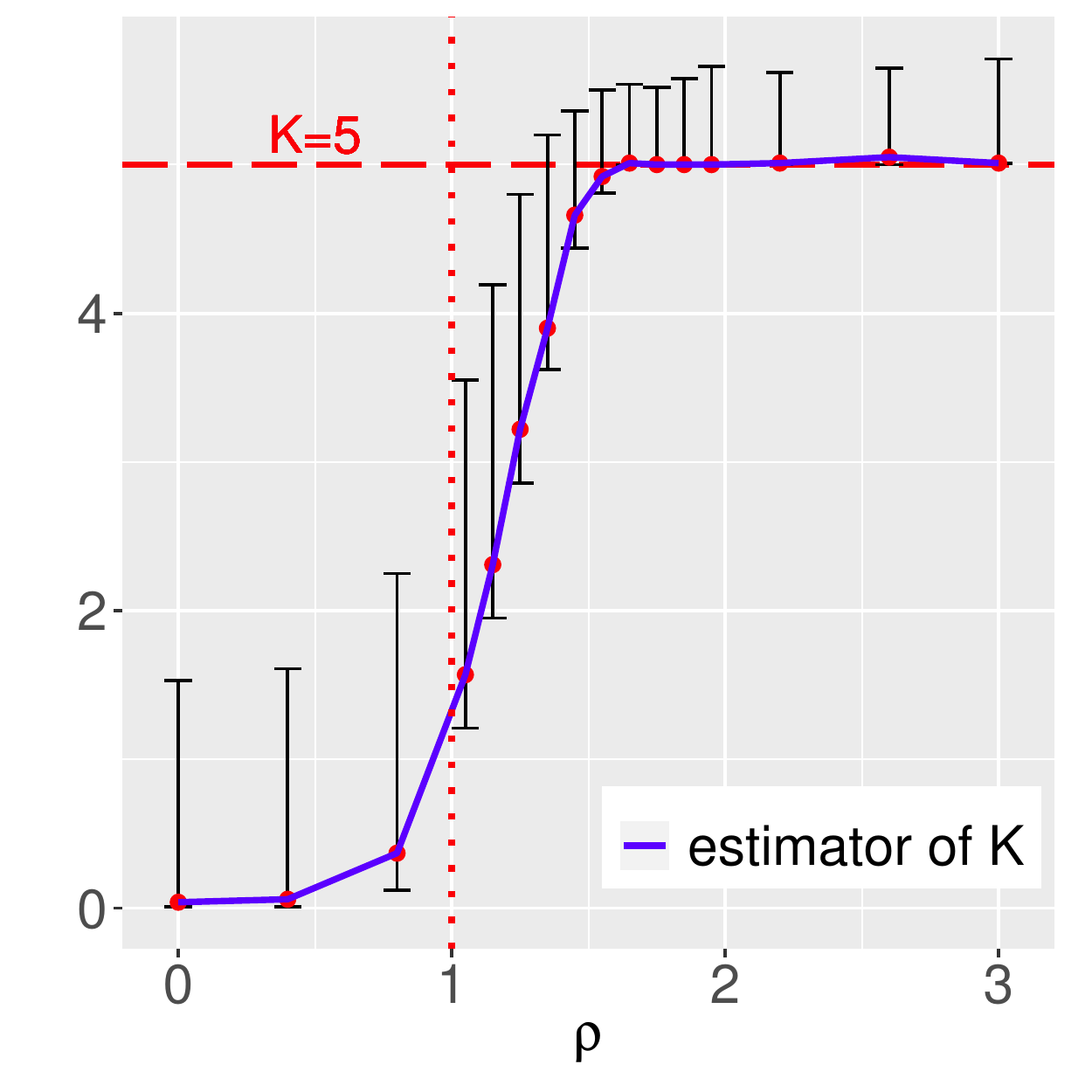} 
\includegraphics[width=0.296\textwidth]{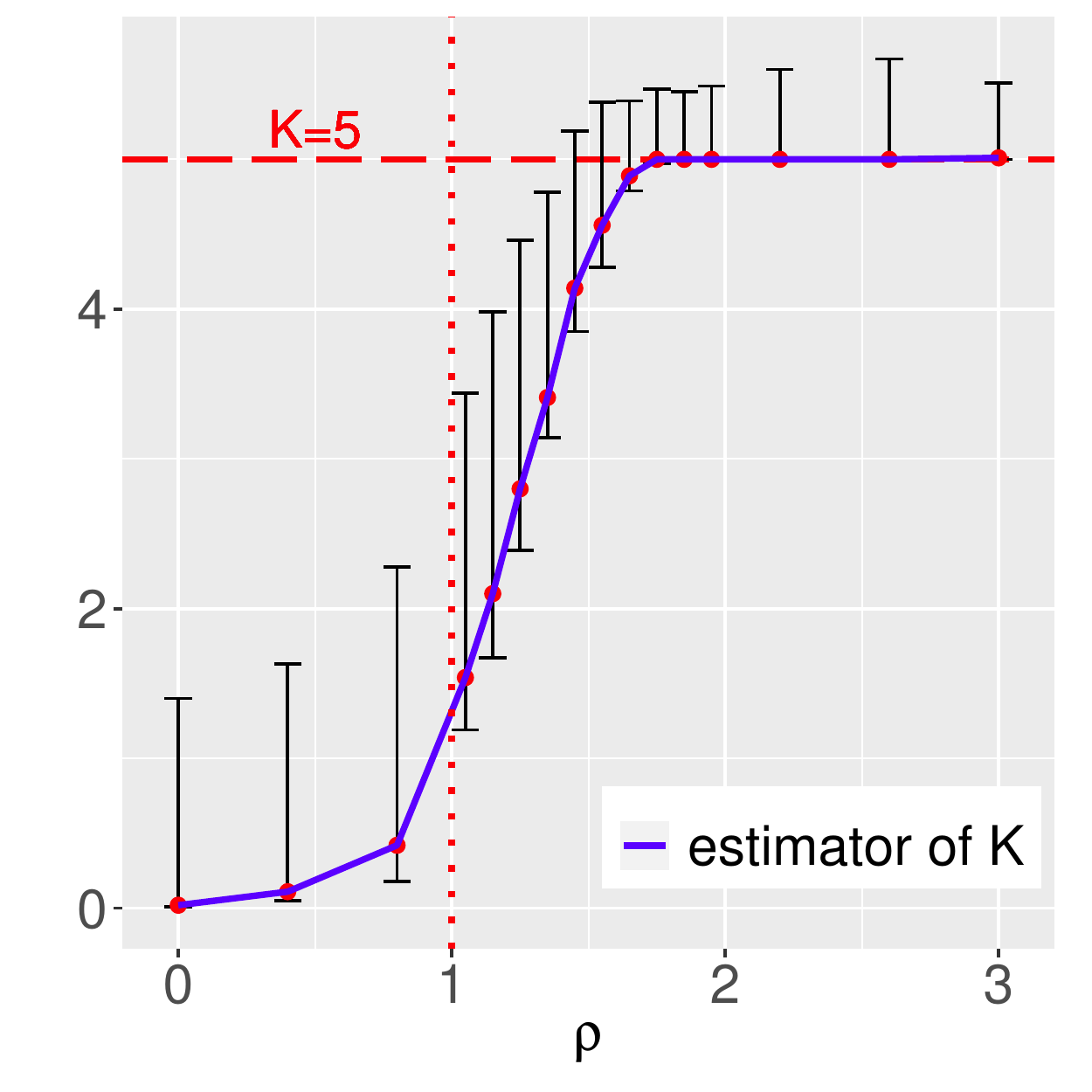}
\includegraphics[width=0.3\textwidth]{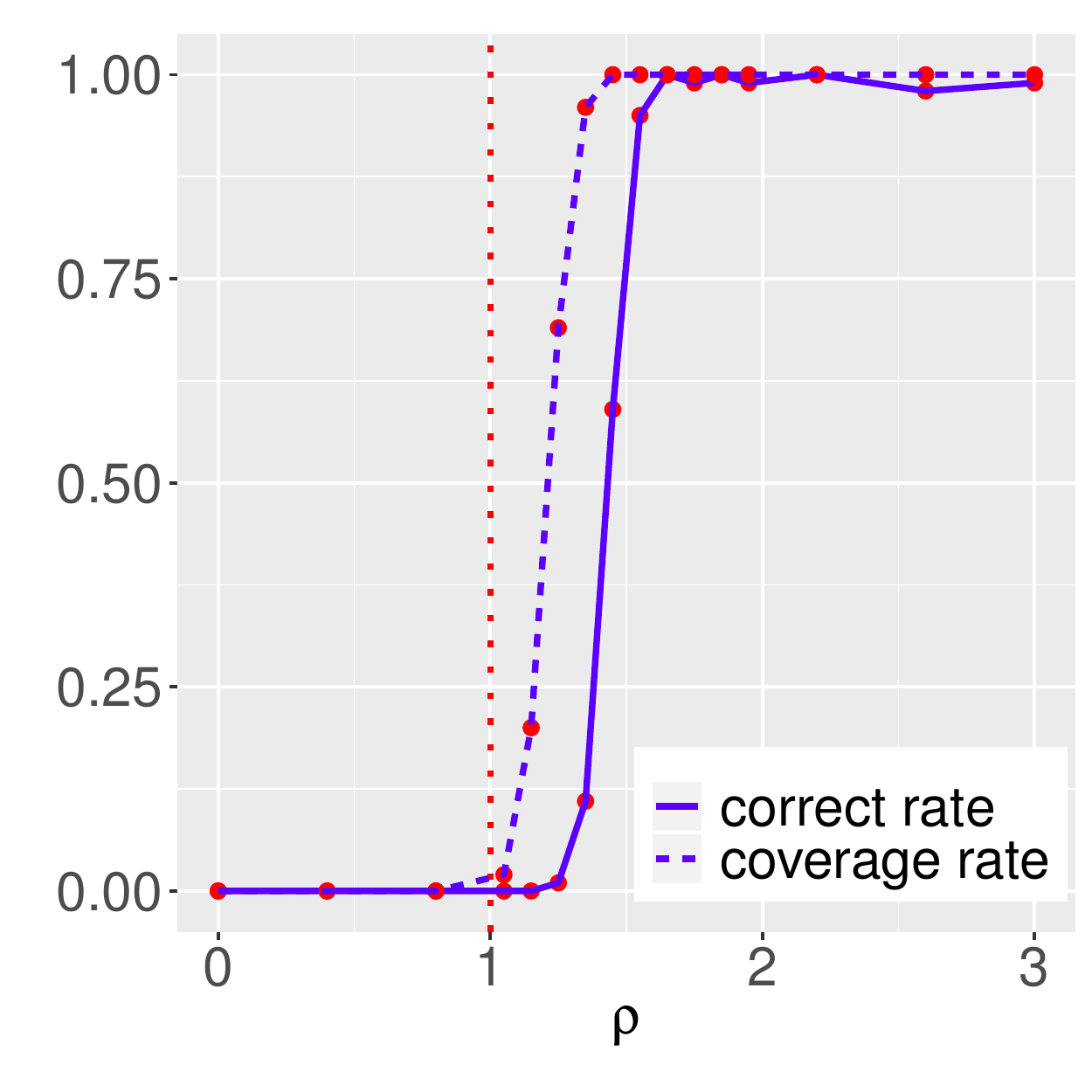}  
\includegraphics[width=0.3\textwidth]{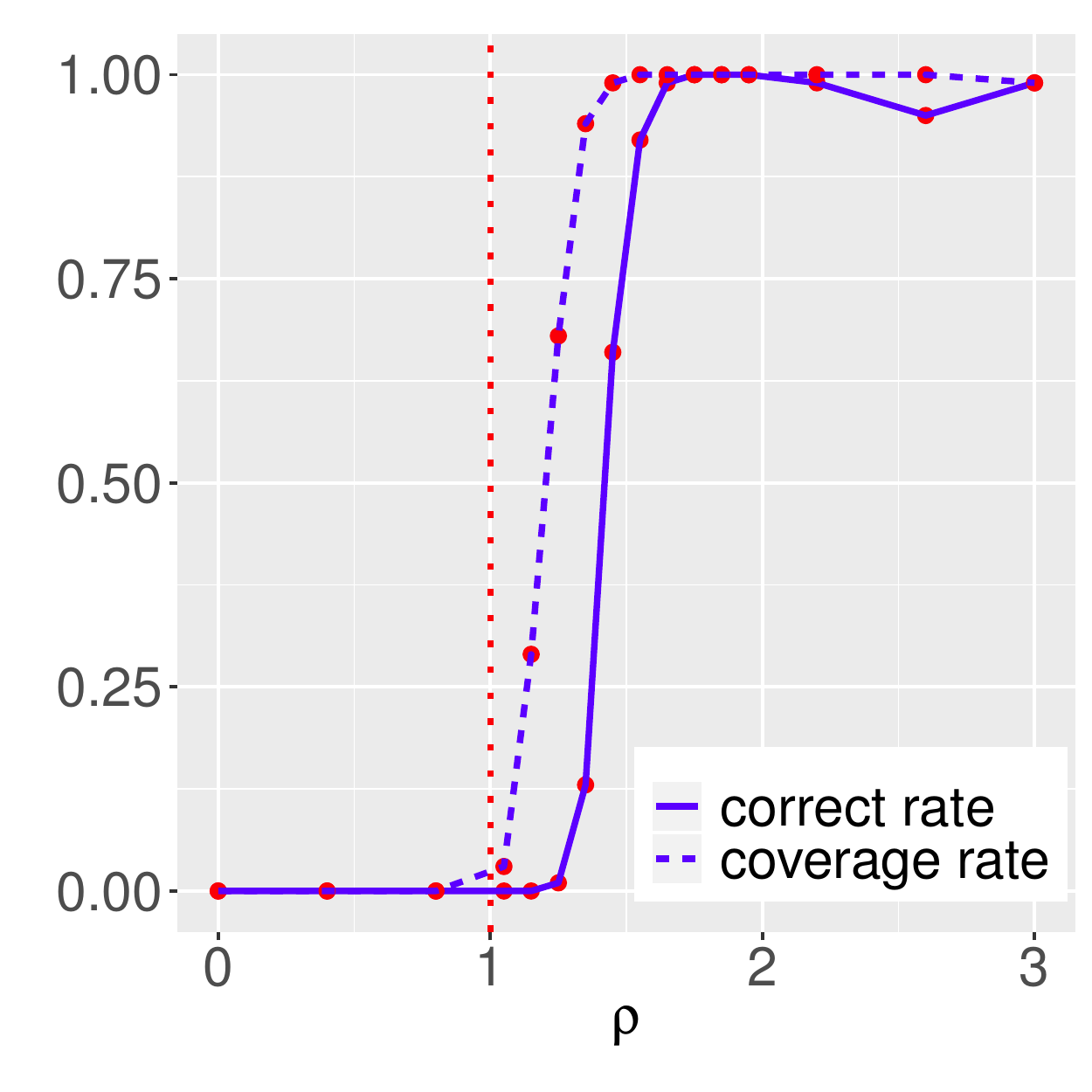}
\includegraphics[width=0.3\textwidth]{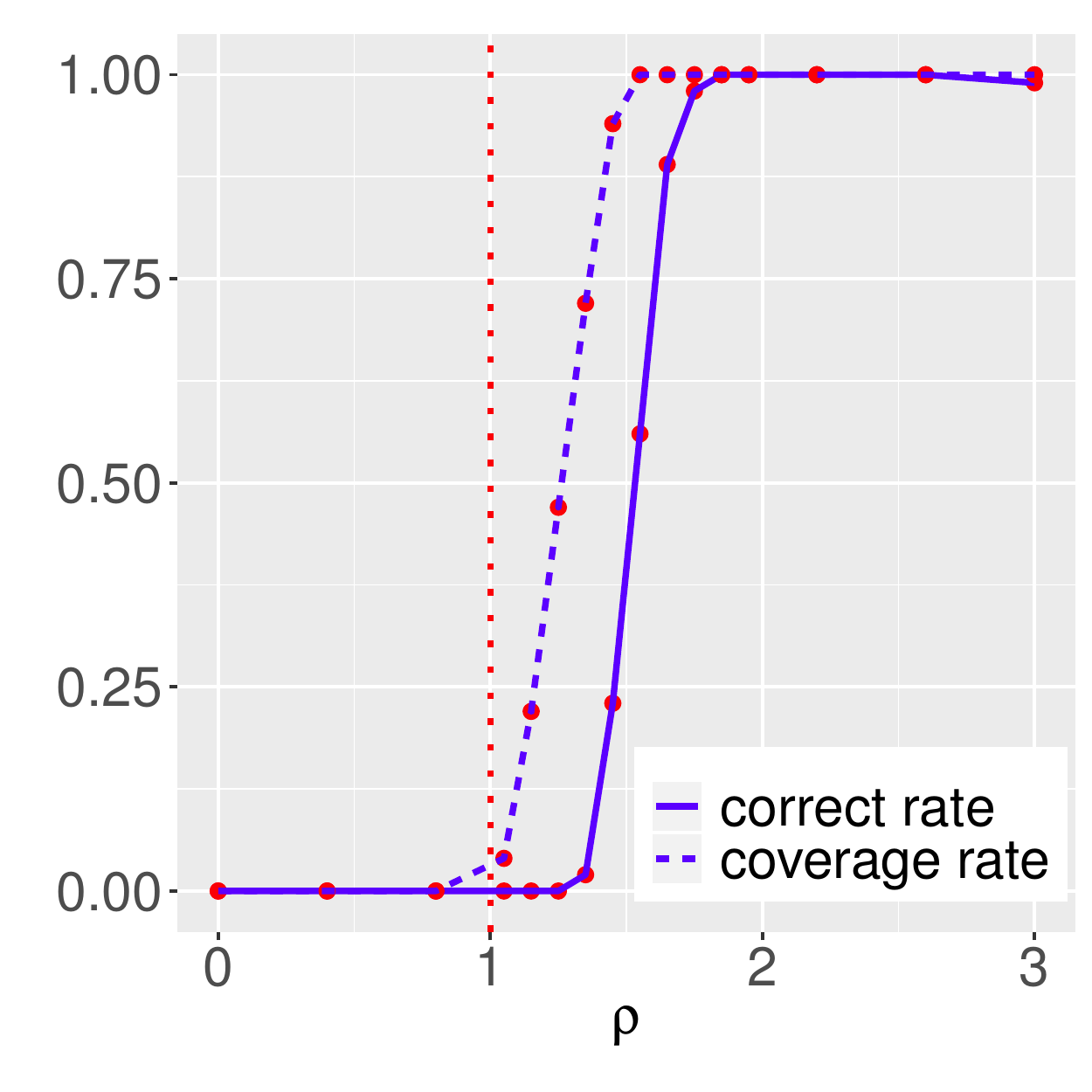}
\caption{Simulation 1: The performance of BEMA0 in a standard spiked model. $K=5$, and $(n,p)$ take the value of $(10000,1000)$, $(1500,5000)$, and $(1500,1500)$ (from left to right). The top three panels show the estimator $\hat{K}$ along with the 95\% confidence upper/lower bound, where each quantity is the average of 100 repetitions. The bottom three panels show the probability of correctly estimating $K$ (correct rate)  and the coverage probabilities of the  $95\%$ confidence intervals (coverage rates). 
In each panel, the x-axis is the value of $\rho$ (see the text for definition), controlling the magnitude of spiked eigenvalues. Our theory states that BEMA0 gives a consistent estimator of $K$ when $\rho$ slightly exceeds 1. This is confirmed by these simulations.} \label{fig:simu-BEMA0}
\end{figure}

Fix $K=5$ and $\sigma^2=1$. We consider three settings, where $(n,p)$ are $(10000,1000)$, $(1500,5000)$, and $(1500, 1500)$, respectively. They cover different cases of size relationship between $p$ and $n$. The eigenvector matrix $\bXi$ is drawn uniformly from the Stiefel manifold (which is the collection of all $p\times K$ matrices that have orthonormal columns). For each of the three settings, we vary the value of $\rho$ and report the average of $\hat{K}$ and upper/lower boundary of a $95\%$ confidence interval, based on $100$ repetitions; the results are in the top three panels of Figure~\ref{fig:simu-BEMA0}. We also report the probability of correctly estimating $K$ (correct rate) and the coverage probability of the $95\%$ confidence interval (coverage rate); see the bottom three panels of Figure~\ref{fig:simu-BEMA0}.


It agrees with our theoretical understanding that $\rho=1$ is the critical phase transition point. When $\rho$ slightly departs from $1$, the coverage rate starts to increase from $0\%$ and quickly reaches the target of $95\%$. The increase of the correct rate is slightly slower, but it reaches $100\%$ before $\rho=1.5$, for all three settings. Our theory suggests that the correct rate is asymptotically 100\% as long as $\rho>1$, but in the finite-sample performance we need a larger $\rho$ to attain a  100\% correct rate. 
Furthermore, as $\rho$ increases, the estimated $\hat{K}$ increases from $0$ to $5$, with a sharp change at around $\rho=1$. The length of the 95\% confidence interval initially decreases with $\rho$ and then stays almost constant.


\paragraph{Simulation 2.}
In this simulation, we compare BEMA0 and BEMA with other methods. We consider both the standard spiked covariance model \eqref{mod-Sigma} and the general spiked covariance model \eqref{mod-Sigma2}. BEMA0 and BEMA are designed for these two settings, respectively. We note that BEMA can also be applied to Model~\eqref{mod-Sigma}, which simply ignores the prior knowledge of equal diagonal in the residual covariance matrix. We thereby also include BEMA in the numerical comparison on the standard spiked covariance model.

Given $(n,p,K,\lambda,\theta)$, we generate data $\bX_i\overset{iid}{\sim}N(0, \bSigma)$, $1\leq i\leq n$, where $\bSigma$ satisfies Model~\eqref{mod-Sigma2} with $\sigma^2=1$ and $\mu_k = \lambda$, for $1\leq k\leq K$. The eigenvector matrix $\bXi$ is drawn uniformly from the Stiefel manifold. We allow $\theta$ to take the value of $\infty$; when $\theta=\infty$, it indicates that $\bSigma$ follows the standard spiked covariance model \eqref{mod-Sigma}. We consider 8 different settings which cover a wide range of parameter values. The results are shown in Table~\ref{tb:experiment2}, where the average $\hat{K}$ and the probability of correctly estimating $K$ (correct rate) are reported based on 500 repetitions.

\begin{table}[!tb]
\scalebox{0.8}{
\begin{tabular}{lllllll}
\toprule
$(n, p, K, \lambda, \theta)$ & BEMA0        & BEMA  & DDPA        & EKC        & Bai\&Ng   & Pass\&Yao \\ \midrule
(100, 500, 5, 9, $\infty$)       & 4.996 (99.6\%) & 4.982 (98.2\%) & 6.102 (41\%)  & 5.552 (57.8\%) & 0 (0\%) & 4.904 (92\%)   \\
(100, 500, 5, 49, $\infty$)      & 5 (100\%)   &  5 (100\%)   & 6.328 (38\%)  & 6.4 (27.4\%) &  5 (100\%)  & 5.012 (98.8\%) \\
(500, 100, 5, 1.5, $\infty$)     & 5 (100\%) & 4.93 (93.0\%) & 6.1 (45.6\%)  & 5.016 (98.4\%) & 0 (0\%)    & 2.784 (43.8\%) \\
(500, 100, 5, 3, $\infty$)       &  5 (100\%)   & 5 (100\%)   & 5.92 (45.4\%)  & 5.056 (94.4\%) & 0 (0\%)    & 4.432 (84.4\%) \\
(100, 500, 5, 15, 3)       & --          &  5.182 (85.2\%) & 9.222 (20.8\%)  & 5.974 (40.2\%) & 0.078 (0\%)  & 5.292 (73.2\%)\\
(100, 500, 5, 50, 3)       & --          & 5.142 (88.4\%) & 9.214 (20.8\%)  & 9.852 (8.6\%)  & 5 (100\%)  & 5.362 (70.4\%) \\
(500, 100, 5, 4.5, 3)      & --          & 4.748 (81.2\%)  & 57.954 (25.4\%) & 5.588 (49.0\%) & 3.392 (39\%)    & 7.624 (5\%) \\
(500, 100, 5, 6, 3)        & --          & 5.018 (98.2\%) & 43.734 (38.8\%) & 6.244 (18.4\%) & 5.002 (99.8\%) & 8.098 (4.2\%)\\ \bottomrule
\end{tabular}}
\caption{Simulation 2: Comparison of different methods in the standard/general spiked model. In these settings, all the spiked eigenvalues are equal to $\lambda$, and the eigenvectors are randomly generated from the Stiefel manifold. The top four rows ($\theta=\infty$) correspond to the standard spiked model, and the bottom four rows correspond to the general spiked model. The number in each cell is the average $\hat{K}$ over 500 repetitions, and the number in brackets is the probability of correctly estimating $K$ (correct rate).}
\label{tb:experiment2}
\end{table}

We have a few observations. First, in the standard spiked covariance model ($\theta=\infty$, top four rows of Table~\ref{tb:experiment2}), BEMA0 has the best performance. Interestingly, BEMA has nearly comparable performance. The reason is that the algorithm will automatically output a very large $\hat{\theta}$, so that the estimator is similar to that of knowing $\theta=\infty$. This suggests that we do not have to choose between BEMA0 and BEMA in practice. 
We can always use BEMA, even when the data come from the standard spiked covariance model. On the other hand, BEMA0 is conceptually simpler and computationally much faster, hence, it is still the better choice if we are confident that the standard spiked covariance model holds. 

Second, in the general spiked covariance model (bottom four rows of Table~\ref{tb:experiment2}), BEMA outperforms DDPA, EKC and Pass\&Yao in all settings, and outperforms Bai\&Ng in two out of four settings. BEMA is the only method whose correct rate is above 80\% in {\it all settings}. 

DDPA requires a delocalization condition. Let $\bXi$ be the $p\times K$ matrix of eigenvectors, and let $\bLambda$ be the diagonal matrix consisting of spiked eigenvalues. The delocalization condition is $\|\bXi\bLambda^{1/2}\|_\infty\to 0$. It prevents eigenvectors from having large entries. This condition is not satisfied here, explaining the unsatisfactory performance of DDPA. Bai\&Ng requires that the spikes are sufficiently large. The larger $p/n$, the higher requirement of spikes. When $p/n=5$ and $\lambda=49$ or when $p/n=0.2$ and $\lambda=6$, Bai\&Ng has a nearly 100\% correct rate. However, as $\lambda$ decreases, the correct rate drops very quickly. EKC uses a thresholding scheme that gives smaller thresholds to lower ranked eigenvalues (e.g., the threshold for $\hat{\lambda}_2$ is smaller than the threshold for $\hat{\lambda}_1$). This method often over-estimates $K$, especially when all the spikes are large (e.g., Row 6 of Table~\ref{tb:experiment2}). Pass\&Yao is developed for the standard spiked model. It has an unsatisfactory performance in the general spiked model (bottom four rows of Table~\ref{tb:experiment2}).

\paragraph{Simulation 3.}
In this simulation, we change the generation process of eigenvectors to satisfy the ``delocalization condition" \citep{dobriban2019deterministic}. 
This condition means $\|\bXi\bLambda^{1/2}\|_\infty$ is sufficiently small, where $\bXi$ is the $p\times K$ matrix consisting of eigenvectors and $\bLambda$ is the diagonal matrix consisting of spiked eigenvalues. 

We adapt the simulation settings in \cite{dobriban2019deterministic} to our general spiked model.
Given $(n,p,K,\theta)$ and $s_1,\ldots,s_K>0$,
 we generate $\bX_i\overset{iid}{\sim}N(0, \bSigma)$, $1\leq i\leq n$, where $\bSigma=\bB\bB^{\top}+\bD$. The matrix $\bD=\mathrm{diag}(\sigma_1^2,\sigma_2^2,\ldots,\sigma^2_p)$ is generated in the same way as in Model~\eqref{mod-Sigma2}, and $\bB$ is a $p\times K$ matrix obtained by first generating a $p\times K$ matrix with independent $N(0,1)$ entries and then re-normalizing each column to have an $\ell^2$-norm equal to $s_k \sqrt{p/n}$. Under this setting, the $L_\infty$-norm of each population eigenvector is only $O(p^{-1/2}\sqrt{\log(p)})$, so the ``delocalization" condition is satisfied. We fix $K=1$ and let $(n,p,s_1,\theta)$ vary. The results are shown in Table~\ref{tb:experiment3}. 
 
Compared with Simulation 2, the performance of DDPA is significantly better. 
BEMA0 and BEMA continue to perform well, indicating that their performance is insensitive to the generating process of eigenvectors. 
This is consistent with our theoretic understanding. In Section~\ref{sec:theory}, we have seen that the success of BEMA0 and BEMA requires no conditions on eigenvectors.  
 
\begin{table}[!tb]
\def~{\hphantom{0}}
\scalebox{0.8}{ 
\begin{tabular}{lllllll}
\toprule
$(n, p, K, s_1, \theta)$ & BEMA0        & BEMA  & DDPA        & EKC        & Bai\&Ng   & Pass\&Yao \\ \midrule
(100, 500, 1, 1, $\infty$)         & 0.988 (96\%) & 0.956 (95.2\%) & 1.086 (88.6\%)  & 1.07 (88.8\%) & 0 (0\%)   & 0.934 (91.8\%)  \\
(100, 500, 1, 3, $\infty$)         & 1.012 (98.8\%)   & 1.008 (99.2\%)   & 1.138 (87\%) & 1.146 (86.4\%) & 1 (100\%)  & 1.036 (96.8\%) \\
(500, 100, 1, 3, $\infty$)       & 1.020 (98\%) & 1 (100\%) & 1.152 (85.6\%)  &  1.056 (94.4\%) & 0 (0\%) & 1.018 (98.2\%)\\
(500, 100, 1, 6, $\infty$)         & 1.014 (98.6\%) & 1 (100\%)   & 1.124 (88.6\%) & 1.12 (88\%) &  1 (100\%)  & 1.014 (98.8\%)   \\
(100, 500, 1, 2, 10)          & --          & 1.096 (90.6\%)  & 1.2 (82.6\%) & 1.102 (90.4\%) & 0.388 (38.8\%) & 1.084 (92.6\%) \\
(100, 500, 1, 6, 10)          & --          & 1.104 (89.8\%)  & 1.226 (79\%) & 1.608 (54.2\%) & 1 (100\%)  & 1.054 (95\%) \\
(500, 100, 1, 6, 3)          & --          & 1.114 (89.2\%)  & 1.062 (95.4\%) & 1.226 (78.2\%)  & 1.008 (99.4\%)    & 3.93 (6.2\%) \\
(500, 100, 1, 12, 3)          & --          & 1.124 (88.0\%)  & 1.042 (97.4\%) & 3.782 (0.8\%) & 1.006 (99.4\%) & 3.672 (9.8\%)\\ \bottomrule
\end{tabular}}
\caption{Simulation 3: Comparison of different methods in the standard/general spiked model, when the eigenvectors are  `delocalized'. Here, $s_1$ controls the magnitude of spiked eigenvalues, where $s_1^2(p/n)$ plays the role of $\lambda$ in Simulation 2.  The top four rows ($\theta=\infty$) correspond to the standard spiked model, and the bottom four rows correspond to the general spiked model. The number in each cell is the average $\hat{K}$, and the number in brackets is the probability of correctly estimating $K$ (correct rate).}
\label{tb:experiment3}
\end{table}

\paragraph{Simulation 4.}
In this simulation, we investigate the case of model misspecification. We still assume that $\bSigma$ is a low-rank matrix plus a residual covariance matrix $\bD$. However, we no longer let $\bD$ be a diagonal matrix. Below, we consider three misspecified models, where $\bD$ is a Toeplitz matrix, a block-wise diagonal matrix, and a sparse matrix, respectively. 
\begin{itemize}
\item In the first model, $\bD(i,j)=(1+|i-j|)^{-t}$, for $1\leq i,j\leq p$. Here, $\bD$ is a Toeplitz matrix with polynomial decays in the off-diagonal. The larger $t$, the closer to a diagonal matrix.  
\item In the second model, $\bD(i,i)=1$ for $1\leq i\leq p$, and $\bD(2j-1,2j)=\bD(2j,2j-1)=b$ for $1\leq j\leq p/2$. $\bD$ is a block-wise diagonal matrix which has many $2\times 2$ diagonal blocks. The smaller $b$, the closer to a diagonal matrix. 
\item In the third model, $\bD(i,i)=1$ for $1\leq i\leq p$, and $\bD(i,j)=\bD(j,i)\sim c\cdot \text{Bernoulli}(0.1)$ for $i \neq j$. The matrix $\bD$ has approximately $0.1p$ nonzero entries in each row. The smaller $c$, the closer to a diagonal matrix. 
\end{itemize}
The low-rank part of $\bSigma$ is generated in the same way as before: We let all $\mu_k$ equal to $\lambda$ and let the eigenvector matrix $\bXi$ be drawn uniformly from the Stiefel manifold, which allows $\bXi$ to have orthonormal columns. 
Fix $(n,p,K)=(500,100,1)$. The results are shown in Table~\ref{tb:misspecify}.  

For each misspecified model, we consider two settings, where $\bD$ is closer to a diagonal matrix in the first setting (Rows 1,3,5 of Table~\ref{tb:misspecify}) than in the second one  (Rows 2,4,6 of Table~\ref{tb:misspecify}). Every method performs better in the first case, suggesting that the diagonal assumption on $\bD$ is indeed critical. In comparison, BEMA is least sensitive to a non-diagonal $\bD$. In Rows 2,4,6 of Table~\ref{tb:misspecify}, the correct rate of BEMA is still above 80\%, while the correct rate of some other methods is only 0\%. Pass\&Yao is the second least sensitive to a non-diagonal $\bD$.

To try to understand this phenomenon, we first note that one can always apply an orthogonal transformation to data vectors $\bX_1,\ldots, \bX_n$, so that the post-transformation data follow a different spiked covariance model whose residual covariance matrix $\widetilde{\bD}$ is a diagonal matrix containing the eigenvalues of $\bD$. This orthogonal transformation is unknown in practice. However, if a method uses the empirical eigenvalues {\it only}, it does not matter whether or not we know this orthogonal transformation, because any orthogonal transformation does not change eigenvalues of the sample covariance matrix and thus it does not change the estimator of $K$. It implies that, for methods that only use eigenvalues, we can treat the misspecified model as if $\bD$ is replaced by the diagonal matrix $\widetilde{\bD}$. Therefore, the surprising robustness of BEMA can be interpreted as the capability of the gamma model \eqref{mod-D} in approximating the eigenvalue structure in $\bD$. The flexibility of this gamma model comes from the parameter $\theta$. In comparison, such strong robustness is not observed for BEMA0, where $\theta$ is fixed as $\infty$.  

The method of DDPA uses empirical eigenvectors in the procedure, thus, it is more sensitive to the diagonal assumption of $\bD$. 
EKC uses eigenvalues only, but its thresholding scheme is too conservative. In these misspecified models, some bulk empirical eigenvalues can get large; EKC gives too small thresholds to non-leading eigenvalues and yields over-estimation of $K$.   

\begin{table}[!tb]
\def~{\hphantom{0}}
\scalebox{0.8}{
\begin{tabular}{llllllll}
\toprule
$\lambda$ & residual covariance  & BEMA0        & BEMA  & DDPA        & EKC        & Bai\&Ng   & Pass\&Yao \\ \midrule
6 & Toeplitz(t=4)          & 1.104 (89.6\%)      & 1 (100\%)       & 1.422 (65.4\%)  & 1.36 (67.4\%) & 1 (100\%)   & 1.06 (94.8\%)  \\
3 & Toeplitz(t=2)         & 9.352 (0\%)    & 1.12 (88.6\%)       & 100 (0\%)       & 15.148 (0\%)    & 0 (0\%)   & 2.46 (24.6\%)\\
6 & block diagonal(b=0.1)  & 1.344 (66.8\%) & 1 (100\%)       & 2.378 (31.6\%)    & 1.854  (33.6\%) & 1 (100\%) & 1.038 (96.6\%)    \\
3 & block diagonal(b=0.2) & 3.764 (0\%)    & 1 (100\%)       & 100 (0\%)       & 6.602 (0\%)    & 0 (0\%)    & 1.12 (89.8\%) \\
6 & sparse(s=0.05)         & 1.784 (30.2\%) & 1.016 (98.4\%)  & 5.024 (9.4\%)     & 2.474 (9.4\%) & 1 (100\%) & 1.084 (91.6\%) \\
3 & sparse(s=0.08)         & 3.348 (0\%)     & 1.036 (96.4\%)     & 97.752 (0\%)       & 5.18 (0\%)    & 0 (0\%)   & 1.58 (47.4\%)\\ \bottomrule
\end{tabular}}
\caption{Simulation 4: Comparison of different methods in three misspecified models, where the residual covariance matrix $\bD$ is a Toeplitz matrix, a block diagonal matrix, and a sparse matrix, respectively. $(n,p,K)=(500, 100, 1)$. The spiked eigenvalue is equal to $\lambda$. For each misspecified model, we consider two settings, where $\bD$ is closer to a diagonal matrix in the first setting (rows 1, 3, 5) than in the second setting (rows 2, 4, 6). The number in each cell is the average $\hat{K}$, and the number in brackets is the probability of correctly estimating $K$ (correct rate).}
\label{tb:misspecify}
\end{table}

\paragraph{Simulation 5.} In this simulation, we tested the robustness of our proposed methods against the choice of $\alpha$ and the distributional assumption on data generation. 
Fix $(n, p, K)=(500, 100, 5)$. We generate $\bX_i= \bXi \omega_i+\epsilon_i$ where $\bXi\in\mathbb{R}^{p\times K}$ is uniformly drawn from the Stiefel manifold, $\omega_i$ are $iid$ drawn from a multivariate zero-mean distribution with covariance matrix $\lambda \textbf{I}_K$, $\epsilon_i$ are $iid$ drawn from a multivariate zero-mean distribution with covariance matrix $\bD$, and $\bD$ is generated in the same way as in Model~\eqref{mod-Sigma2} with $\sigma^2=1$ and $\theta\in \{\infty,3\}$. We consider three settings where the entries of $\omega_i$ and $\epsilon_i$ are Gaussian, random sign, or Laplace variables   (centered and re-scaled to match the required variance), respectively. The results are in shown Table~\ref{tb:robust}. 

For the standard spiked covariance model (top 3 rows of Table~\ref{tb:robust}), the results are very similar for different distributions. For the general spiked covariance model (bottom 3 rows of Table~\ref{tb:robust}), the performance of BEMA increases/decreases when the data have lighter/heavier tails, but the difference is within a reasonable range. Our theory only requires a mild distributional assumption (Assumption~\ref{cond:moments}), which is validated by this simulation. 

The choice of $\alpha$ decides the fraction of bulk eigenvalues used to estimate $(\sigma^2,\theta)$. The larger $\alpha$, we restrict to a narrower range of eigenvalues deep into the bulk. The performance of BEMA is similar for $\alpha\in \{0.2, 0.3\}$ and slightly worse for $\alpha=0.1$. In the asymptotic theory, $\alpha$ can be chosen as any constant, but for good finite-sample performance we need $(\tilde{p}\alpha-K)$ to be properly large, where $\tilde{p}=n\wedge p$. In practice, if $\tilde{p}$ is extremely large, the choice of $\alpha$ has a negligible effect; if $\tilde{p}$ is only moderately large, we recommend choosing a large $\alpha$ so that we are confident that $\tilde{p}\alpha$ is significantly larger than $K$.

\begin{table}[!tb]
\def~{\hphantom{0}}
\scalebox{0.8}{
\begin{tabular}{llllllll}
\toprule
distribution & $(\lambda, \theta)$  & BEMA0 (0.1)     & BEMA0 (0.2)  & BEMA0 (0.3)        & BEMA (0.1)     & BEMA (0.2)    & BEMA (0.3)   \\ \midrule
Gaussian & $(1.5,\infty)$          & 5 (100\%)      & 5 (100\%)        & 5 (100\%)   & 4.95 (95\%)  & 4.93 (93\%)    & 4.904 (90.4\%)   \\
Random sign & $(1.5,\infty)$          & 4.996 (99.6\%)      & 4.996 (99.6\%)        & 4.998 (99.8\%)   & 4.972 (97.2\%)  & 4.96 (96\%)    & 4.94 (94\%)   \\
Laplace & $(1.5,\infty)$         & 4.998 (99.8\%)     & 4.998 (99.8\%)        & 4.998 (99.8\%)        & 4.914 (91.4\%)    & 4.9 (90\%)    & 4.88 (88\%) \\
Gaussian & $(1.5, 3)$          & --      & --        & --   & 4.518 (69\%)  & 4.748 (81.2\%)    & 4.76 (81\%)   \\
Random sign & $(4.5,3)$         & --  & --   & --      & 4.678 (78.4\%)  & 4.818 (85\%)  & 4.9 (85.4\%)  \\
Laplace & $(4.5,3)$        & --      & --     & --       & 4.352 (56.8\%)    & 4.634 (73.8\%)   & 4.656 (74.8\%)\\ \bottomrule
\end{tabular}}
\caption{Simulation 5: The robustness of BEMA0 and BEMA under non-Gaussian data and different values of $\alpha$. Data are generated from the factor model with Gaussian/random-sign/Laplace factors and noise. $K=5$, and all the spiked eigenvalues are equal to $\lambda$. BEMA0 and BEMA are implemented with $\alpha\in\{0.1,0.2,0.3\}$ (denoted as BEMA0 ($\alpha$)/BEMA ($\alpha$) in the table). The number in each cell is the average $\hat{K}$, and the number in brackets is the probability of correctly estimating $K$ (correct rate).}
\label{tb:robust}
\end{table}

\section{Real applications}\label{sec:realdata}
We apply BEMA to two real datasets. We compare our method with EKC \citep{braeken2017empirical}, Bai\&Ng \citep{bai2002determining}, DDPA and its variants \citep{dobriban2019deterministic}, and Pass\&Yao \citep{passemier2014estimation}. DDPA has 3 versions: DPA is a deterministic implementation of parallel analysis \citep{horn1965rationale}; DDPA is an improvement of DPA aiming to resolve the issue of ``eigenvalue shadowing," that is, an extremely large spiked eigenvalue shadows the other spiked eigenvalues and causes an under-estimation of $K$; DDPA+ is a robust version of DDPA recommended for real data analysis. We include all three versions in comparison.   


\subsection{The Lung Cancer data} \label{subsec:Lung}
The Lung Cancer dataset was collected and cleaned by \cite{gordon2002translation}. The original data set contains the expression data of 12,533 genes and 181 subjects. The subjects divide into two groups, the diseased group and the normal group. 
\cite{jin2016influential} processed this data set by removing genes that are not differentially expressed across subject groups and resulted in a new data matrix with $(p,n)=(251,181)$. The selection of these 251 ``influential genes" used no information of true groups, including the number of groups. We use this processed data matrix, because the original data matrix contains too many features (genes) that are irrelevant to the clustering structure, where no method gives meaningful results. It was argued in \cite{jin2016influential} that this data matrix follows a clustering model. As a result, the covariance matrix has $(K_0-1)$ spiked eigenvalues, where $K_0$ is the number of clusters. Here, the ground-truth is $K_0=2$, i.e., the true number of spiked eigenvalues is $K=1$.

\begin{figure}[!tb]
\centering
\begin{subfigure}[b]{.6\textwidth}
\includegraphics[width=.48\textwidth, trim=10 0 0 0, clip=true]{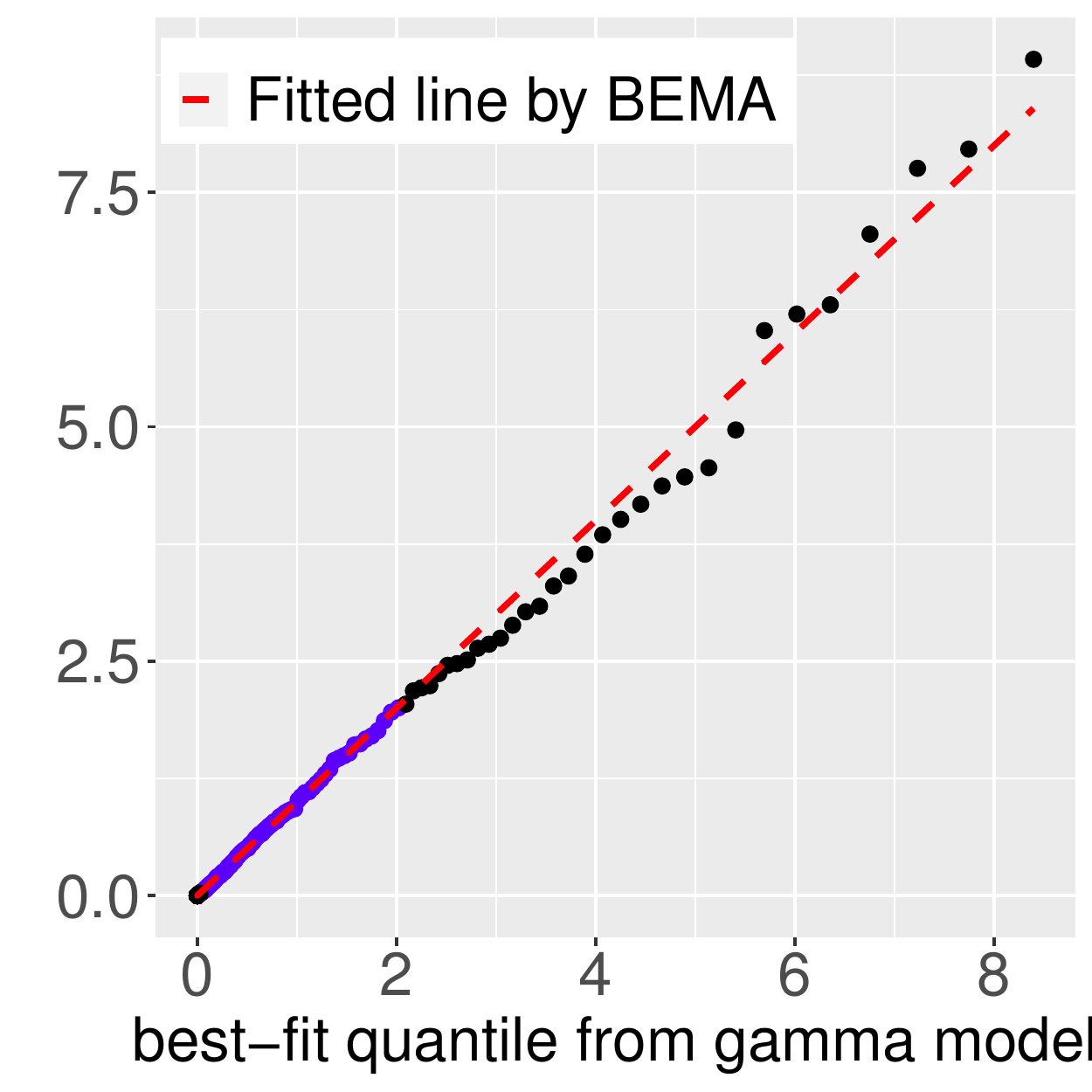}
\includegraphics[width=.482\textwidth, trim=10 0 0 0, clip=true]{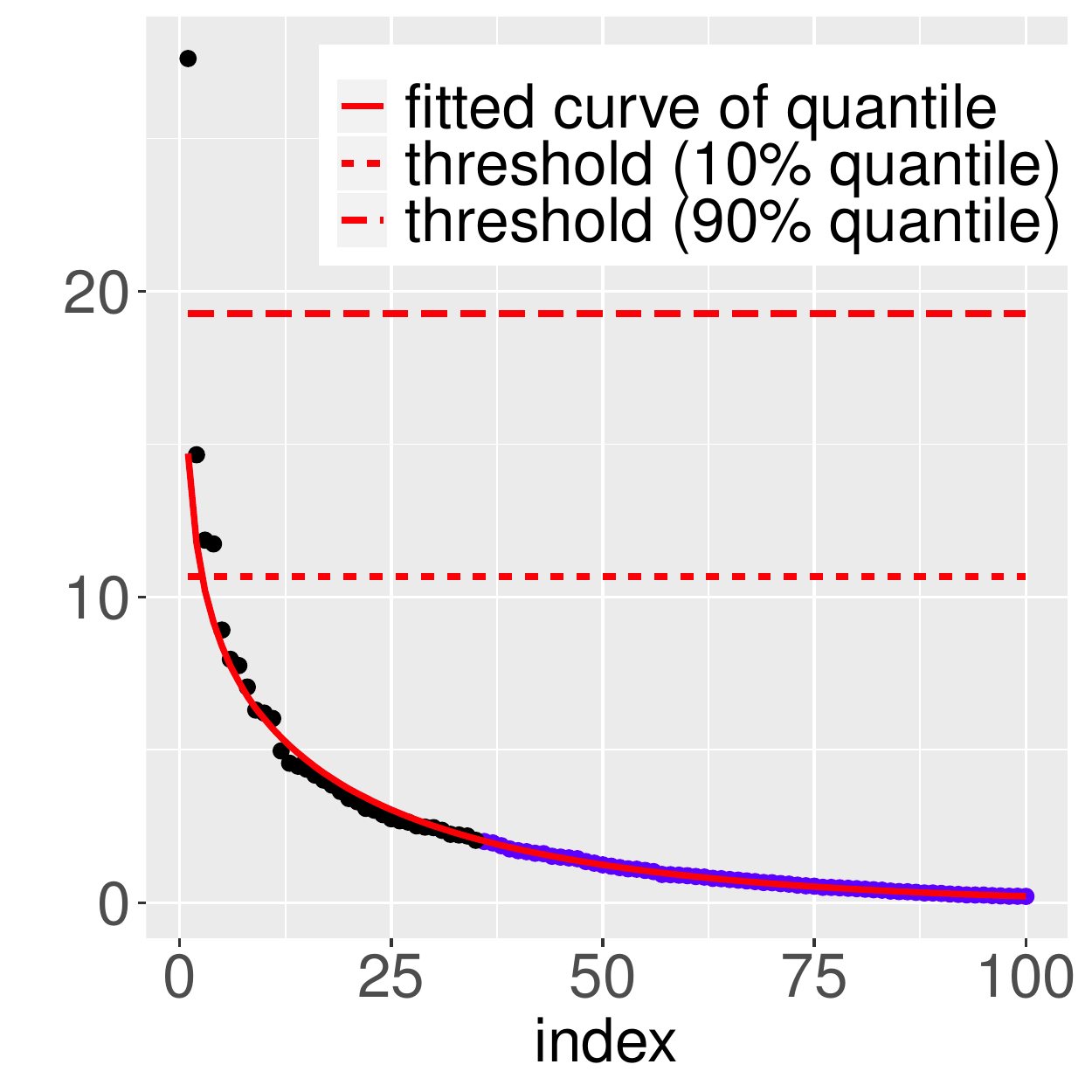}
\vspace{10pt}
\caption{\footnotesize The goodness-of-fit of BEMA on the Lung Cancer data. The left panel plots $\hat{\lambda}_k$ versus $\bar{F}_{\gamma_n}^{-1}(k/\tilde{p};1, \hat{\theta})$ (which is quantile of the theoretical limit of ESD with estimated $\theta$), where the first 4 eigenvalues are removed for better visualization. It fits well a line crossing the origin. The right panel plots $\hat{\lambda}_k$ versus $k$, where the red solid curve is $\bar{F}_{\gamma_n}^{-1}(k/\tilde{p};\hat{\sigma}^2, \hat{\theta})$ versus $k$. The curve fits the bulk eigenvalues (blue dots). These two plots together suggest that the spiked covariance model \eqref{mod-Sigma2} is suitable for this dataset.}
\vspace{10pt}
\end{subfigure}
\begin{subfigure}[b]{.39\textwidth}
\centering
\includegraphics[width=.7\textwidth]{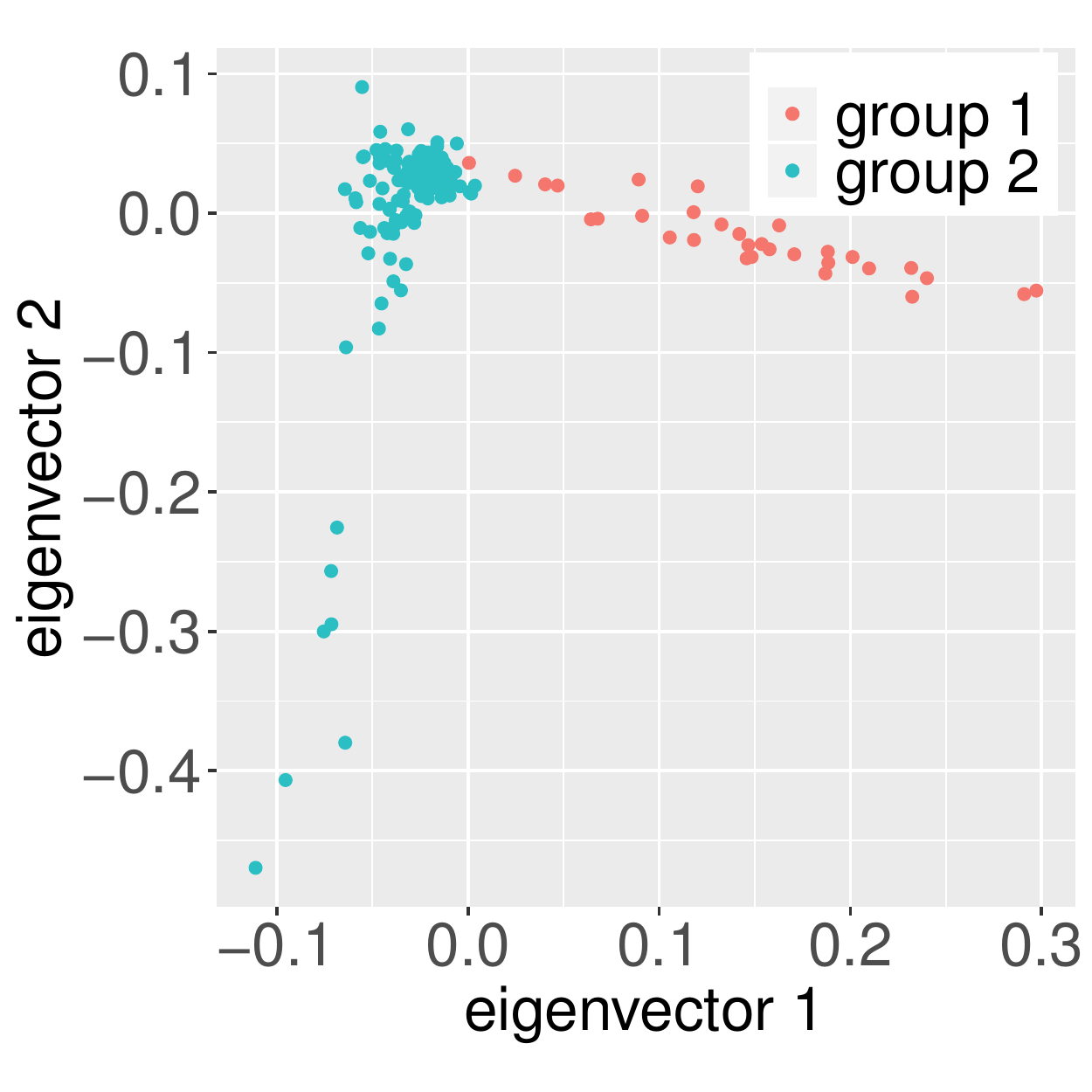}\\
\includegraphics[width=.7\textwidth]{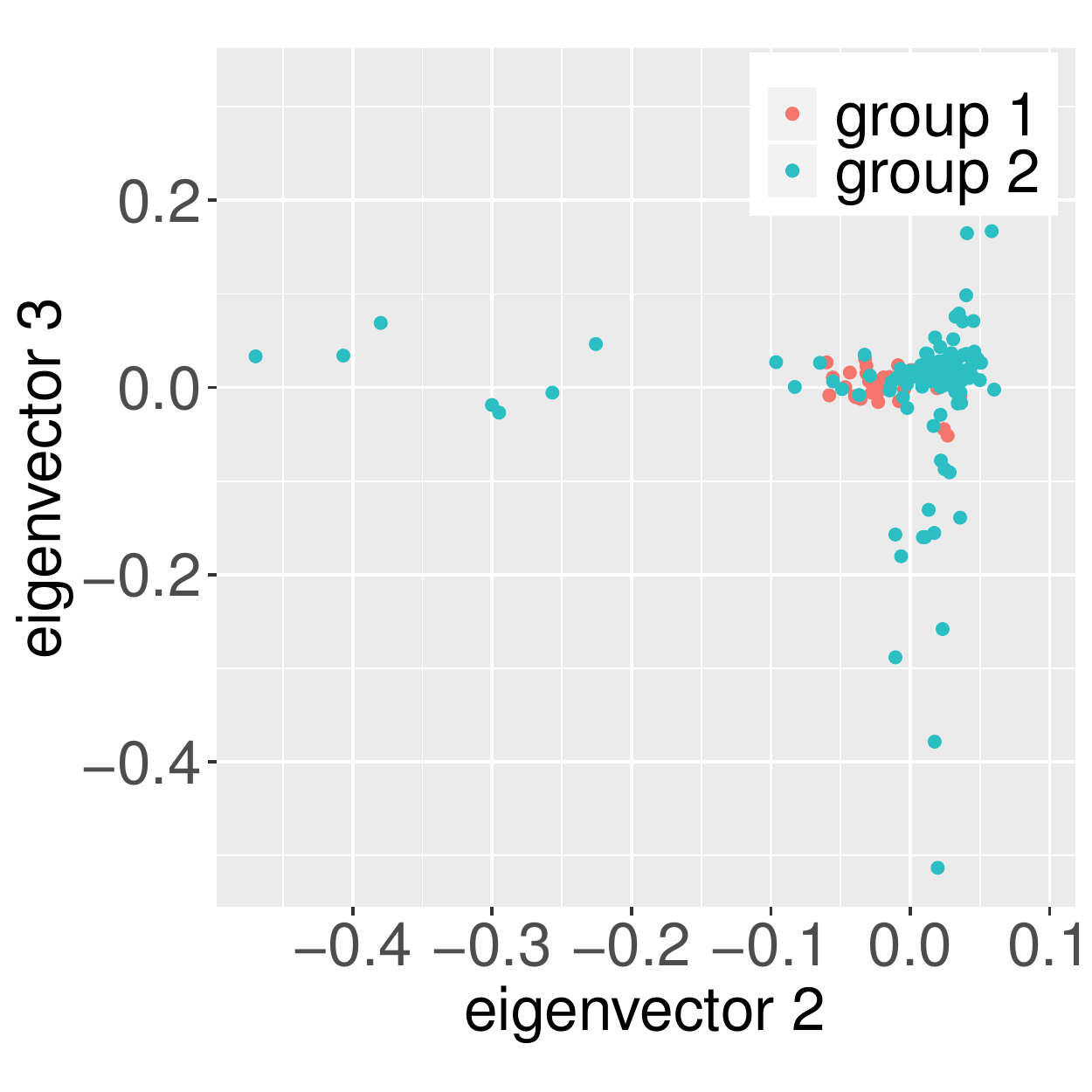}
\caption{\footnotesize The plots of singular vectors of $\bX$.}
\end{subfigure}
\caption{Results for the Lung Cancer data.}
 \label{fig:lungcancer}
\end{figure}

We apply BEMA with $(\alpha, \beta, M)=(0.2, 0.1, 500)$, i.e., $60\% (=1-2\alpha)$ of the bulk eigenvalues in the middle range are used to estimate model parameters, the probability of over-estimating $K$ is controlled by 0.1, and 500 Monte Carlo samples are used to determine the ultimate threshold for eigenvalues. The BEMA algorithm outputs $(\hat{\theta}, \hat{\sigma}^2)=(0.288, 0.926)$. In Figure~\ref{fig:lungcancer}(a), we check the goodness-of-fit. If the proposed spiked covariance model \eqref{mod-Sigma2} is suited for the data, we expect to see $\hat{\lambda}_k\approx \hat{\sigma}^2\cdot \bar{F}^{-1}_{\gamma_n}(k/\tilde{p}; 1, \hat{\theta})$, except for a few small $k$. The left panel of Figure~\ref{fig:lungcancer}(a) plots $\hat{\lambda}_k$ versus $\bar{F}^{-1}_{\gamma_n}(k/\tilde{p}; 1, \hat{\theta})$, suggesting a good fit to a line crossing the origin. The right panel contains the scree plot, i.e., $\hat{\lambda}_k$ versus $k$. We also plot the curve of $\bar{F}^{-1}_{\gamma_n}(k/\tilde{p};\hat{\sigma}^2,\hat{\theta})$ versus $k$. This curve is a good fit to the scree plot in the middle range. These plots suggest that Model~\eqref{mod-Sigma2} is well-suited for this dataset.

\begin{table}[!tb]
\centering
\scalebox{0.8}{
\begin{tabular}{l|llllllll|l}
\toprule
 &  BEMA\ \ & BEMA0\ \ & EKC\ \ \ & Bai\&Ng\ & Pass\&Yao &  DDPA\ \ & DPA\ \  & DDPA+\    & truth\   \\ \hline
Lung Cancer Data & {\bf 1} & 27 & 56 & 180 & 8 & 180 & {\bf 1}  & 11 &1   \\
1000 Genomes Data & {\bf 28} & 67 & 2503 & 4 &  {\bf 28} & 85 & 20  & 4 & 25 \\ \bottomrule
\end{tabular}}
\caption{Comparison of different estimators of $K$ using two real data sets: the lung cancer gene expression data and the 1000 Genome data of genome-wide common genetic variants. For BEMA and BEMA0, the choices of tuning parameters are described in the text. In the Appendix, we report the results with various choices of tuning parameters, which are very stable.}
\label{tb:realdata}
\end{table}

The estimator of $K$ by BEMA is $\hat{K}=1$, which is exactly the same as the ground truth. This is the output of the algorithm by setting $\beta=0.1$. Using the argument in Section~\ref{subsec:main-CI}, this is also a confidence lower bound for $K$. By setting $\beta=0.9$ in the algorithm, we get a confidence upper bound which is $4$. This gives an 80\% confidence interval for $K$ as $[1,4]$. Figure~\ref{fig:lungcancer}(b) contains the scatter plots of the left singular vectors of $X$, colored by the true group label. The first singular vector clearly contains information for separating two groups, but other singular vectors also contain some information. This explains why the confidence upper bound is larger than $1$.

The comparison with other methods is summarized in Table~\ref{tb:realdata}. 
The behavior of EKC is consistent with our observation in simulations. In this dataset, the eigenvalues of the residual covariance matrix vary widely (this can be seen from the estimated $\theta$ by BEMA, $\hat{\theta}=0.288$, which is far from $\infty$), and EKC gives too small threshold to non-leading eigenvalues. The behavior of Bai\&Ng is different from what we observe in simulations. Note that we have to use the effective $p$ after the data processing by \cite{jin2016influential}, where the dimension reduces from 12,533 to 251. As a result, the penalty in Bai\&Ng is weaker than that in simulations, and so the method significantly over-estimates $K$. Pass\&Yao also over-estimates $K$.
Among DDPA and its variants, DPA performs the best. A possible reason is that DPA does not use empirical eigenvectors and is more stable than DDPA and DDPA+. 

Different from all other methods, BEMA not only outputs an estimator of $K$ but also yields a fitted model, $\mathrm{Gamma}(\hat{\theta},\, \hat{\theta}/\hat{\sigma}^2)=\mathrm{Gamma}(0.288, 0.311)$, for eigenvalues of the residual covariance matrix. This can be useful for many other statistical inference tasks.

\subsection{The 1000 Genomes data} \label{subsec:1000G}
The 1000 Genomes Phase 3 whole genome sequencing dataset \citep{10002015global} consists of the genotypes of $2504$ subjects for over 84.4 million variants. We restrict the analysis to common variants with minor allele frequencies greater than 0.01. There are $26$ self-reported ethnicity groups, coming from five super-populations: African (AFR), Ad Mixed American (AMR), East Asian (EAS), European (EUS), and South Asian (SAS). 

In view of high linkage disequilibrium (LD) among some variants, which can distort the eigenvector and eigenvalue structure \citep{patterson2006population}, we first performed LD pruning. We used an independent pair-wise LD pruning, with window size 1000, step size 50 and a threshold 0.02 for R-squared. Restricting to LD pruned markers, we obtain a data matrix with $p=24,248$ and $n =2,504$. 
The number of spiked eigenvalues equals to the number of true ancestry groups minus one \citep{patterson2006population}. We treat the self-reported ethnicity groups as the ground truth, which gives $K=25$. 

\begin{figure}[!tb]
\hspace*{-15pt}\begin{subfigure}[b]{.52\textwidth}
\includegraphics[width=0.472\textwidth,trim=25 0 8 0, clip=true]{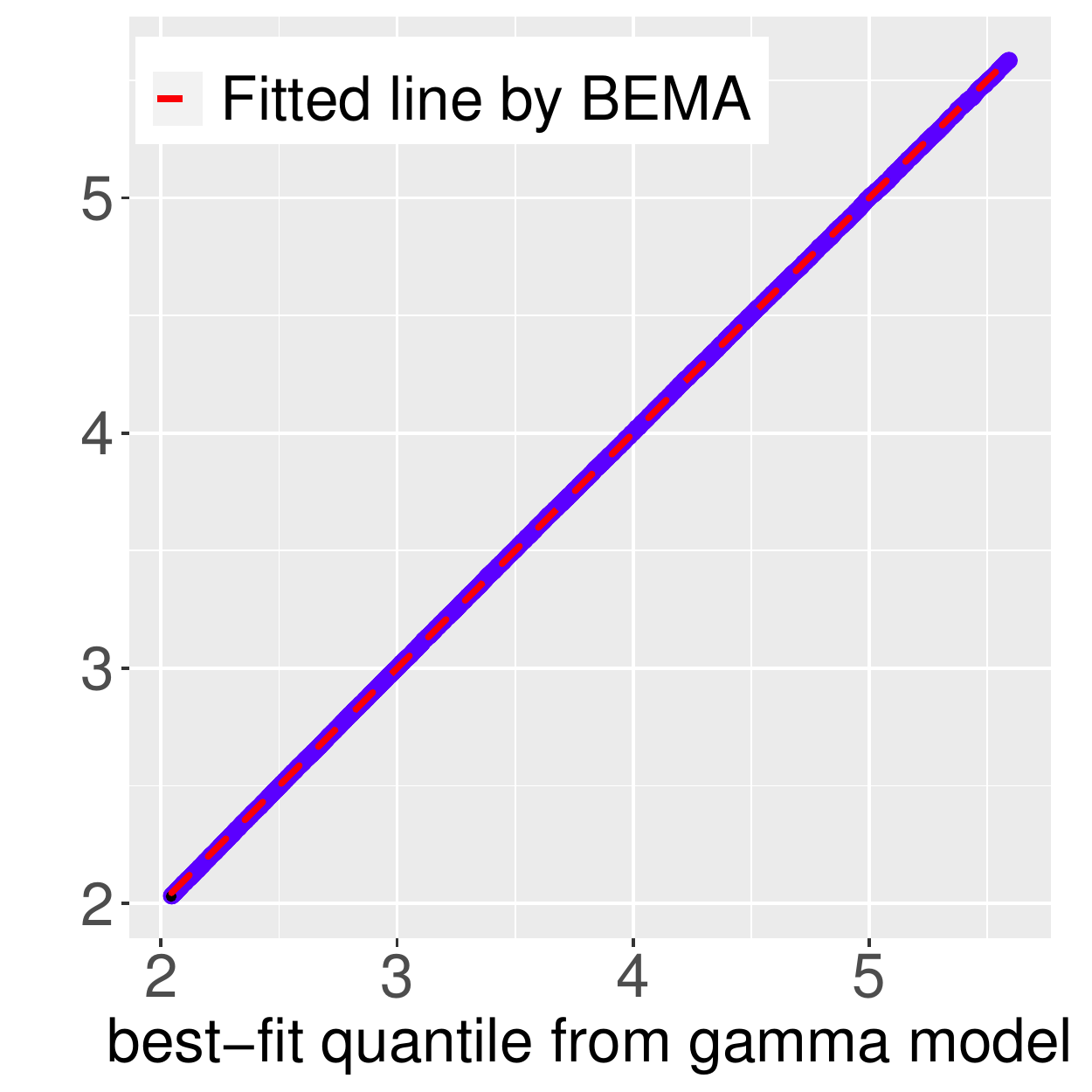} 
\includegraphics[width=0.515\textwidth, trim=20 8 15 2, clip=true]{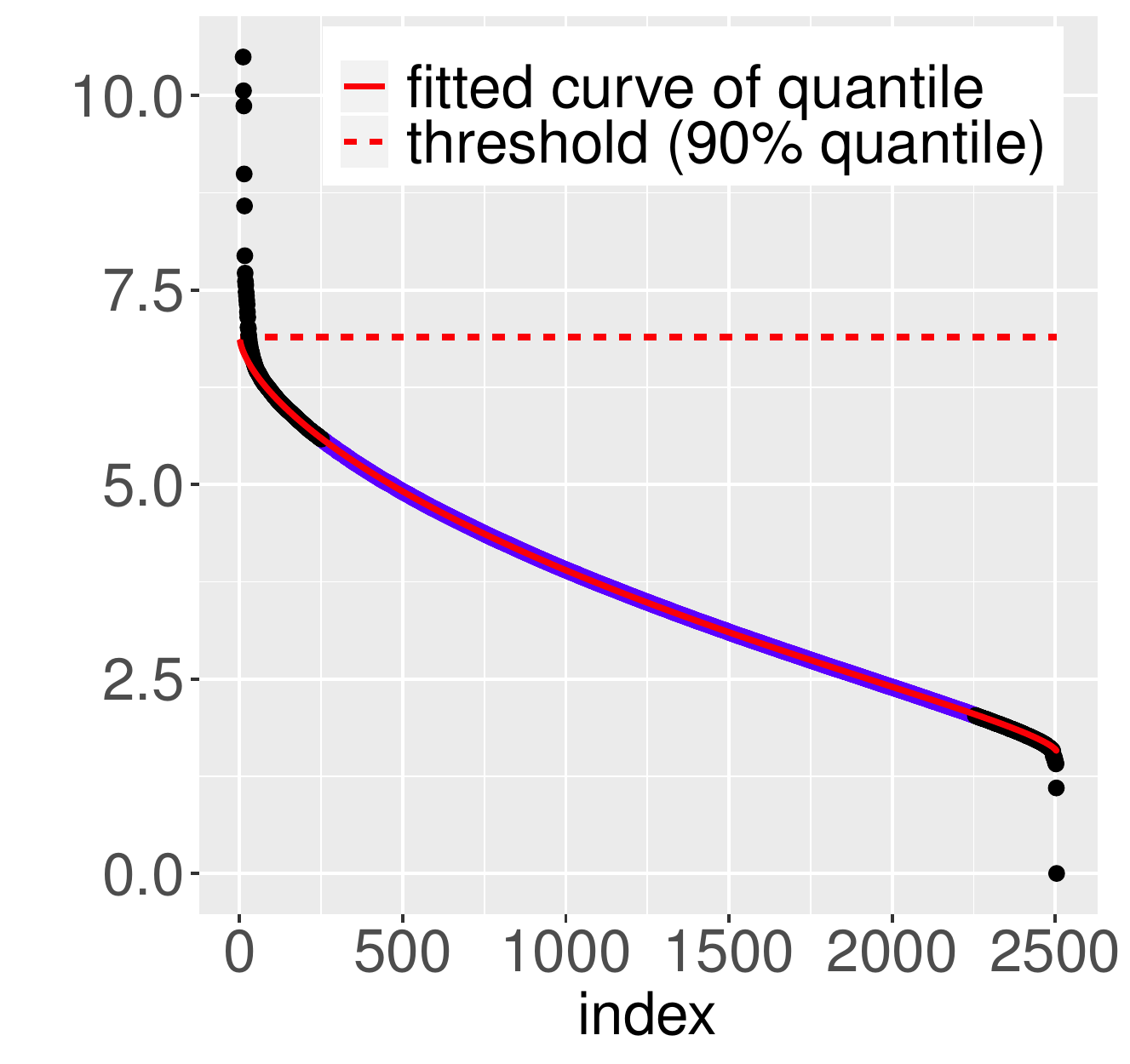}
\vspace{5pt}
\caption{\footnotesize The goodness-of-fit of BEMA on the 1000 Genomes data. The left panel is the plot of $\hat{\lambda}_k$ versus $\bar{F}_{\gamma_n}^{-1}(k/\tilde{p};1, \hat{\theta})$ (which is quantile of the theoretical limit of ESD with estimated $\theta$) for $\alpha n\leq k\leq (1-\alpha) n$. It fits well a line crossing the origin. The right panel plots $\hat{\lambda}_k$ versus $k$, where the red solid curve is the curve of $\bar{F}_{\gamma_n}^{-1}(k/\tilde{p};\hat{\sigma}^2, \hat{\theta})$ versus $k$; for better visualization, this curve is only plotted for $11\leq k\leq 2504$. It fits well the bulk eigenvalues (blue dots).
These two plots suggest that the spiked covariance model \eqref{mod-Sigma2} is suitable for this dataset.}  
\vspace{45pt}
\end{subfigure}
\hspace{10pt}
\begin{subfigure}[b]{.55\textwidth}
\includegraphics[width=0.492\textwidth,trim=0 10 30 20, clip=true]{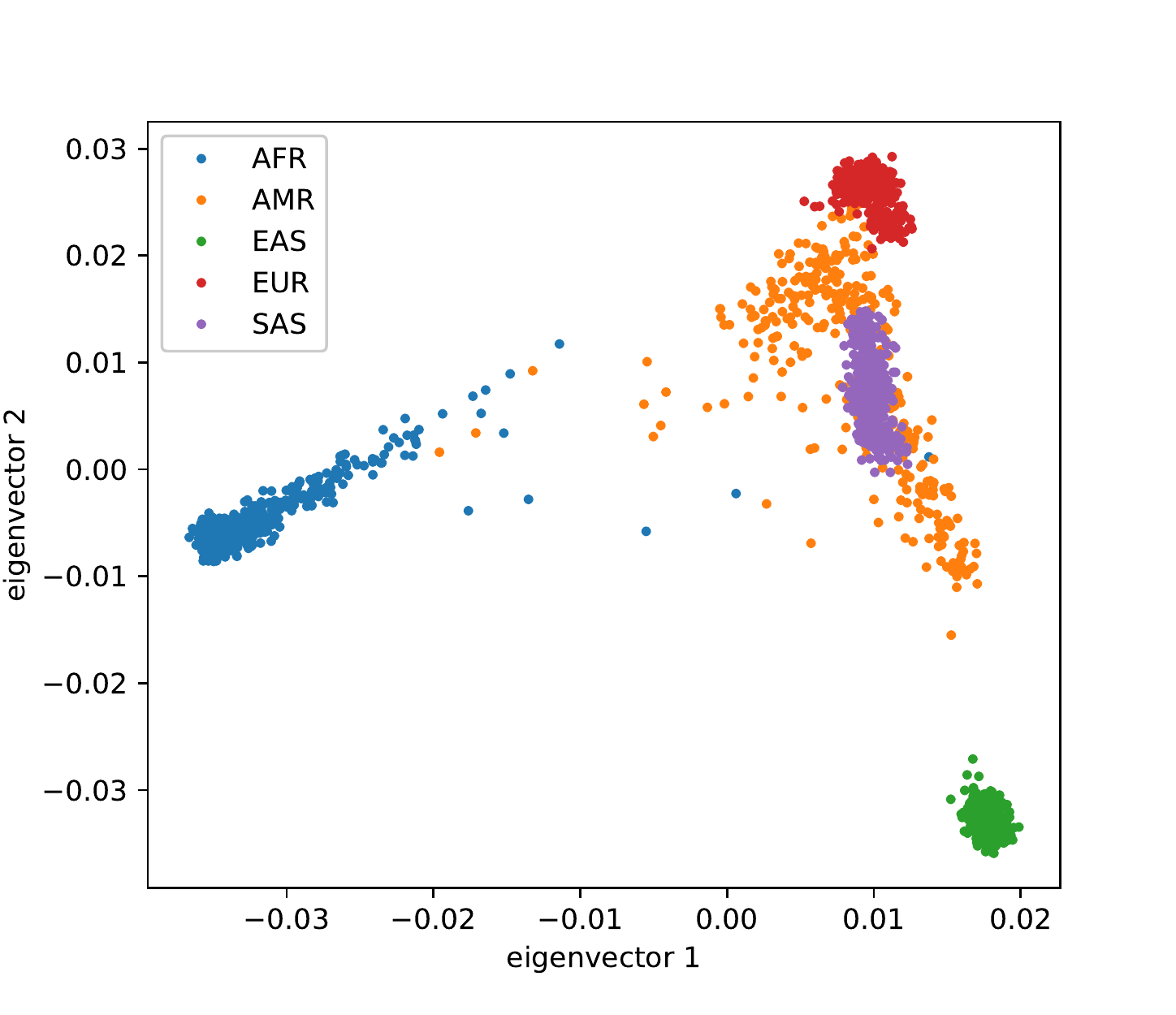}
\includegraphics[width=0.492\textwidth, trim=0 10 30 20, clip=true]{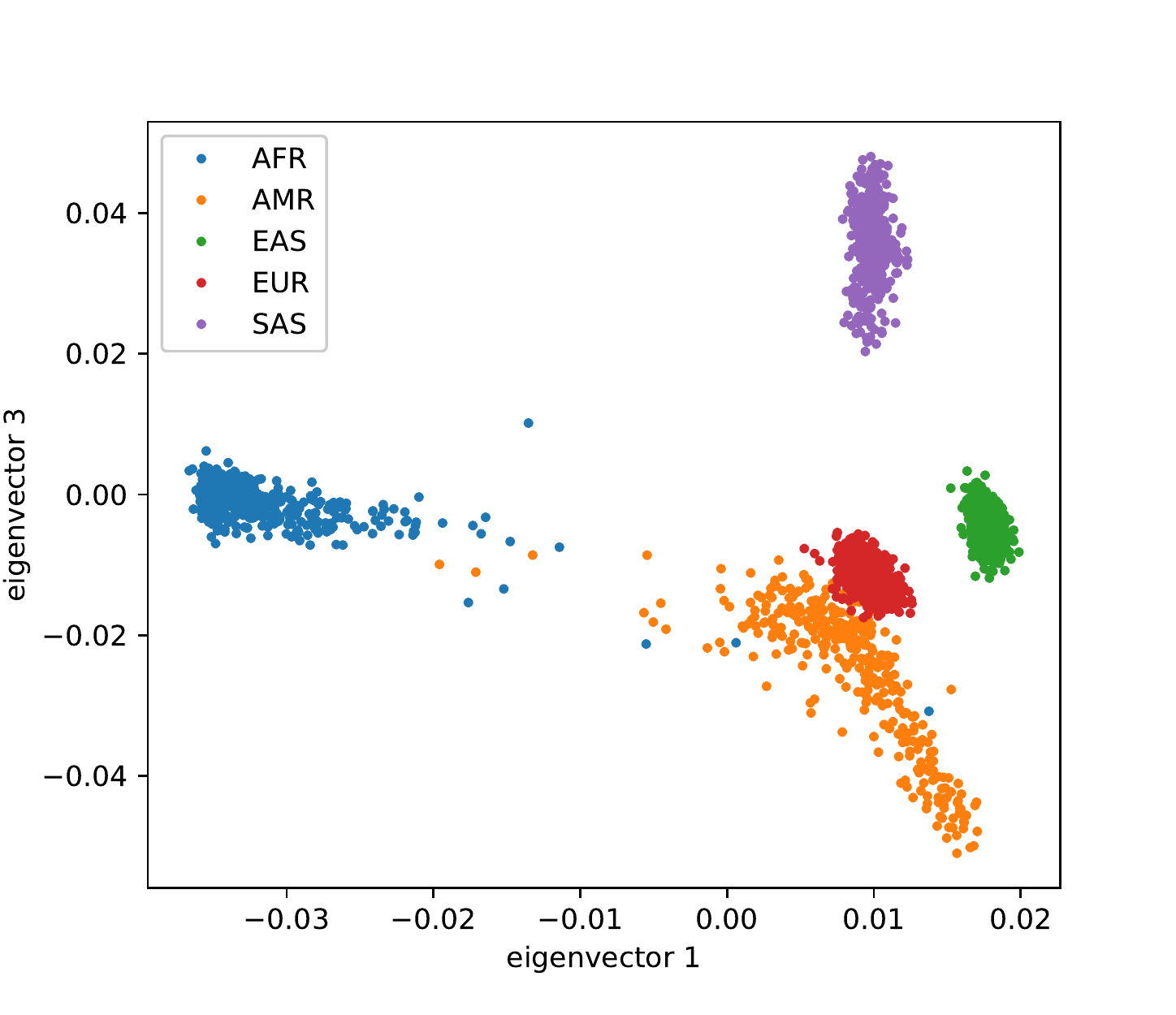}\\
\includegraphics[width=0.492\textwidth, trim=0 10 30 20, clip=true]{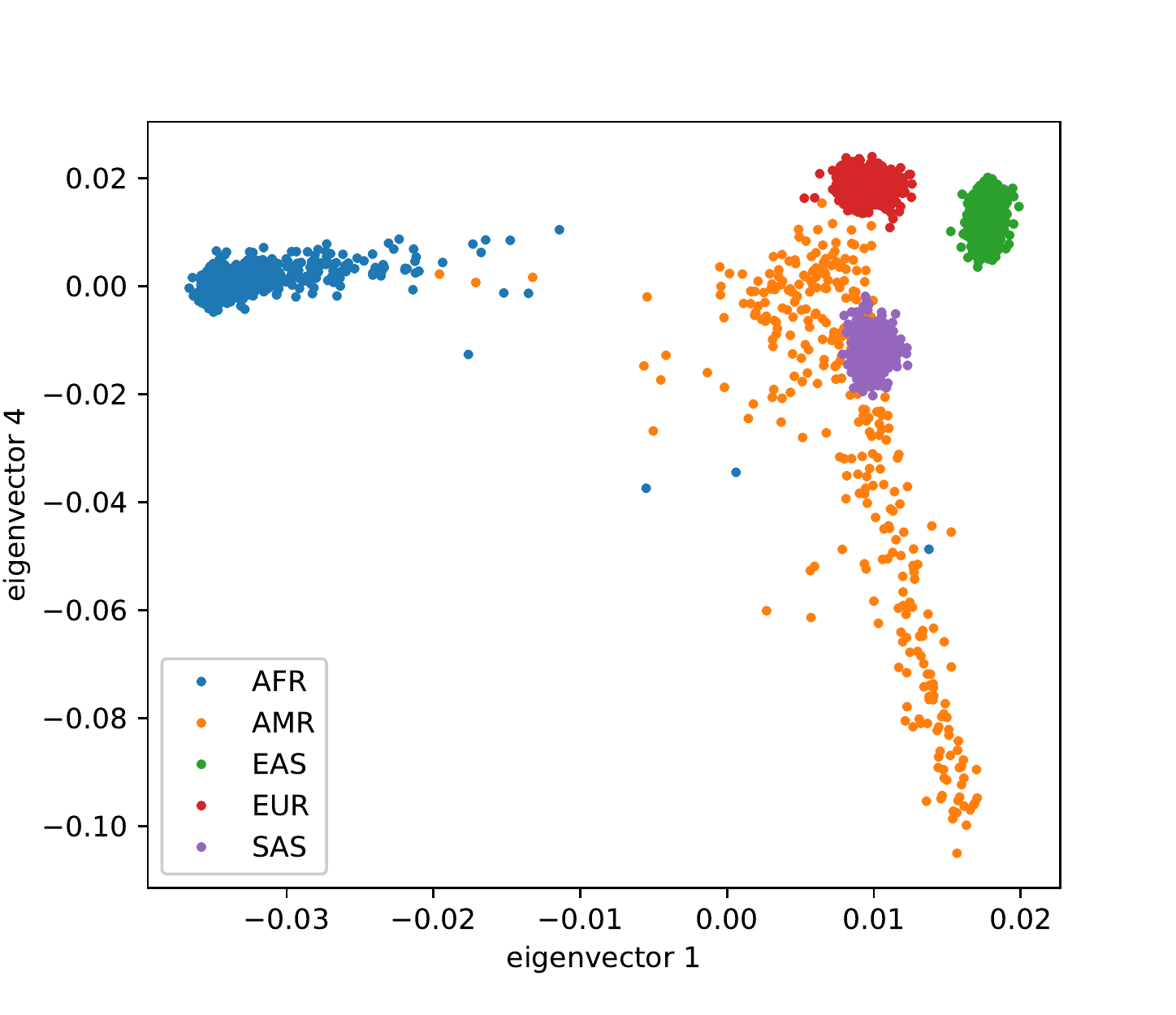}
\includegraphics[width=0.492\textwidth, trim=0 10 30 20, clip=true]{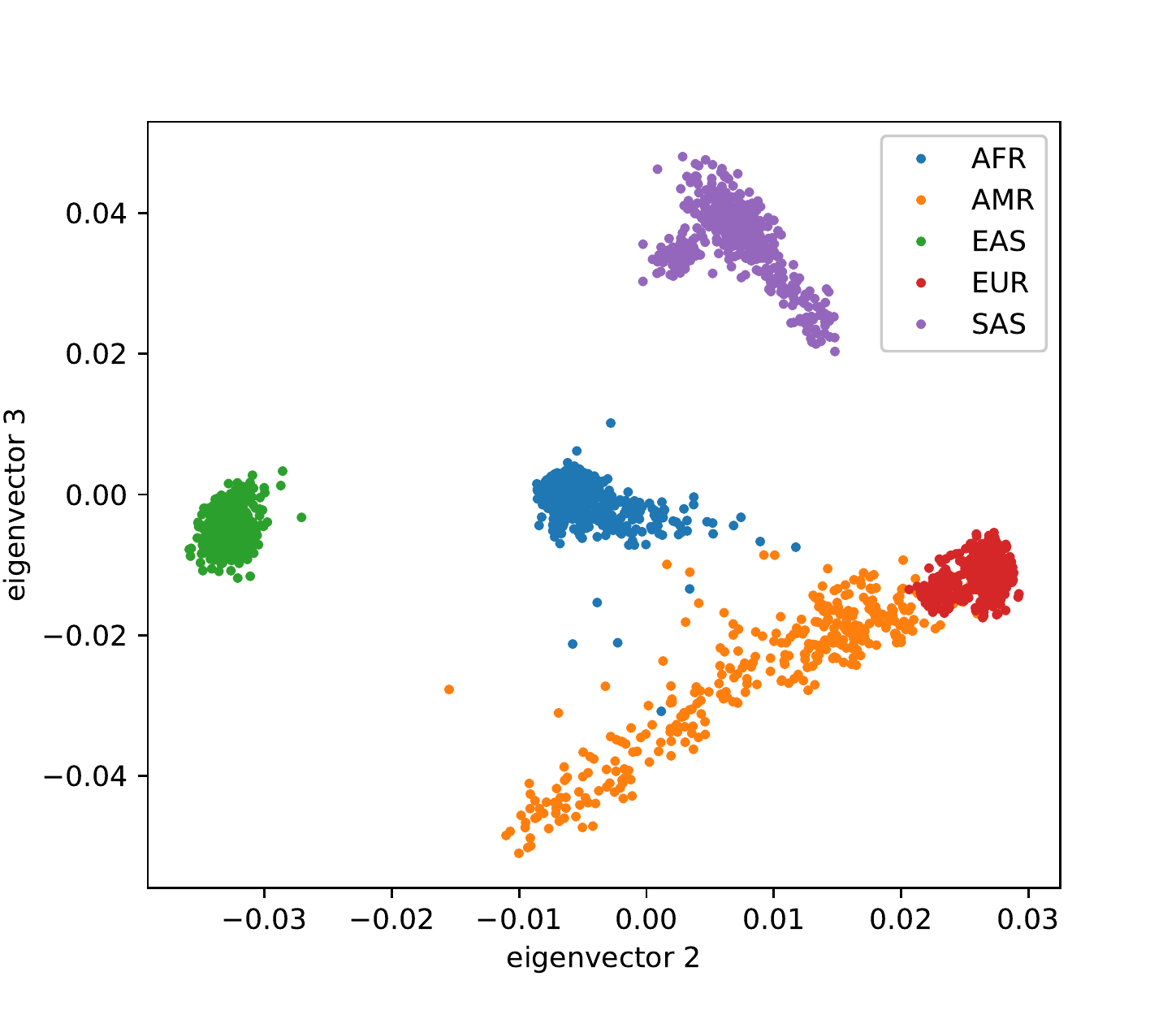}\\
\includegraphics[width=0.492\textwidth, trim=0 10 30 20, clip=true]{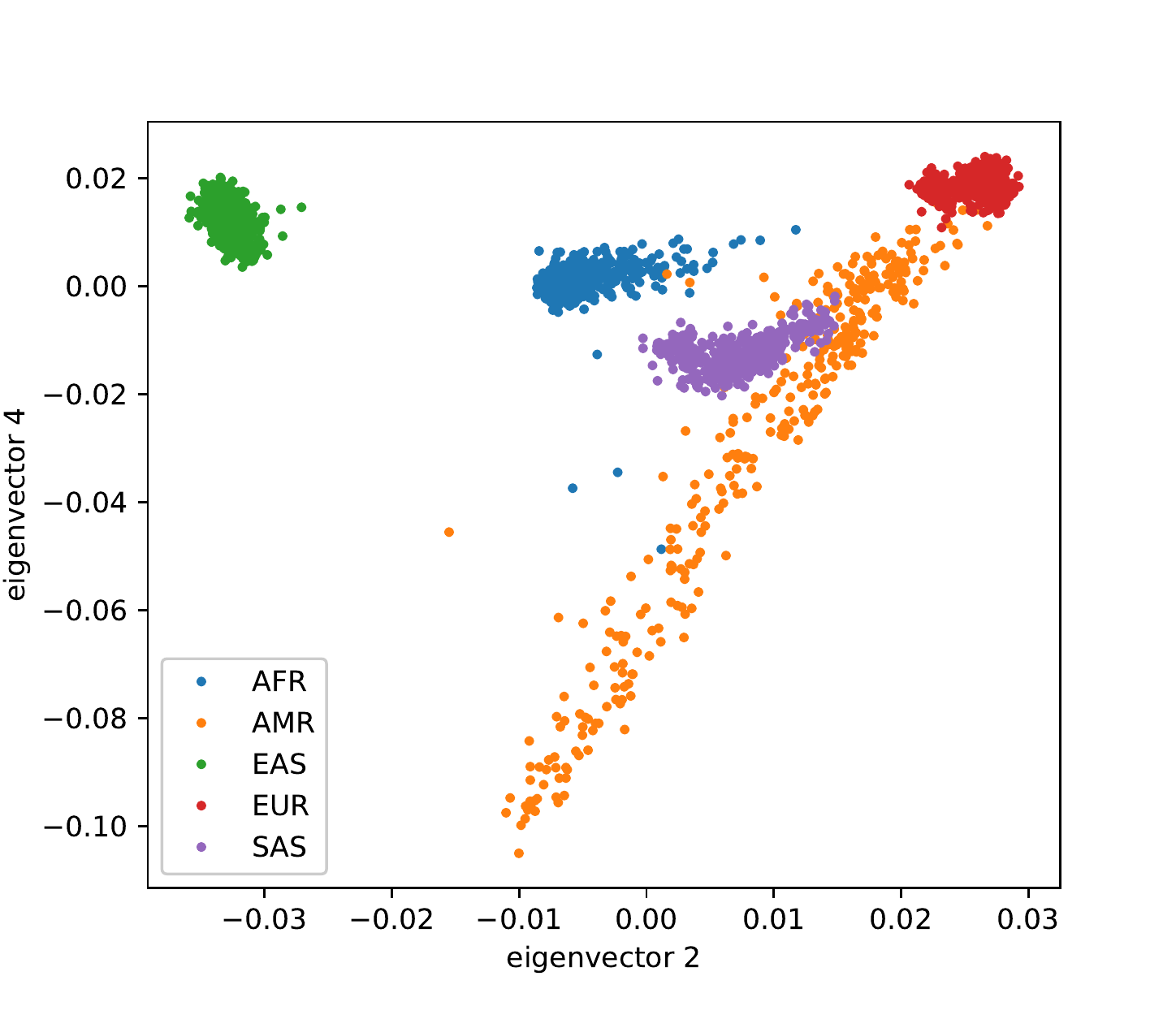}
\includegraphics[width=0.492\textwidth, trim=0 10 30 20, clip=true]{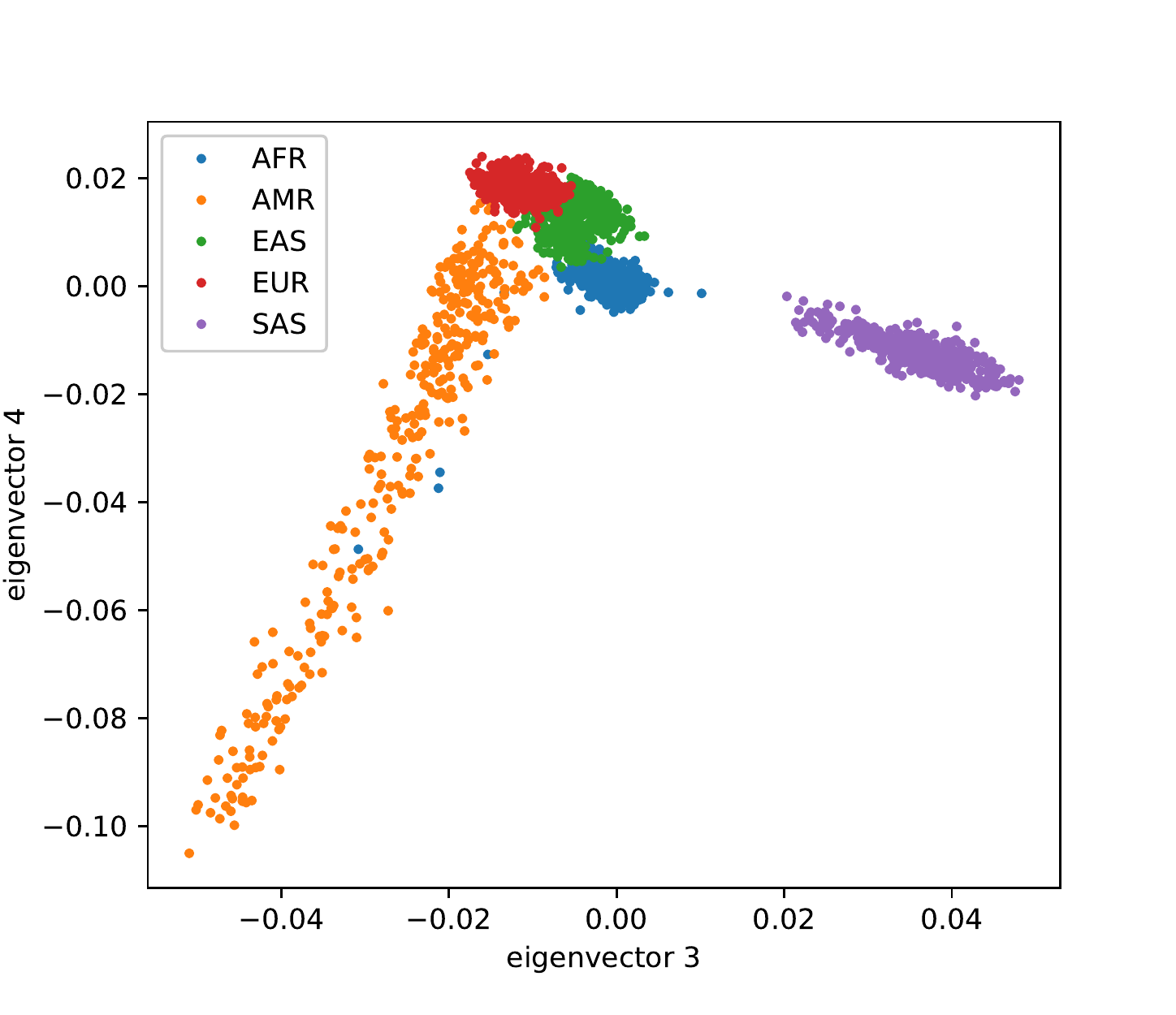}
\caption{\footnotesize The plots of singular vectors of $\bX$.}
\end{subfigure}
\caption{Results for analysis of the 1000 Genomes data .} \label{fig:1000G}
\end{figure}

We apply BEMA with $(\alpha, \beta, M)=(0.1, 0.1, 500)$. First, we check the goodness-of-fit. BEMA outputs $(\hat{\theta},\hat{\sigma}^2)=(4.256,0.377)$. Figure~\ref{fig:1000G}(a) shows the Q-Q plot and the scree plot, with reference curves from the BEMA fitting. The  meaning of these plots is the same as described in Section~\ref{subsec:Lung} and is also explained in the caption of this figure, which we do not repeat here. The conclusion is that our proposed spiked covariance model \eqref{mod-Sigma2} is an excellent fit to this dataset. 

The estimated model for eigenvalues of the residual covariance matrix is $\mathrm{Gamma}(\hat{\theta},\, \hat{\theta}/\hat{\sigma}^2)=\mathrm{Gamma}(4.256, 11.3)$. We note that the variance of the genotype on each SNP is $2q(1-q)$, where $q$ is the null Minor Allele Frequency (MAF) of this SNP. We thus interpret the BEMA fitting as follows: After the ancestry effect is removed, the null MAFs $q_j$ (on LD pruned SNPs) satisfy that $2q_j(1-q_j)\overset{iid}{\sim}\mathrm{Gamma}(4.256, 11.3)$. The mean and standard deviation of this gamma distribution is $0.377$ and $0.18$, respectively.

Next, we look at the estimation of $K$. The BEMA algorithm outputs $\hat{K}=28$, which is very close to the ground truth $K=25$. The 98\% confidence interval of $K$ is $[27,31]$. 

A comparison with other methods is summarized in Table~\ref{tb:realdata}.
EKC and DDPA significantly over-estimate $K$, and Bai\&Ng and DDPA+ significantly under-estimate $K$. DPA gives $\hat{K}=20$, which is relatively close to the ground truth. 
 BEMA and Pass\&Yao both give $\hat{K}=28$, which is closest to the ground truth.
Pass\&Yao assumes that all $\sigma_j^2$ are equal. In this data set, BEMA estimates the standard deviation of $\sigma_j^2$ to be 0.18, which is relatively small. This explains why Pass\&Yao also performs well.
  



\begin{figure}[!tb]
\hspace*{-10pt}
\begin{subfigure}{.4\textwidth}
\centering
\includegraphics[width=.8\textwidth]{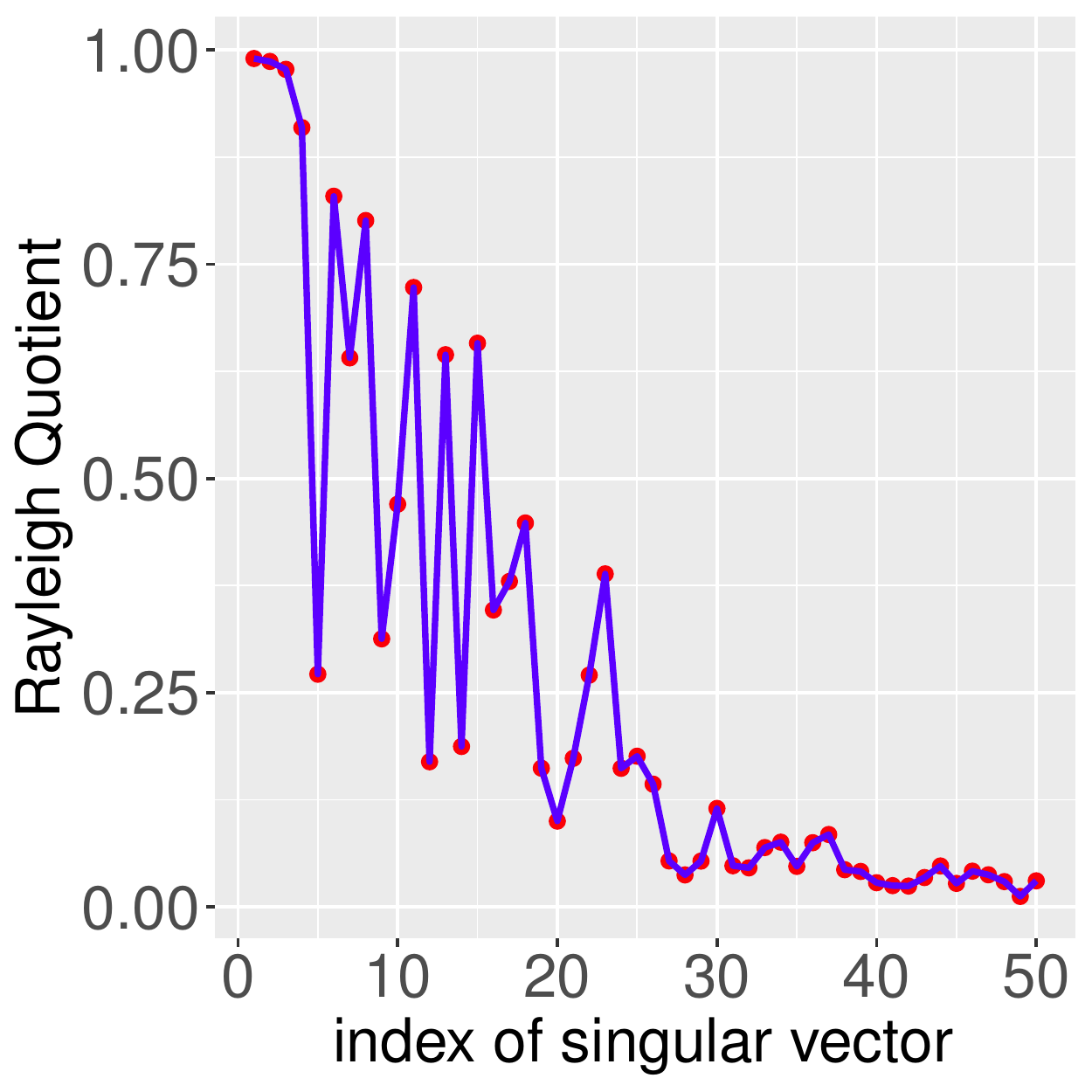}
\caption{\footnotesize The association between singular vectors of $\bX$ and the true ethnicity labels.}
\end{subfigure}
\hspace{15pt}
\begin{subfigure}{.65\textwidth}
\includegraphics[width=1\textwidth, trim=100 30 100 75, clip=true]{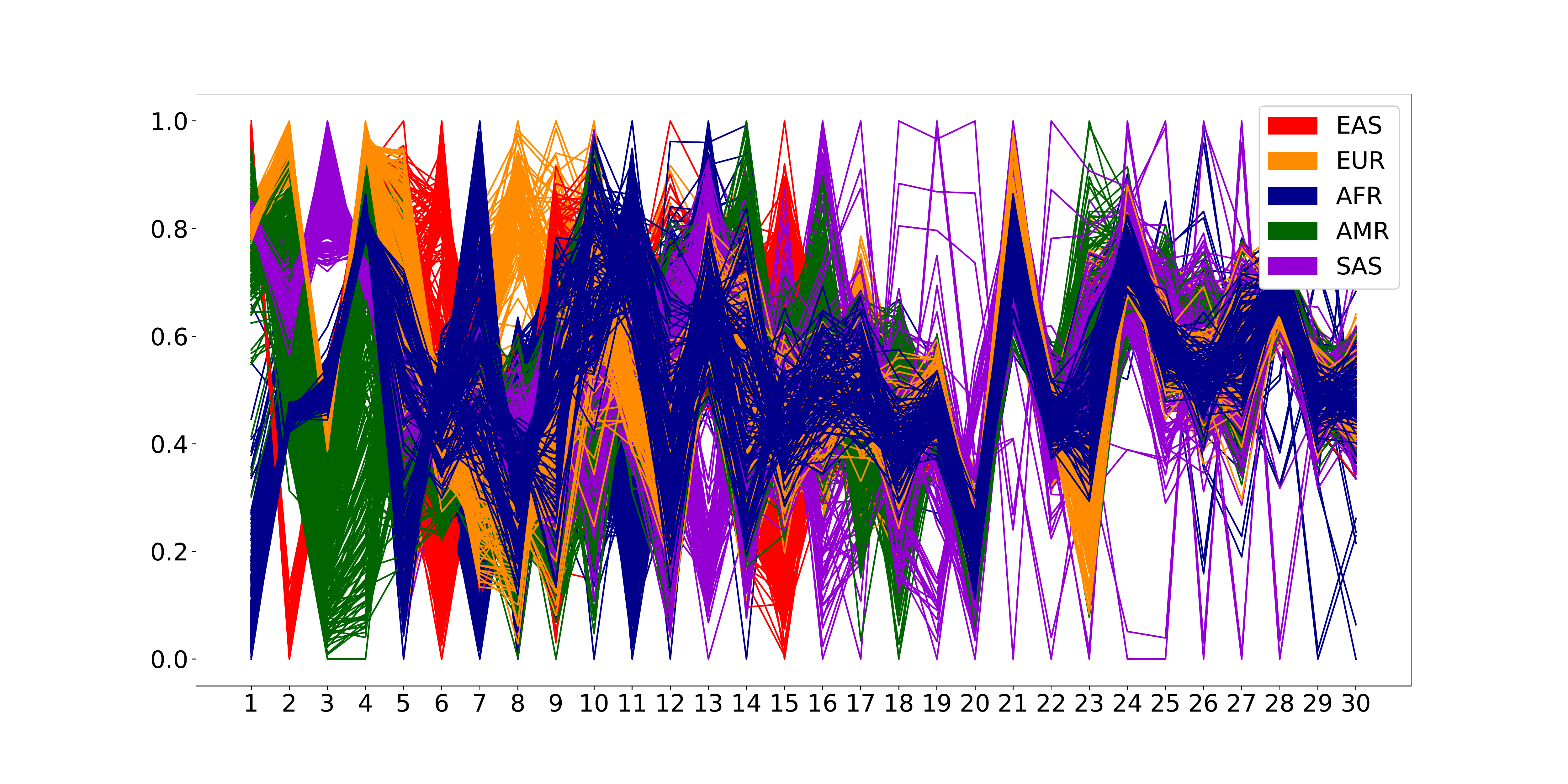}
\caption{\footnotesize The parallel coordinate plot of singular vectors, color-coded by five super-populations.}
\end{subfigure}\\
\vspace{30pt}
\begin{subfigure}{1\textwidth}
\includegraphics[width=.495\textwidth, trim=100 60 100 0, clip=true]{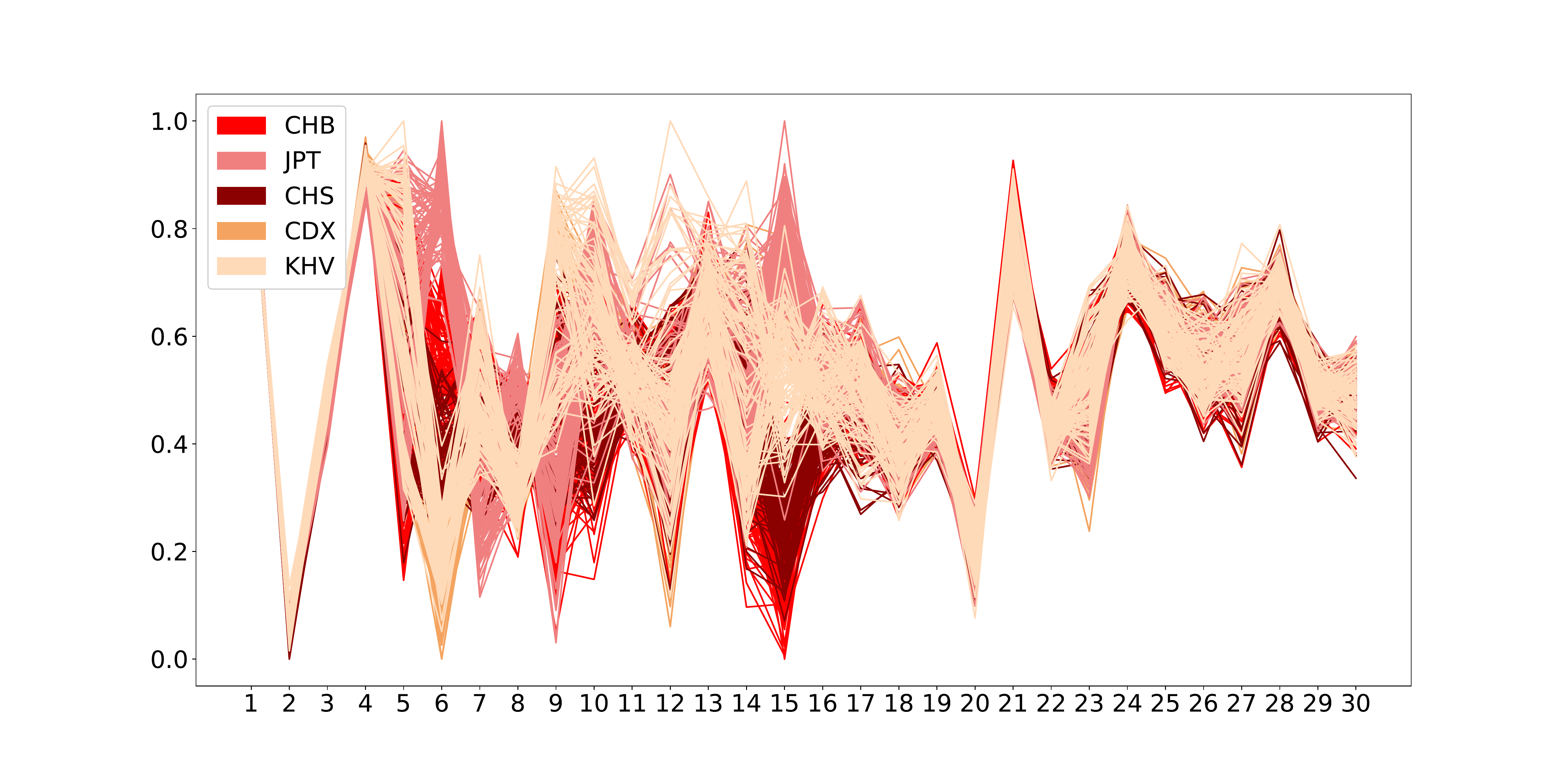}
\includegraphics[width=.495\textwidth, trim=100 60 100 0, clip=true]{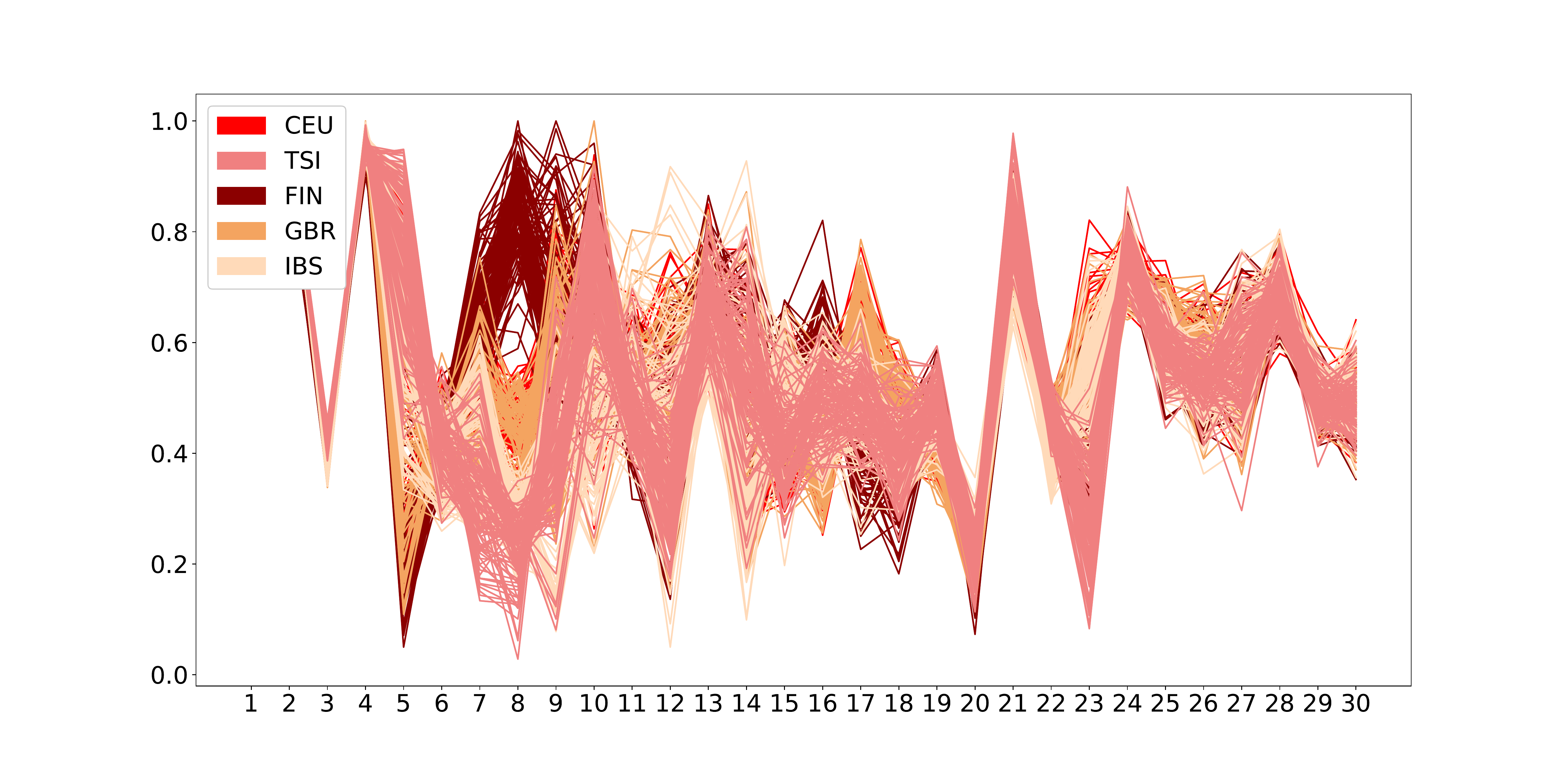}
\includegraphics[width=.495\textwidth, trim=100 60 100 0, clip=true]{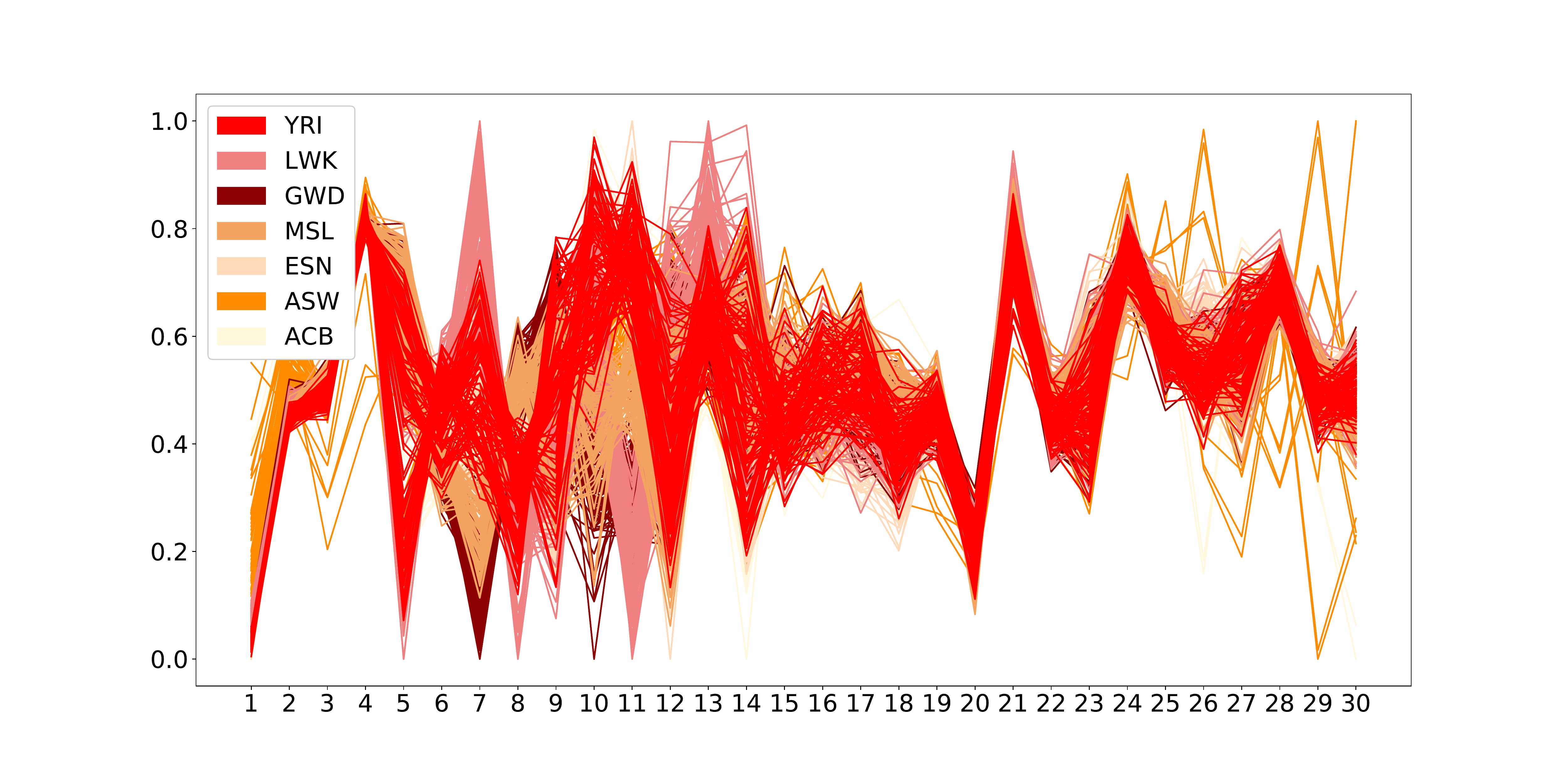}
\includegraphics[width=.495\textwidth, trim=100 60 100 0, clip=true]{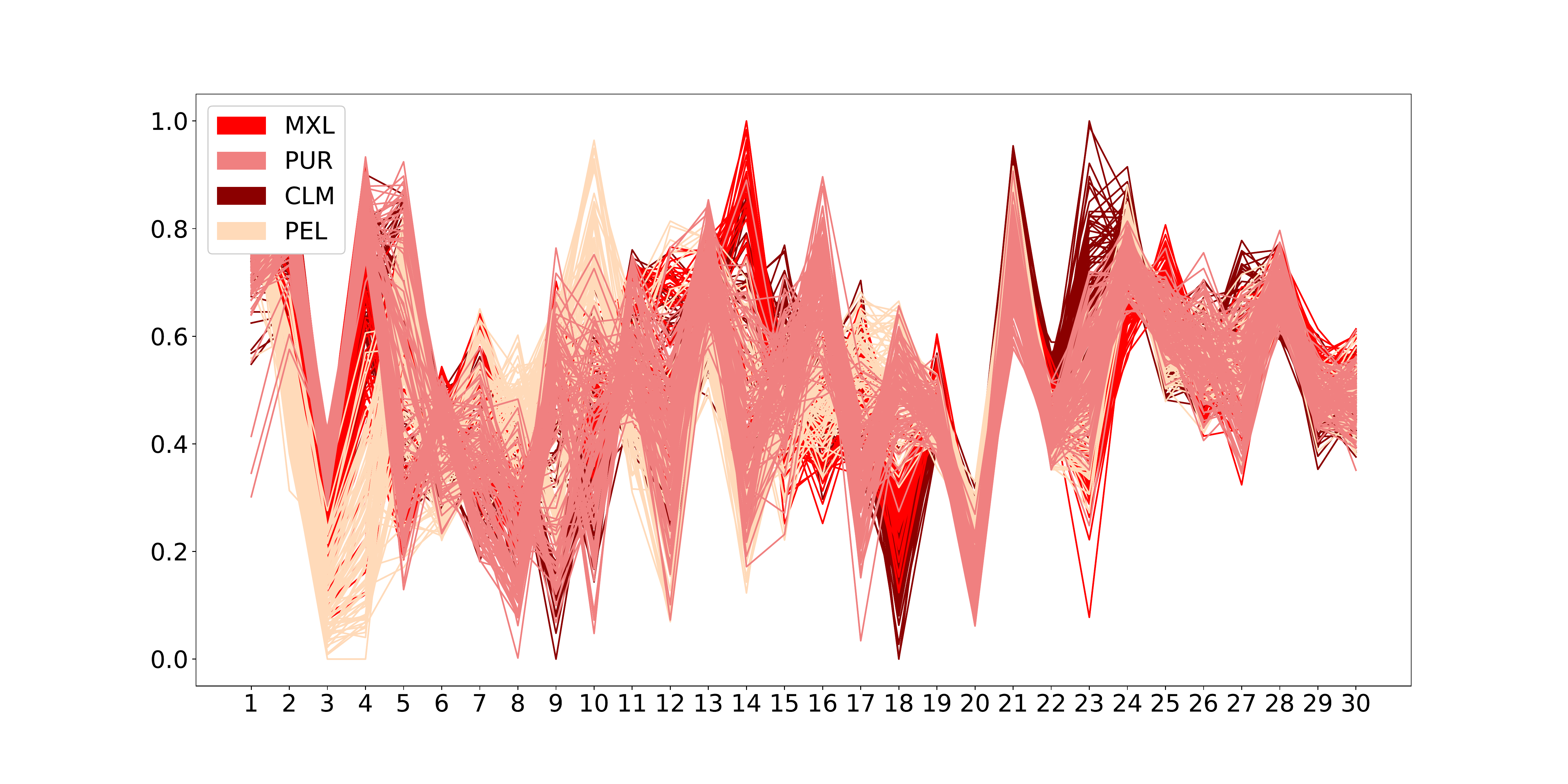}
\includegraphics[width=.495\textwidth, trim=100 60 100 0, clip=true]{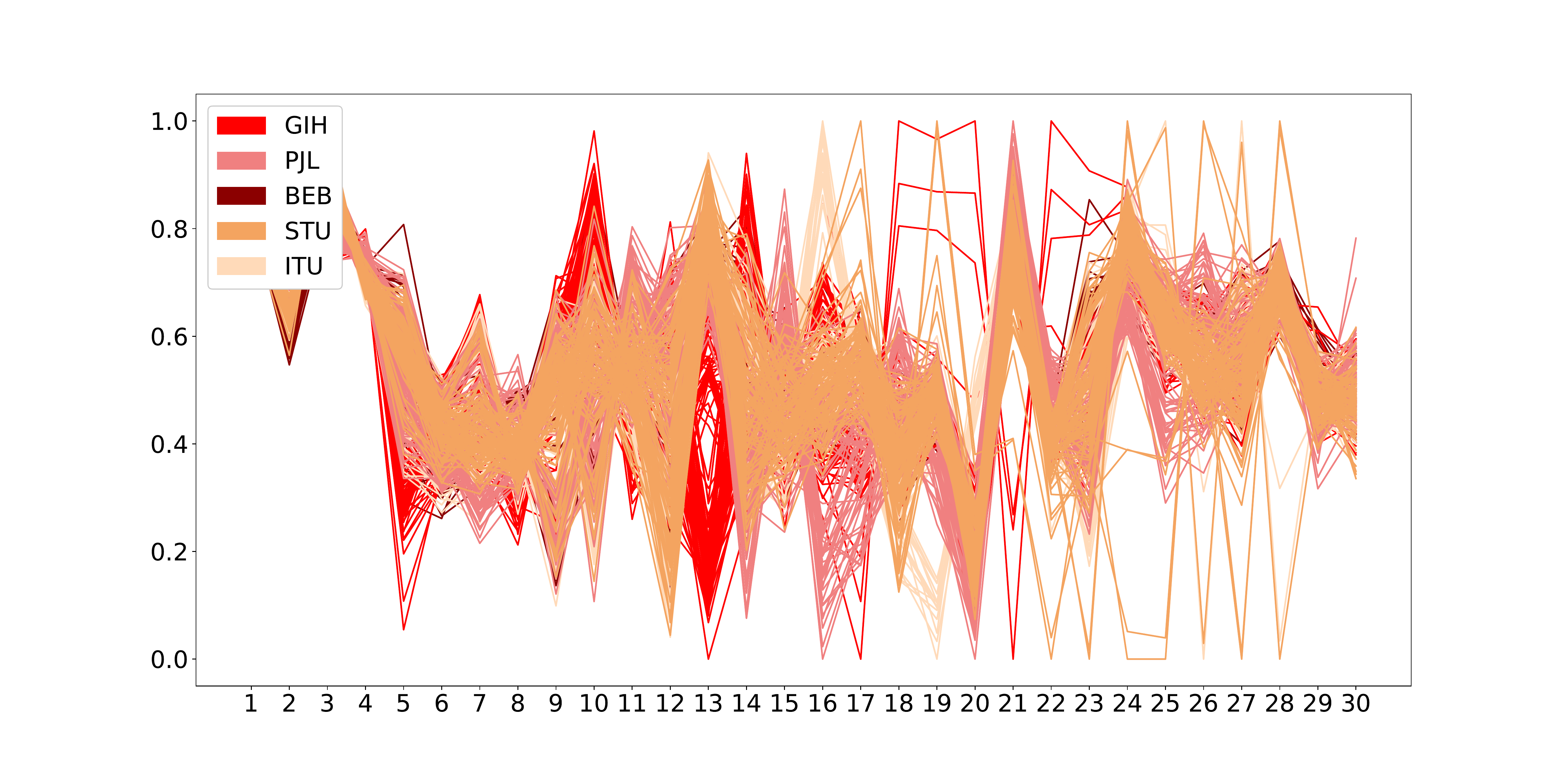}
\caption{\footnotesize The parallel coordinate plots of singular vectors for each super-population, color-coded by the ethnicity groups within each super-population. The five super-populations are EAS (top left), EUR (top right), AFR (middle left), AMR (middle right), and SAS (bottom left). The sub-population labels used in the legends of can be found in \cite{10002015global}. } 
\end{subfigure}
\vspace*{-30pt}
\caption{Interpretation of results for the 1000 Genomes data.} \label{fig:1000G-add}
\end{figure}

Last, we validate the results by investigating the singular vectors of $\bX$. 
We first measure the association between each singular vector and the true ethnicity labels by the Rayleigh quotient \citep{horn2012matrix}. Let $\hat{\boldeta}_k\in\mathbb{R}^n$ be the $k$th left singular vector of the centralized data matrix. We treat its entries as $n$ data points and compute the ratio of between-cluster-variance and within-cluster-variance, denoted as $RQ_k$. A larger $RQ_k$ indicates that $\hat{\boldeta}_k$ is more correlated with the true ethnicity labels.  Figure~\ref{fig:1000G-add}(a) plots $RQ_k$ versus $k$. The first a few singular vectors have very high association with the ethnicity labels. These singular vectors capture the super population structure. The pairwise scatter plots of the first 4 singular vectors are contained in Figure~\ref{fig:1000G}(b), which show clearly that super populations are well separated on these singular vectors. Besides the first few singular vectors, the remaining singular vectors capture more of the sub-structure within each super population. Figure~\ref{fig:1000G-add}(b) is the parallel coordinate plot. 
In Figure~\ref{fig:1000G-add}(c), we re-generate parallel coordinate plots by restricting to each super population. Within the super population AMR, there is still separation of ethnicity groups for $k$ as large as $27$. This explains why BEMA outputs a $\hat{K}$ that is slightly larger than the ground truth.

\section{Discussion} \label{sec:discussion}

We propose a new method for estimating the number of spiked eigenvalues in a large covariance matrix. The novelty of our method lies in a systematic approach to incorporating bulk eigenvalues in the estimation of $K$. Under a working model which assumes the diagonal entries of the residual covariance matrix are {\it iid} drawn from a Gamma distribution,  we fit a parametric curve on bulk eigenvalues. The estimated parameters of this curve are then used to decide a threshold for top eigenvalues and produce an estimator of $K$. We study the theoretical properties of our method under a standard spiked covariance model, and show that our estimator requires weaker conditions for consistent estimation of $K$ compared with the existing methods. We examine the performance of our method using both simulated data and two real data sets. Our empirical results show that the proposed method outperforms other competitors in a variety of scenarios.

Our approach is conceptually connected to the empirical null \citep{efron2004large} in multiple testing. The empirical null imposes a working model (e.g., a normal distribution) on $Z$-scores of individual null hypotheses and estimates the parameters of this distribution from a large number of $Z$-scores. The fitted null model is then used to correct $p$-values and help identify the non-null hypotheses. Similarly, we impose a working model (i.e., a Gamma distribution) on non-spiked population eigenvalues and estimate the parameters of this distribution from a large number of bulk empirical eigenvalues. The fitted null model is then used to assist estimation of $K$. From this perspective, our method  
can be regarded as a {\it conceptual} application of 
the empirical null approach to eigenvalues. Meanwhile, our setting is much more complicated than that in multiple testing. The bulk eigenvalues are highly correlated, and their marginal distribution has no explicit form. 
These impose great challenges on algorithm design and theoretical analysis.

For the theoretical study, we first analyze the special case of $\theta=\infty$. This corresponds to the well-known standard spiked covariance model \citep{Johnstone2001}, which has attracted many theoretical interests. Our theory contributes to this literature with an explicit error bound on estimating $\sigma^2$ and consistency theory on estimating $K$. The theoretical study for a general $\theta$ that corresponds to the setting of heterogeneous residual variances is of great interest but is technically challenging. Instead, we study a proxy model where the population eigenvalues are $iid$ drawn from a truncated Gamma distribution. Under this model we derive error bounds for $(\hat{\sigma}^2,\hat{\theta})$ and prove the consistency of $\hat{K}$ with mild conditions. The analysis uses advanced results in random matrix theory \citep{bloemendal2016principal,knowles2017anisotropic,ding2020spiked}.


The method can be extended in multiple directions. Here we assume that the diagonal entries of the residual covariance matrix are from a Gamma distribution. It can be generalized to other parametric distributions. In Section~\ref{subsec:theory-general}, we have already seen a variant of our method by using a truncated Gamma distribution, which assumption helps eliminate extremely large variances for the residuals. We can also use a mixture of Gamma distributions to accommodate heterogeneous feature groups. Our main algorithm can be easily adapted to such cases. 
When the distribution family is unknown, we may combine our method with the techniques in nonparametric density estimation. The thresholding scheme in our method can also be modified. We currently apply a single threshold to all eigenvalues. Alternatively, we may use different thresholds for different eigenvalues. One proposal is to use the $(1-\beta)$-quantile of the distribution of $\hat{\lambda}_k^*$ in the null model \eqref{SQMgamma-threshold} as a threshold for $\hat{\lambda}_k$. We leave these extensions to future work. 

In the numerical experiments, our method exhibits robustness to model misspecification. It is suggested by Simulation 4 of Section~\ref{sec:simulation} that our method continues to work when the residual covariance matrix is a Toeplitz matrix, or a block-wise diagonal matrix,  or a sparse matrix. A theoretical understanding to this phenomenon will be useful. As stated in Section~\ref{sec:simulation}, we have observed empirically that there always exist $(\sigma^2, \theta)$ such that the theoretical limit of ESD induced by the Gamma model \eqref{mod-D} can accurately approximate the theoretical limit of ESD induced by a Toeplitz or  block-wise diagonal or sparse covariance matrix. It remains an interesting question  on how to justify it theoretically. We leave it to future work.

\section*{Appendix}
\begin{appendix}

\section{\texttt{GetQT} algorithms} \label{sec:getQT}
We present details of the \texttt{GetQT} algorithms used in BEMA. Under the general spiked covariance model \eqref{mod-Sigma2}, the empirical spectral distribution (ESD) 
converges to a fixed distribution $F_{\gamma}(x;\sigma^2, \theta)$.
Write $\gamma_n=p/n$. The purpose of the algorithm \texttt{GetQT}($y,\gamma_n,\theta$) is as follows: Fixing $\sigma=1$, given any $\theta>0$ and $y\in [0,1]$, it outputs the $y$-upper-quantile of the distribution $F_{\gamma_n}(x; 1, \theta)$.

\subsection{The Monte Carlo simulation algorithm \texttt{GetQT1}} \label{subsec:GetQT1}

As explained in Section~\ref{subsec:main-RMT}, $F_{\gamma_n}(\cdot; 1, \theta)$ is also the theoretical limit of the ESD under the following null covariance model:
\beq \label{gamma-model}
\bSigma=\mathrm{diag}(\sigma^2_1,\ldots,\sigma^2_p), \qquad \mbox{where}\;\;  \sigma^2_k\overset{iid}{\sim}\mathrm{Gamma}(\theta,\theta). 
\eeq
We can simulate data from \eqref{gamma-model} and use its ESD as a numerical approximation to $F_{\gamma_n}(\cdot; 1, \theta)$.

Write $\tilde{p}=\min\{n,p\}$ and $y=k/\tilde{p}$. When the population covariance matrix satisfies \eqref{gamma-model}, the $k$th eigenvalue of the sample covariance matrix, $\hat{\lambda}_k$, is asymptotically close to the $y$-upper-quantile of $F_{\gamma_n}(\cdot; 1, \theta)$. We thereby use the mean of $\hat{\lambda}_k$, obtained by sampling the data matrix multiple times, to estimate the desired quantile. We note that model \eqref{gamma-model} only specifies how to sample $\bSigma$, but it does not specify how to sample $\bX_i$'s. Due to universality theory of eigenvalues \citep[Section 3.3]{knowles2017anisotropic}, the choice of distribution of $\bX_i$'s does not matter. For convenience, we  sample $\bX_i$'s from multivariate normal distributions. See Algorithm 3.


\RestyleAlgo{boxed}
\begin{algorithm}[!htb]
{\bf Algorithm 3}. GetQT1.
\vspace*{2pt}

{\it Input}: $n$, $p$, $\theta$, $k$, and an integer $B$. \\
{\it Output}: An estimate of the $(k/\tilde{p})$-upper-quantile of $F_{\gamma_n}(\cdot; 1, \theta)$.
\vspace*{5pt}

\begin{enumerate} \itemsep +2pt
\item For $b=1,2,\ldots,B$, repeat: First generate $\bSigma^{(b)}$ from \eqref{gamma-model}, and then generate $\bX_i^{(b)}\overset{iid}{\sim}N(0,\bSigma^{(b)})$, $1\leq i\leq n$. Write $\bX^{(b)}=[\bX_1^{(b)},\ldots,\bX_n^{(b)}]^{\top}\in\mathbb{R}^{n\times p}$. Construct the sample covariance matrix $\bS^{(b)}=(1/n)(\bX^{(b)})^{\top}\bX^{(b)}$ and obtain its $k$th eigenvalue $\hat{\lambda}_k^{(b)}$. 
\item Output  $\frac{1}{B}\sum_{b=1}^B \hat{\lambda}_k^{(b)}$ as the estimated $(k/\tilde{p})$-upper-quantile. 
\end{enumerate}  
\end{algorithm}

In the practical implementation, we use the following strategies to further reduce computation time and memory use: (i) When $n$ is smaller than $p$, we no longer construct the $p\times p$ covariance matrix $\bS^{(b)}$. Instead, we construct an $n\times n$ matrix $(1/n)\bX^{(b)}(\bX^{(b)})^{\top}$. This matrix shares the same nonzero eigenvalues as $\bS^{(b)}$ but requires much less memory in eigen-decomposition. This strategy is especially useful for genomic data, where $n$ is typically much smaller than $p$. (ii) In the main algorithm, Algorithm 2, $\texttt{GetQT1}$ is applied multiple times to compute the $(k/\tilde{p})$-upper-quantile for a collection of $k$. We let the sampling step, Step 1 above, be shared across different values of $k$: For each $b=1,2,\ldots,B$, we obtain and store $\hat{\lambda}_k^{(b)}$ for all values of $k$; next, in Step 2, we output the estimated $(k/\tilde{p})$-upper-quantile simultaneously for all values of $k$. This strategy can significantly reduce the actual running time.


\subsection{The deterministic algorithm \texttt{GetQT2}} \label{subsec:GetQT2}
This algorithm directly uses the definition of $F_{\gamma_n}(\cdot;, 1,\theta)$. Let $H_\theta(t)$ be the CDF of $\mathrm{Gamma}(\theta,\theta)$. Given a positive sequence $\xi_n$ such that $\xi_n\to 0$ as $n\to\infty$, let $m_n(y)=m_n(y,\xi_n, \gamma_n,\theta) \in \mathbb{C}^+$ be the unique solution to the equation 
\begin{equation}\label{stieltjes}
y+\mathrm{i}\, \xi_n =-\frac{1}{m_n}+\gamma_n\int \frac{t}{1+t m_n}dH_{\theta}(t). 
\end{equation}
Then, the density of $F_{\gamma_n}(\cdot; 1, \theta)$, denoted by $f_{\gamma_n}(y; 1, \theta)$, is approximated by
\beq \label{density}
 \hat{f}^*_{\gamma_n}(y; 1, \theta) =  \frac{1}{\pi (\gamma_n\wedge 1)}\, \Im( m_n(y, \xi_n, \gamma_n,\theta)),  
\eeq
where $\Im(\cdot)$ denotes the imaginary part of a complex number.  The choice of $\xi_n$ needs to satisfy $\xi_n\gg n^{-1}$, in order to guarantee that the approximation is not governed by stochastic fluctuations \citep{knowles2017anisotropic}. We choose $\xi_n=n^{-2/3}$ for convenience.


The above motivates a three-step algorithm. 
\begin{enumerate} \itemsep +2pt
\item Fix a grid $y_1<y_2<\ldots<y_N$. Solve equation \eqref{stieltjes} to obtain $ m_n(y_j)$ for $1\leq j\leq N$.
\item Use equation \eqref{density} to obtain $\hat{f}^*_{\gamma_n}(y_j; 1, \theta)$, for $1\leq j\leq N$. Obtain the whole density curve $\hat{f}_{\gamma_n}(y; 1,\theta)$ by linear interpolation. 
\item Find $q$ such that $\int_q^{(1+\sqrt{\gamma_n})^2}\hat{f}_{\gamma_n}(z; 1,\theta)dz = y$. Output $q$ as the estimated $y$-upper-quantile. 
\end{enumerate}

\noindent
Step 2 is straightforward. Step 3 is also easy to implement, since $\hat{f}_{\gamma_n}(y; 1,\theta)$ is a piece-wise linear function. Below, we describe Step 1 with more details.

\RestyleAlgo{boxed}
\begin{algorithm}[!tb]
{\bf Algorithm 4}. GetQT2. 
\vspace*{2pt}

{\it Input}: $n$, $p$, $\theta$, and $y\in [0,1]$. \\
{\it Output}: An estimate of the $y$-upper-quantile of $F_{\gamma_n}(\cdot; 1, \theta)$.
\vspace*{5pt}

\noindent
Step 1: Write $\tilde{p}=n\wedge p$ and $\gamma_n=p/n$. Fix a grid $y_1<y_2<\ldots y_{N-1}<y_N$. For each $1\leq j\leq N$, compute $\hat{m}_n(y)$ as follows:

\begin{itemize} \itemsep -2pt
\item  For a tuning parameter $\delta>0$, construct the set of grid points in $\mathbb{R}\times \mathbb{R}^+$: 
\[
\hspace*{-3em}    S_{y,\gamma_n,\delta}=\bigl\{(a,b):a=k\delta, b=\ell\delta,\ k,\ell\in \mathbb{Z},\  (a-1/y_j)^2+b^2\leq \gamma_n/y_j^2, \ a<(\gamma_n-1)/2y_j\bigr\}.
\]
\item For each $(a,b)\in S_{y,\gamma_n,\delta}$ and  $\xi_n=n^{-2/3}$, compute
\[
\Delta(a,b) = \Big|y+ \mathrm{i}\, \xi_n + \frac{1}{m}-\gamma_n\int \frac{t}{1+tm}dH_{\theta}(t)\Big|, 
\]
where $H_{\theta}(t)$ is the CDF of $\mathrm{Gamma}(\theta,\theta)$. The integral above can be computed via standard Monte Carlo approximation (by sampling data from $\mathrm{Gamma}(\theta,\theta)$). 

\item Find $(\hat{a},\hat{b})=\mathrm{argmin}_{(a,b)\in S_{y,\gamma_n,\delta}}\Delta(a,b)$. Let $\hat{m}(y)=\hat{a}+\hat{b}\mathrm{i}$. 
\end{itemize}

\vspace*{5pt}

\noindent
Step 2: Let $\hat{f}_{\gamma_n}(y_j; 1,\theta)=\frac{1}{\pi (\gamma_n\wedge 1)}\, \Im( \hat{m}(y))$, for $1\leq j\leq N$. For any $y_{j-1}<z<y_j$, let 
\[
\hat{f}_{\gamma_n}(z; 1,\theta) = \frac{y_j-z}{y_j-y_{j-1}}\hat{f}_{\gamma_n}(y_{j-1}; 1,\theta)+\frac{z-y_{j-1}}{y_j-y_{j-1}}\hat{f}_{\gamma_n}(y_j; 1,\theta). 
\]

\vspace*{5pt}

\noindent
Step 3: Find $q$ such that $\int_q^{(1+\sqrt{\gamma_n})^2}\hat{f}_{\gamma_n}(z; 1,\theta) = y$. Output $q$ as the estimated $y$-upper-quantile. 

\end{algorithm}

In Step 1, fix $y$ and write $m=a+b\mathrm{i}$, where $\mathrm{i}=\sqrt{-1}$, and $a\in \mathbb{R}$ and $b\in \mathbb{R}^+$ are the real and imaginary parts of $m$, respectively. We aim to find $(a,b)$ so that $m$ solves the complex equation \eqref{stieltjes}.  Pretending that $\xi_n=0$, 
the equation \eqref{stieltjes} can be re-written as a set of real equations: \footnote{The second equation is obtained by letting the imaginary part of both hand sides of \eqref{stieltjes} be equal. The first equation is obtained by letting the real part of both hand sides of \eqref{stieltjes} be equal and then substituting $\frac{a}{a^2+b^2}$ by $a$ times the second equation.}
\[
\begin{cases}
y=\gamma_n\int\frac{t}{1+2at+(a^2+b^2)t^2}dH_{\theta}(t),\cr
\frac{1}{a^2+b^2}=\gamma_n\int\frac{t^2}{1+2at+(a^2+b^2)t^2}dH_{\theta}(t), 
\end{cases} \quad \Longleftrightarrow \qquad 
\begin{cases}
2ay = \gamma_n\int\frac{2at}{1+2at+(a^2+b^2)t^2}dH_{\theta}(t),\cr
1=\gamma_n\int\frac{(a^2+b^2)t^2}{1+2at+(a^2+b^2)t^2}dH_{\theta}(t). 
\end{cases}
\]
First, by combining the above equations with $\gamma_n= \gamma_n\int\frac{1+2at+(a^2+b^2)t^2}{1+2at+(a^2+b^2)t^2}dH_{\theta}(t)$, we have 
\[
    \gamma_n-1-2ay=\gamma_n\int\frac{1}{1+2at+(a^2+b^2)t^2}dH_{\theta}(t) >0.
\]
It yields that $a<(\gamma_n-1)/2y$. Second, by Cauchy-Schwarz inequality, $\bigl[\int\frac{t}{1+2at+(a^2+b^2)t^2}dH_{\theta}(t)\bigr]^2\leq \int\frac{1}{1+2at+(a^2+b^2)t^2}dH_{\theta}(t)\cdot \int\frac{t^2}{1+2at+(a^2+b^2)t^2}dH_{\theta}(t)$. It follows that
\[
y^2\leq (\gamma_n-1-2ay)\cdot \frac{1}{a^2+b^2}. 
\]
Re-arranging the terms gives $(a-1/y)^2+b^2\leq \gamma_n/y^2$. So far, we have obtained a feasible set of $(a,b)$ for the solution of \eqref{stieltjes}  when $\xi_n=0$: 
\beq \label{feasible}
    S_{y,\gamma_n}=\bigl\{(a,b):(a-1/y)^2+b^2\leq \gamma_n/y^2, \ a<(\gamma_n-1)/2y\bigr\}.
\eeq
 Since $\xi_n$ is very close to $0$, we use the same feasible set when solving \eqref{stieltjes}.
Observing that $S_{y,\gamma_n}$ is bounded,  
we solve equation \eqref{stieltjes} by a grid search on this feasible set. See Algorithm 4.

\subsection{Comparison} \label{subsec:GetQTsimu}
\begin{figure}[!t]
\centering
\includegraphics[width=0.8\textwidth, trim=40 30 40 30, clip=true]{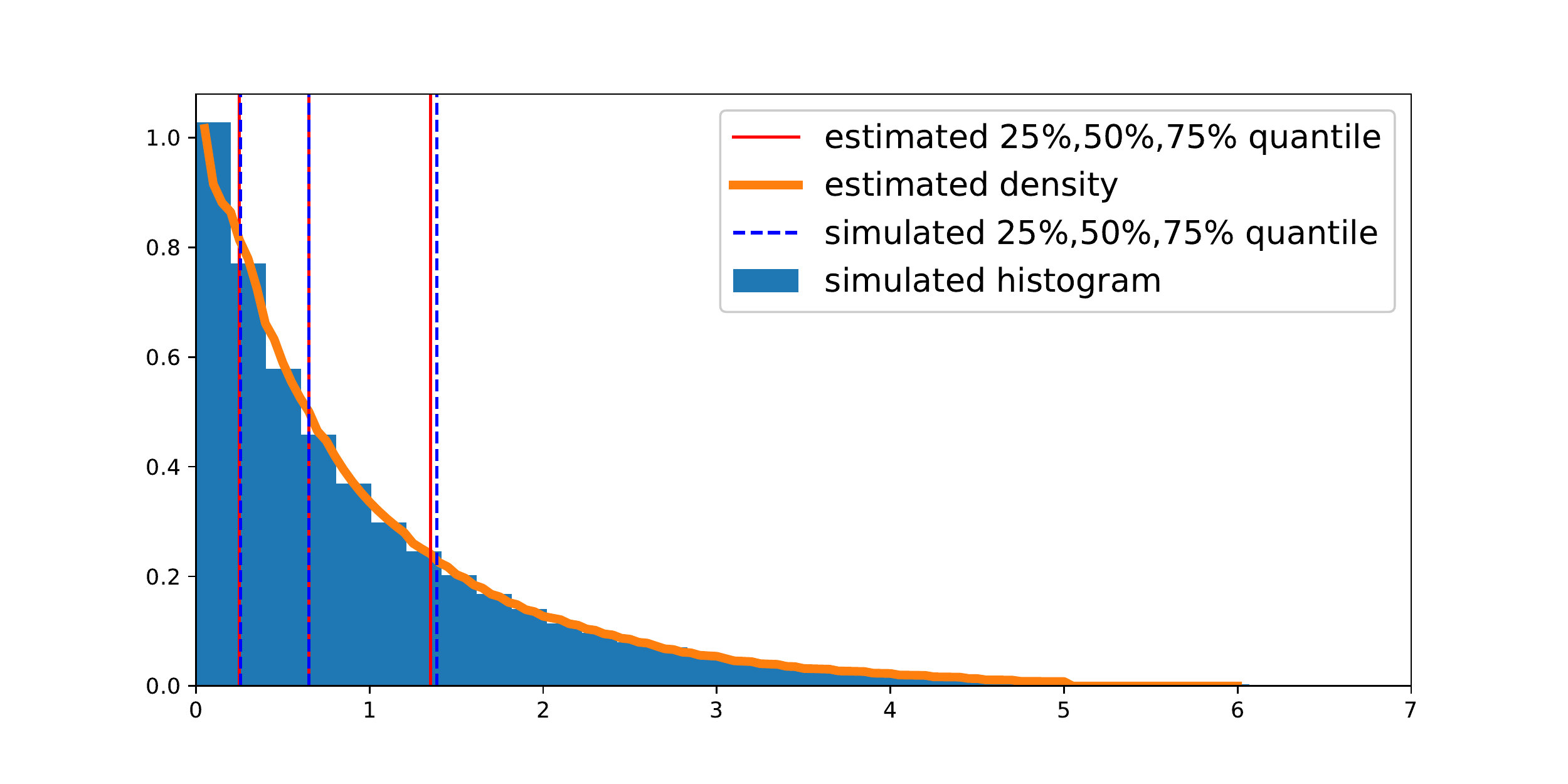}  

\caption{Comparison of two \texttt{GetQT} algorithms. The simulated histogram is from \texttt{GetQT1}, and the density curve is estimated by \texttt{GetQT2}.} \label{fig:GetQT}
\end{figure}

We compare the performance of two \texttt{GetQT} algorithms on a numerical example where $(n, p, \theta)=(10000, 1000, 1)$. The results are in Figure~\ref{fig:GetQT}. To generate this figure, first, we simulate eigenvalues $\{\hat{\lambda}_k^{(b)}\}_{1\leq k\leq p,1\leq b\leq B}$ as in Step 1 of \texttt{GetQT1}, where $B=20$, and plot the histogram of eigenvalues. Next, we plot the estimated density $\hat{f}_{\gamma_n}(y; 1,\theta)$ from \texttt{GetQT2} (tuning parameter is $\delta=0.05$). The estimated density fits the histogram well, suggesting that the steps in \texttt{GetQT2} for estimating $f_{\gamma_n}(y; 1,\theta)$ are successful. 
Furthermore, the estimated quantiles from two algorithms are very close to each other.

In terms of numerical performance, the two \texttt{GetQT} algorithms are similar. We now discuss the computing time. The main computational cost of \texttt{GetQT1} comes from computing the eigenvalues of $\bS^{(b)}$ at each iteration. As we have mentioned in the end of Section~\ref{subsec:GetQT1}, if $p<n$, we conduct eigen-decomposition on an $p\times p$ matrix; if $n<p$, we conduct eigen-decomposition on an $n\times n$ matrix. Therefore, as long as $\min\{n,p\}$ is not too large, \texttt{GetQT1} is fast.  

Compared with \texttt{GetQT1}, the advantage of \texttt{GetQT2} is that it does not need to compute any eigen-decomposition. As a result, when $\min\{n,p\}$ is large, \texttt{GetQT2} is much faster than \texttt{GetQT1} (and \texttt{GetQT2} also requires less memory use). 
The computational cost of \texttt{GetQT2} is proportional to the number of grid points in the algorithm, governed by the tuning parameter $\delta$. Sometimes, we need to choose $\delta$ sufficiently small to guarantee the accuracy of computing $\hat{m}(y,\gamma_n,\theta)$, which significantly increases the cost of grid search. Our experience suggests that \texttt{GetQT2} is faster than \texttt{GetQT1} only in the case that $\min\{n,p\}$ is larger than $10^4$.

\subsection{Modifications under Model~\eqref{mod-D-truncated}}
Section~\ref{subsec:theory-general} introduces Model~\eqref{mod-D-truncated}, as a proxy of Model~\eqref{mod-D}, to facilitate the theoretical analysis. In Model~\eqref{mod-D-truncated}, the diagonal entries of $\bD$ are {\it iid} generated from a truncated Gamma distribution. 
In Section~\ref{subsec:theory-general}, we described how to adapt Algorithm 2 to this setting, where the key is to modify \texttt{GetQT} so that it can compute the $y$-upper-quantile of the distribution $F_{\gamma}(\cdot;1,\theta,T_1,T_2)$, for any given $y$ and $(\theta,T_1,T_2)$.  

To modify \texttt{GetQT1}, we note that $F_{\gamma_n}(\cdot; 1, \theta, T_1, T_2)$ is the theoretical limit of the ESD under the null covariance model:
\beq \label{gamma-model-truncated}
\bSigma=\mathrm{diag}(\sigma^2_1,\ldots,\sigma^2_p), \qquad \mbox{where}\;\;  \sigma^2_k\; \overset{iid}{\sim}\; \mathrm{TruncGamma}(\theta,\theta, T_1, T_2). 
\eeq
We can simulate data from \eqref{gamma-model-truncated} and use its ESD as a numerical approximation to $F_{\gamma_n}(\cdot; 1, \theta, T_1, T_2)$. 
In Algorithm 3, we only need to modify Step 1 so that $\bSigma^{(b)}$ is generated from \eqref{gamma-model-truncated}.  

To modify \texttt{GetQT2}, we solve \eqref{stieltjes} with $H_\theta(t)$ replaced by $H_{\theta,T_1,T_2}(t)$, where $H_{\theta,T_1,T_2}(\cdot)$ is the CDF of $\mathrm{TruncGamma}(\theta,\theta,T_1, T_2)$. 
We note that the feasible set in \eqref{feasible} is derived without using the explicit form of $H_\theta(t)$, so it continues to apply. 
In Algorithm 4, we only need to modify the definition of $\Delta(a,b)$ to 
\[
\Delta(a,b) = \Big|y+ \mathrm{i}\, \xi_n + \frac{1}{m}-\gamma_n\int \frac{t}{1+tm}dH_{\theta,T_1,T_2}(t)\Big|, 
\]
and the other steps remain the same.

\section{Proofs}  \label{sec:proofs}

\subsection{Proof of Theorem~\ref{thm:sigma}}
Let $z_k=\hat{\lambda}_k-\sigma^2 q_k$, for all $1\leq k\leq \tilde{p}$. It follows that
\[
\hat{\sigma}^2 =  \frac{\sum_{\alpha\tilde{p}\leq k\leq (1-\alpha)\tilde{p}}q_k(\sigma^2 q_k + z_k)}{\sum_{\alpha\tilde{p}\leq k\leq (1-\alpha)\tilde{p}}q_k^2} =\sigma^2 +  \frac{\sum_{\alpha\tilde{p}\leq k\leq (1-\alpha)\tilde{p}}q_k z_k}{\sum_{\alpha\tilde{p}\leq k\leq (1-\alpha)\tilde{p}}q_k^2}. 
\]
It follows that 
\[
|\hat{\sigma}^2 - \sigma^2|\leq \underbrace{\frac{\sum_{\alpha\tilde{p}\leq k\leq (1-\alpha)\tilde{p}}|q_k|}{\sum_{\alpha\tilde{p}\leq k\leq (1-\alpha)\tilde{p}}q_k^2}}_{\equiv B_{n,p}(\alpha)}\times \max_{\alpha\tilde{p}\leq k\leq (1-\alpha)\tilde{p}} |z_k|. 
\]
We recall that $q_k$ is the $(k/\tilde{p})$-upper-quantile of a standard Machenko-Pastur distribution associated with $\gamma_n=p/n$. Note that $p/n\to\gamma$ and $\alpha\leq k/\tilde{p}\leq 1-\alpha$, where $\gamma>0$ and $\alpha\in (0,1/2)$ are constants. It follows immediately that there is a constant $C_1=C_1(\alpha,\gamma)$ such that $B_{n,p}(\alpha)\leq C_1$. As a result,
\beq \label{proof-1}
|\hat{\sigma}^2 - \sigma^2|\leq C_1 \max_{\alpha\tilde{p}\leq k\leq (1-\alpha)\tilde{p}}|\hat{\lambda}_k - \sigma^2 q_k|. 
\eeq

We bound the right hand side of \eqref{proof-1}. By Assumption~\ref{cond:factorM}, the data vectors $\bX_1,\bX_2,\ldots,\bX_n$ are obtained from a random matrix $\bY=[\bY_1,\bY_2,\ldots,\bY_n]^{\top}\in\mathbb{R}^{n\times p}$, where the entries of $\bY$ are independent variables with zero mean and unit variance.  Given $\bY$, define $\bX^*_1,\bX^*_2,\ldots,\bX^*_n$ by 
\[
\bX^*_i(j) = \sigma\cdot \bY_i(j), \qquad 1\leq i\leq n,1\leq j\leq p. 
\]
Then, $\bX^*_1, \ldots,\bX^*_n$ follow a ``null" model that is similar to the factor model in Assumption~\ref{cond:factorM} but corresponds to $K=0$. Let $\bS^*$ be the sample covariance matrix of $\bX^*_1,\ldots,\bX^*_n$. Then, $\bS^*$ serves as a reference matrix for $\bS$.
The {\it eigenvalue sticking} result says that eigenvalues of $\bS$ ``stick" to eigenvalues of the reference matrix. The precise statement is as follows: Let $\hat{\lambda}_1^*>\hat{\lambda}_2^*>\ldots>\hat{\lambda}^*_{\tilde{p}}$ be the nonzero eigenvalues of $\bS^*$. When the entries of $\bY$ satisfy the regularity conditions stated in Theorem~\ref{thm:sigma}, by Theorem 2.7 of \cite{bloemendal2016principal}, there is a constant $C_2=C_2(\alpha, \gamma,\sigma^2)$ such that, for any $\epsilon>0$ and $s>0$, 
\beq \label{proof-2}
\mathbb{P}\Bigl\{  \max_{(\alpha/2)\tilde{p}\leq j\leq (1-\alpha/2)\tilde{p}} |\hat{\lambda}_{j+K_1}-\hat{\lambda}^*_j| > C_2 n^{-(1-\epsilon)} \Bigr\}\leq n^{-s}, 
\eeq
where $K_1$ is the total number of spiked eigenvalues in Model \eqref{mod-Sigma} such that $\lambda_k=\sigma^2(\sqrt{\gamma}+\tau_k)$ for some $\tau_k\geq n^{-1/3}$. It remains to study $\hat{\lambda}_j^*$. Its large deviation bound can be found in \cite{PillaiYin} (also, see Theorem 3.3 of \cite{ke2016detecting}). There is a constant $C_3=C_3(\alpha, \gamma,\sigma^2)>0$ such that, for any $\epsilon>0$ and $s>0$,  
\beq \label{proof-3}
\mathbb{P}\Bigl\{ \max_{(\alpha/2)\tilde{p}\leq j\leq (1-\alpha/2)\tilde{p}} |\hat{\lambda}^*_j -\sigma^2 q_j|> C_3 n^{-(1-\epsilon)} \Bigr\} \leq n^{-s}.  
\eeq
Furthermore, since $K_1\leq K$ and $K$ is fixed, there is a constant $C_4=C_4(\gamma, K)$ such that
\beq \label{proof-4}
\max_{(\alpha/2)\tilde{p}\leq j\leq (1-\alpha/2)\tilde{p}}|q_j - q_{j+K_1}|\leq C_4 n^{-1}. 
\eeq
Combining \eqref{proof-2}-\eqref{proof-4} gives that, for any $\epsilon>0$ and $s>0$, 
\[
\mathbb{P}\Bigl\{ \max_{(\alpha/2) \tilde{p}\leq j\leq (1-\alpha)\tilde{p}} |\hat{\lambda}_{j+K_1} - \sigma^2 q_{j+K_1} |> Cn^{-(1-\epsilon)}\Bigr\} \leq n^{-s}. 
\]
We plug it into \eqref{proof-1}. The claim follows immediately. \qed

\subsection{Proof of Theorem~\ref{thm:K}}
Denote by $T_{n,p}(\hat{\sigma}^2, \beta_n)$ the threshold used in Algorithm 1. It satisfies that 
\beq\label{proofK-00}
T_{n,p}(\hat{\sigma}^2,\beta_n)=\hat{\sigma}^2[(1+\sqrt{\gamma}_n)^2  + \omega_n], \qquad \mbox{where}\;\; \omega_n = O(n^{-2/3}t_{1-\beta_n}). 
\eeq 
Here, $t_{1-\beta_n}$ is the $(1-\beta_n)$-quantile of Tracy-Widom distribution. Note that $\tau_n\gg n^{-1/3}$. We can choose $\beta_n\to\infty$ appropriately slow such that $1\ll t_{1-\beta_n}\ll n^{2/3}\min\{\tau_n^2, 1\}$. It follows that 
\beq\label{proofK-0}
n^{-2/3}\; \ll \; \omega_n \; \ll\; \min\bigl\{\tau_n^2,\; 1\bigr\}. 
\eeq

First, we derive a lower bound for $\hat{\lambda}_K$ and show that $\hat{K}\geq K$ with probability $1-o(1)$. Recall that $\lambda_k$ denotes the $k$th largest eigenvalue of $\bSigma$. In view of Model~\eqref{mod-Sigma}, it is true that $\lambda_k=\mu_k+\sigma^2$ for $1\leq k\leq K$ and $\lambda_k=\sigma^2$, for $K<k\leq p$. Introduce 
\[
\lambda^*_k=\lambda_k \Bigl(1+\frac{\gamma_n}{\lambda_k/\sigma^2- 1}\Bigr), \qquad 1\leq k\leq K.  
\]
Write $\delta_k=\lambda_k/\sigma^2-1$, for $k=1,2,\ldots,K$. 
Let $g(t)=(1+t)(1+\gamma_n/t)$. Then, 
\[
\lambda_k^*=\sigma^2\cdot g(\delta_k), \qquad 1\leq k\leq K. 
\] 
The function $g$ satisfies that $g(\sqrt{\gamma_n})=(1+\sqrt{\gamma_n})^2$ and $g'(t)\geq 1-\sqrt{\gamma_n}/t$. Hence, it is monotone increasing in $(\sqrt{\gamma_n},\infty)$. For any $\tau>0$ and $t>\sqrt{\gamma_n}+\tau$, we have $g(t)\geq g(\sqrt{\gamma_n})+g'(\sqrt{\gamma_n}+\tau)\cdot \tau \geq (1+\sqrt{\gamma_n})^2+\tau^2/(\sqrt{\gamma}_n+\tau)$. It follows that 
\beq\label{proofK-1}
\lambda^*_K\geq \sigma^2 \Bigl[(1+\sqrt{\gamma_n})^2 + \frac{ \delta_K^2}{\sqrt{\gamma_n}+\delta_K}\Bigr]. 
\eeq 
At the same time, by Theorem 2.3 of \cite{bloemendal2016principal}, with probability $1-o(1)$, 
\beq \label{proofK-2}
|\hat{\lambda}_K - \lambda^*_K | \leq C_2  \sigma^2 n^{-1/2}
\begin{cases}
\delta_K^{1/2}, &\mbox{if }\delta_K<1,\\
1+\delta_K/(1+\sqrt{\gamma_n}), & \mbox{if }\delta_K\geq 1,
\end{cases}
\eeq 
for a constant $C_2>0$. 
If $\delta_K\geq 1$, then \eqref{proofK-1} implies $\lambda_K^*-\sigma^2(1+\sqrt{\gamma_n})^2\geq C_3 \sigma^2 \delta_K$, for a constant $C_3>0$, and \eqref{proofK-2} yields that $|\hat{\lambda}_K - \lambda^*_K | \leq C_2 \sigma^2 (1+\delta_K) n^{-1/2}$. It follows that 
\[
\hat{\lambda}_K -\sigma^2(1+\sqrt{\gamma_n})^2\; \geq\; (C_3/2)\cdot \sigma^2 \delta_K\; \geq\;  (C_3/2)\cdot \sigma^2. 
\] 
If $\delta_K<1$, then \eqref{proofK-1} yields that $\lambda_K^*-\sigma^2(1+\sqrt{\gamma_n})^2\geq C_4 \sigma^2 \delta^2_K$, for a constant $C_4>0$, and \eqref{proofK-2} yields that $|\hat{\lambda}_K - \lambda^*_K | \leq C_2 \sigma^2 \delta^{1/2}_K n^{-1/2}$. It follows that 
\[
\hat{\lambda}_K -\sigma^2(1+\sqrt{\gamma_n})^2\; \geq\; C_4\sigma^2\delta^2_K - \frac{C_2\sigma^2\delta^2_K}{\sqrt{n\delta_K^3}}\; \geq \; (C_4/2)\cdot \sigma^2 \delta^2_K,
\] 
where the last inequality is because $\delta_K\geq\tau_n\gg n^{-1/3}$. We combine the two cases and note that $\delta_K\geq \tau_n$. It gives that 
\[
\mathbb{P}\Bigl\{ \hat{\lambda}_K \geq \sigma^2 \bigl[(1+\sqrt{\gamma_n})^2 + C \min\{\tau_n^2,\, 1\}\bigr] \Bigr\}= 1-o(1). 
\]
Furthermore, by Theorem~\ref{thm:sigma}, $|\hat{\sigma}^2-\sigma^2|\prec n^{-1}\ll \min\{\tau_n^2,\, 1\}$. Hence, we can replace $\sigma^2$ by $\hat{\sigma}^2$ in the above equation, i.e., 
\beq \label{proofK-3}
\mathbb{P}\Bigl\{ \hat{\lambda}_K \geq \hat{\sigma}^2 \bigl[(1+\sqrt{\gamma_n})^2 + C \min\{\tau_n^2,\, 1\}  \bigr]\Bigr\}= 1-o(1). 
\eeq
We compare $\hat{\lambda}_K$ with the threshold in \eqref{proofK-00}. Since $\omega_n\ll\min\{\tau_n^2, 1\}$, it is implied from \eqref{proofK-3} that $\hat{\lambda}_K$ exceeds this threshold with probability $1-o(1)$. Therefore, 
\[
\mathbb{P}\Bigl\{\hat{K}\geq K \Bigr\}=1-o(1). 
\]

Next, we derive an upper bound for $\hat{\lambda}_{K+1}$ and show that $\hat{K}\leq K$ with probability $1-o(1)$. We apply Theorem 2.3 of \cite{bloemendal2016principal} again: For any $\epsilon>0$ and $s>0$, 
\beq \label{proofK-4}
\mathbb{P}\Bigl\{ \hat{\lambda}_{K+1}-\sigma^2(1+\sqrt{\gamma_n})^2 \leq \sigma^2 n^{-(2/3-\epsilon)}\Bigr\}=1-o(1). 
\eeq
Since $\omega_n\gg n^{-2/3}$, we can take $\epsilon$ arbitrarily small to make $n^{-(2/3-\epsilon)}\leq \omega_n/2$. We also apply the large deviation bound for $\hat{\sigma}^2$ in Theorem~\ref{thm:sigma} to replace $\sigma^2$ by $\hat{\sigma}^2$. It follows immediately that
\beq \label{proofK-5}
\mathbb{P}\Bigl\{  \hat{\lambda}_{K+1} \leq \hat{\sigma}^2\bigl[ (1+\sqrt{\gamma}_n)^2 +\omega_n/2 \bigr]  \Bigr\} = 1- o(1). 
\eeq
We compare $\hat{\lambda}_{K+1}$ with the threshold in \eqref{proofK-00}. It is seen that $\hat{\lambda}_{K+1}$ is below this threshold with probability $1-o(1)$. Therefore, 
\[
\mathbb{P}\Bigl\{\hat{K}\leq K \Bigr\}=1-o(1). 
\]
The claim follows immediately.  \qed

\subsection{Proof of Theorem~\ref{thm:generalM-error}}
Throughout this proof, we let $C$ be a generic constant, whose meaning may vary from occurrence to occurrence. Let $F_{\gamma}(\cdot;\sigma^2,\theta,T_1,T_2)$ be the theoretical limit of ESD, whose definition is given in Lemma~\ref{lem:eigenvalue-rigidity}. We replace $\gamma$ by $\gamma_n=p/n$ in this definition, write $\bar{F}_{\gamma_n}=1-F_{\gamma_n}$ and let $q_i(\sigma^2, \theta)=\bar{F}_{\gamma_n}^{-1}(y;\sigma^2,\theta,T_1,T_2)$ denote the $(i/\tilde{p})$-upper-quantile of this distribution, where $\tilde{p}=n\wedge p$. We use $(\sigma_0^2,\theta_0)$ to denote the true parameters. Write $s_n=\lceil \alpha\tilde{p}\rceil$ and
\[
\hat{R}(\sigma^2, \theta)=  \sum_{s_n\leq i\leq \tilde{p}-s_n}[ \hat{\lambda}_i - q_i(\sigma^2, \theta)]^2, \qquad R(\sigma^2,\theta) =  \sum_{s_n\leq i\leq \tilde{p}-s_n}[ q_i(\sigma_0^2,\theta_0^2) - q_i(\sigma^2, \theta)]^2. 
\]
Let $\Delta=\sum_{s_n\leq i\leq \tilde{p}-s_n}|\hat{\lambda}_i-q_i(\sigma_0^2,\theta_0)|^2$. By direct calculations and Cauchy-Schwarz inequality,  
\begin{align*} 
 |\hat{R}(\sigma^2, \theta)-R(\sigma^2, \theta)|& \leq  2 \sum_{s_n\leq i\leq \tilde{p}-s_n} |q_i(\sigma_0^2, \theta_0)-q_i(\sigma, \theta)|\cdot |\hat{\lambda}_i-q_i(\sigma_0^2, \theta_0)|\cr
 & \qquad + \sum_{s_n\leq i\leq \tilde{p}-s_n}|\hat{\lambda}_i-q_i(\sigma_0^2, \theta_0)|^2 \cr
  &\leq  2 \sqrt{R(\sigma^2,\theta)}\sqrt{\Delta} + \Delta. 
\end{align*}
It follows that $\hat{R}(\sigma^2,\theta)\leq R(\sigma^2,\theta)+2 \sqrt{R(\sigma^2,\theta)}\sqrt{\Delta} + \Delta = \bigl( \sqrt{R(\sigma^2,\theta)}+\sqrt{\Delta}\bigr)^2$. In the above inequality, we can switch $\hat{R}(\sigma^2,\theta)$ and $R(\sigma^2,\theta)$ and similarly  derive that $R(\sigma^2,\theta)\leq \bigl( \sqrt{\hat{R}(\sigma^2,\theta)}+\sqrt{\Delta}\bigr)^2$. As a result,
\beq \label{thm-general-1}
\Bigl|\sqrt{\hat{R}(\sigma^2,\theta)}-\sqrt{R(\sigma^2,\theta)}\Bigr|\leq \sqrt{\Delta}. 
\eeq
We now bound $\Delta$. 
By Lemma~\ref{lem:eigenvalue-rigidity}, for all $K<i\leq \tilde{p}$, 
\[
| \hat{\lambda}_i -  q_i( \sigma_0^2, \theta_0) | 
 \prec  [i\wedge (\tilde{p}+1-i)]^{-1/3} n^{-2/3}. 
\]
We note that the stochastic dominance in Lemma~\ref{lem:eigenvalue-rigidity} can be made `uniform' over $i$; i.e., the integer $N(\epsilon,s)$ in Definition~\ref{def:stochastic-dominance} is shared by all $K<i\leq \tilde{p}$ \citep{knowles2017anisotropic}. Hence, summing over $i$ preserves `stochastic dominance.' Additionally, $\sum_{i=s_n}^{\tilde{p}/2} i^{-2/3}n^{-4/3}\leq Cn^{-1}\bigl[ \frac{1}{\tilde{p}}\sum_{i=s_n}^{\tilde{p}/2}(i/\tilde{p})^{-2/3}\bigr]\leq Cn^{-1} \int_{s_n/n}^{1/2} x^{-2/3}dx \leq Cn^{-1}$. 
Combining the above arguments gives 
\begin{align*}
\sum_{s_n\leq i\leq \tilde{p}-s_n}|\hat{\lambda}_i-q_i(\sigma_0^2, \theta_0)|^2   &\prec\;   \sum_{s_n \leq i\leq \tilde{p}-s_n} [i\wedge (\tilde{p}+1-i)]^{-2/3} n^{-4/3}\cr
&\prec \; \sum_{s_n\leq i\leq \tilde{p}/2}i^{-2/3}n^{-4/3} \;\; \prec\;\; n^{-1}. 
\end{align*}
This gives $\Delta\prec n^{-1}$. 
We plug it into \eqref{thm-general-1} to get
\beq \label{thm-general-2}
\Bigl|\sqrt{\hat{R}(\sigma^2,\theta)}-\sqrt{R(\sigma^2,\theta)}\Bigr| \prec n^{-1/2}. 
\eeq
Since $\Delta$ does not depend on $(\sigma^2,\theta)$, the `stochastic dominance' here is uniform for all $(\sigma^2,\theta)\in {\cal J}_{\sigma^2}\times {\cal J}_\theta$. We claim that there exists a constant $c_0>0$ such that for any $(\sigma^2 ,\theta)$ in ${\cal J}_{\sigma^2}\times {\cal J}_\theta$, 
\beq \label{thm-general-3}
R(\sigma^2,\theta) \geq c_0 n\cdot \bigl[(\sigma^2-\sigma_0^2)^2+(\theta-\theta_0)^2\bigr]. 
\eeq
Note that $R(\sigma_0^2,\theta_0)=0$. 
Combining it with \eqref{thm-general-2}-\eqref{thm-general-3} gives
\[
\sqrt{\hat{R}(\sigma_0^2, \theta_0)} \prec n^{-1/2}, \qquad  \sqrt{c_0  n}\sqrt{(\hat{\sigma}^2-\sigma_0^2)^2+(\hat{\theta}-\theta_0)^2} \leq \sqrt{\hat{R}(\hat{\sigma}^2,\hat{\theta})} + O_{\prec}(n^{-1/2}),
\]
where a random variable is $O_{\prec}(b_n)$ if its absolute value is $\prec b_n$. 
Since $(\hat{\sigma}^2, \hat{\theta})$ minimizes $\hat{R}(\sigma^2,\theta)$, we have $\hat{R}(\hat{\sigma}^2,\hat{\theta})\leq \hat{R}(\sigma_0^2, \theta_0)\prec n^{-1}$. It follows that 
\[
\sqrt{(\hat{\sigma}^2-\sigma_0^2)^2+(\hat{\theta}-\theta_0)^2} \prec n^{-1}. 
\]
This proves the claim. 

What remains is to show \eqref{thm-general-3}. Define the quantile function $h_{\sigma^2,\theta}(\alpha) = \bar{F}^{-1}_{\gamma_n}(\alpha; \sigma^2,\theta,T_1,T_2)$. Then, $q_i(\sigma^2,\theta)=h_{\sigma^2,\theta}(i/\tilde{p})$. We can re-write
\[
R (\sigma^2,\theta)=\sum_{i=s_n}^{\tilde{p}-s_n} \bigl[h_{\sigma^2,\theta}(i/\tilde{p}) - h_{\sigma_0^2,\theta_0}(i/\tilde{p})\bigr]^2. 
\] 
Introduce $R^*(\sigma^2, \theta) = \tilde{p}\int_0^1 [h_{\sigma^2,\theta}(\alpha) -h_{\sigma_0^2,\theta_0}(\alpha)]^2d\alpha$. Then, $\tilde{p}^{-1}R(\sigma^2, \theta)$ is the Riemann approximation of the integral $\tilde{p}^{-1}R^*(\sigma^2, \theta)$. Note that $s_n/\tilde{p}=o(1)$. Furthermore, $h_{\sigma^2, \theta}(\alpha)$ is uniformly square integrable for $(\sigma^2,\theta)\in {\cal J}_{\sigma^2}\times {\cal J}_{\theta}$ (the proof is very similar to the analysis of $C_2$ below; we thus omit it). Hence, the Riemann approximation error is negligible. Particularly, there exists a constant $c_1\in (0,1)$ such that  
\beq \label{thm-general-4}
R(\sigma^2,\theta) \geq c_1\cdot R^*(\sigma^2, \theta).
\eeq
It suffices to study $R^*(\sigma^2,\theta)$. The next lemma is proved in Section~\ref{subsec:proof-lem-cdf-quantile}. 
\begin{lem} \label{lem:cdf-quantile}
Let $F(x)$ be a distribution on $(0,\infty)$ with a continuous density $f(x)$. Let $\bar{F}(x)=1-F(x)$, $h_F(\alpha)=\bar{F}^{-1}(\alpha)$, and $\mu_m(f)=\int x^m f(x)dx$, $m\geq 1$. For another distribution $G(x)$ on $(0,\infty)$ with a continuous density $g(x)$, we define $\bar{G}(x)$, $h_G(\alpha)$, and $\mu_m(g)$ similarly. Suppose $\int x^2|\bar{F}(x)-\bar{G}(x)|dx<\infty$. Let $\check{g}(x,y)=\max_{z\in [x,y]\cup [y,x]}g(z)$ for $x, y\in (0,\infty)$. We assume that    
$C_1 \equiv \int_0^1 \bigl[\frac{\check{g}(h_F(\alpha), h_G(\alpha))}{f(h_F(\alpha))}\bigr]^2d\alpha<\infty$ and $C_2\equiv \int_0^1 \bigl[\frac{h_F(\alpha)\check{g}(h_F(\alpha), h_G(\alpha))}{f(h_F(\alpha))}\bigr]^2d\alpha <\infty$. Then, 
\[
\int_0^1 [h_G(\alpha)-h_F(\alpha)]^2d\alpha\geq \frac{|\mu_1(f)-\mu_1(g)|^2}{4C_1} , \qquad  \int_0^1 [h_G(\alpha)-h_F(\alpha)]^2d\alpha\geq \frac{|\mu_2(f)-\mu_2(g)|^2}{4C_2}. 
\]  
\end{lem}

\vspace{10pt}

\noindent
We apply Lemma~\ref{lem:cdf-quantile} to $F(\cdot)= F_{\gamma_n}(\cdot;\sigma_0^2,\theta_0, T_1, T_2)$ and $G(\cdot)=F_{\gamma_n}(\cdot;\sigma^2,\theta,T_1,T_2)$. Define 
\[
\mu_1(\sigma^2,\theta) = \int x\, dF_{\gamma_n}(x;\sigma^2,\theta,T_1, T_2), \qquad  \mu_2(\sigma^2,\theta) = \int x^2\, dF_{\gamma_n}(x;\sigma^2,\theta,T_1, T_2).
\]
We now show that the quantities $C_1, C_2$ in Lemma~\ref{lem:cdf-quantile} are uniformly upper bounded by constants for all $(\sigma^2,\theta)\in {\cal J}_{\sigma}^2\times J_{\theta}$. We only study $C_2$, and the analysis of $C_1$ is similar. By \cite{knowles2017anisotropic,ding2020spiked}, the support of $F_{\gamma_n}(\cdot;\sigma^2,\theta,T_1,T_2)$ is in a compact subset of $(0,\infty)$, and the density is upper bounded by a constant; these constants are uniform for $(\sigma^2,\theta)\in {\cal J}_{\sigma^2}\times {\cal J}_{\theta}$. It follows that 
\[
C_2\leq C \int_0^1 \Bigl[\frac{1}{f(h_F(\alpha))}\Bigr]^2d\alpha=\int \frac{1}{f^2(x)}f(x)dx = \int \frac{1}{f(x)}dx.
\]
Here we have used a change of variable $x=h_F(\alpha)$, where $\alpha=1-F(x)$ and $d\alpha=f(x)dx$. We then apply Theorem 3.3 of \cite{ji2020regularity}. Note that $F(\cdot)=F_{\gamma_n}(\cdot; \sigma_0^2,\theta_0,T_1,T_2)$ is the free multiplicative convolution between a truncated Gamma distribution and the standard MP distribution. These two distributions are compacted supported and have power law behavior on left/right ends. The conditions in Theorem 3.3 of \cite{ji2020regularity} are satisfied for $t^{\mu}_{\pm}=0$ (truncated Gamma) and $t^{\nu}_{\pm}=1/2$ (MP law). By that theorem, the density of $F(\cdot)$ has a square-root decay at the left/right edge: Let $[b^-, b^+]$ be the support of $F(\cdot)$; then, $C^{-1}\leq f(x)/\sqrt{(x-b^-)(b^+-x})\leq C$ for $x\in [b^-, b^+]$. It yields hat 
\[
C_2\leq \int_{b^-}^{b^+} \frac{C}{\sqrt{(x-b^-)(b^+-x)}}dx = O(1). 
\]
We have verified that $C_1$ and $C_2$ in Lemma~\ref{lem:cdf-quantile} are uniformly upper bounded. As a result,  
\beq \label{thm-general-5}
R^*(\sigma^2,\theta) \geq Cn\Bigl(\bigl|\mu_1(\sigma^2,\theta)-\mu_1(\sigma_0^2,\theta_0)\bigr|^2+ \bigl|\mu_2(\sigma^2,\theta)-\mu_2(\sigma_0^2,\theta_0)\bigr|^2\Bigr). 
\eeq 

Below, we study $\mu_1(\sigma^2,\theta)$ and $\mu_2(\sigma^2,\theta)$. Note that $\mathrm{Gamma}(\theta,\theta/\sigma^2,\sigma^2T_1,\sigma^2T_2)$ is equivalent to $\sigma^2\cdot \mathrm{Gamma}(\theta,\theta,T_1,T_2)$. Then, the distributions $F_{\gamma_n}(\cdot;\sigma^2,\theta,T_1,T_2)$ and $F_{\gamma_n}(\cdot;1, \theta, T_1,T_2)$ also have such a connection.
This implies $\mu_1(\sigma^2,\theta)=\sigma^2\cdot \mu_1(1,\theta)$ and $\mu_2(\sigma^2,\theta)=\sigma^4\cdot \mu_2(1,\theta)$. Define
\[
\kappa(\theta) = \mu_2(\sigma^2,\theta)/[\mu_1(\sigma^2,\theta)]^2.
\]
Consider a mapping $M$ from $\mathbb{R}^2$ to $\mathbb{R}^2$, where $M(x,y)=(x, y/x^2)$ . It maps $(\mu_1(\sigma^2,\theta), \mu_2(\sigma^2,\theta))$ to $(\mu_1(\sigma^2,\sigma^2), \kappa(\theta))$. The Jacobian matrix is
\[
\begin{bmatrix}
1 & 0\\
-2y/x^3 & 1/x^2
\end{bmatrix}.
\]
When $(\sigma^2,\theta)\in {\cal J}_{\sigma^2}\times {\cal J}_{\theta}$, the vector $(\mu_1(\sigma^2, \theta), \mu_2(\sigma^2,\theta))$ is in a compact set. The spectral norm of Jacobian is uniformly upper bounded. It follows that
\begin{align}  \label{thm-general-6}
& \bigl|\mu_1(\sigma^2,\theta)-\mu_1(\sigma_0^2,\theta_0)\bigr|^2+\bigl|\mu_2(\sigma^2,\theta)-\mu_2(\sigma_0^2,\theta_0)\bigr|^2\cr
\geq\;\; & C\Bigl(\bigl|\mu_1(\sigma^2,\theta)-\mu_1(\sigma_0^2,\theta_0)\bigr|^2+\bigl|\kappa(\theta)-\kappa(\theta_0)\bigr|^2\Bigr). 
\end{align}
We then study $\mu_1(\sigma^2,\theta)$ and $\kappa(\theta)$. Denote by $\hat{F}(\cdot;\sigma^2,\theta,T_1,T_2)$ the ESD when $(\sigma^2,\theta)$ are true parameters. Write $\hat{\mu}_1(\sigma^2,\theta)=\int xd\hat{F}(x;\sigma^2,\theta,T_1,T_2)$ and $\hat{\mu}_2(\sigma^2,\theta)=\int x^2 d\hat{F}(x;\sigma^2,\theta,T_1,T_2)$. The converges of ESD to its theoretical limit yields  that $|\hat{\mu}_1(\sigma^2,\theta)-\mu_1(\sigma^2,\theta)|\to 0$ and $|\hat{\mu}_2(\sigma^2,\theta)-\mu_2(\sigma^2,\theta)|\to 0$ in probabiliy. In fact, we have a stronger result \citep{knowles2017anisotropic}:
\beq \label{thm-general-7(1)}
\bigl|\mathbb{E}[\hat{\mu}_1(\sigma^2,\theta)]-\mu_1(\sigma^2,\theta)\bigr|\prec n^{-1}, \qquad \bigl|\mathbb{E}[\hat{\mu}_2(\sigma^2,\theta)]-\mu_2(\sigma^2,\theta)\bigr|\prec n^{-1}. 
\eeq
Here the expectation is with respect to the null model (i.e., $K=0$) with true parameters $(\sigma^2,\theta)$. 
The left hand sides above are non-stochastic quantities, and ``$\prec n^{-1}$" is interpreted as ``$\leq n^{-1+\epsilon}$ for any $\epsilon>0$." 
Since $\mu_1(\sigma^2, \theta)$ and $\mu_2(\sigma^2,\theta)$ are uniformly upper/lower bounded, it follows that  
\beq \label{thm-general-7(2)}
\Bigl|\hat{\kappa}(\theta)- \frac{\mathbb{E}[\hat{\mu}_2(\sigma^2,\theta)]}{\bigl(\mathbb{E}[\hat{\mu}_1(\sigma^2,\theta)]\bigr)^2}\Bigr|\prec n^{-1}. 
\eeq
By definition, we can also write $\hat{\mu}_1=\frac{1}{\tilde{p}}\sum_{i=1}^{\tilde{p}}\hat{\lambda}_i=\frac{1}{\tilde{p}}\mathrm{tr}(\bS)$ and $\hat{\mu}_2=\frac{1}{\tilde{p}}\sum_{i=1}^{\tilde{p}}\hat{\lambda}^2_i=\frac{1}{\tilde{p}}\|\bS\|_F^2$, where $\bS=\frac{1}{n}\bX^{\top}\bX$ is the sample covariance matrix under the null model of $K=0$. By Assumption~\ref{cond:factorM}, $\bX=\bY\bSigma^{1/2}$, where $\bY$ contains $iid$ zero-mean, unit variance entries. Note that our purpose here is to approximate the moments of the theoretical limit of ESD, and we are flexible to choose the eigenvectors in $\bSigma$. We choose $\bxi_k$ as the $k$th standard basis, and so $\bSigma=\mathrm{diag}(\sigma_1^2,\sigma_2^2,\ldots,\sigma_p^2)$. By direct calculations, 
\begin{align*}
\mathbb{E}[\hat{\mu}_1(\sigma^2,\theta)] &= \frac{1}{n\tilde{p}}\mathbb{E}\biggl[\sum_{j=1}^p\Bigl(\sum_{i=1}^n \sigma^2_j Y_{ij}^2\Bigr) \biggr]=(\gamma_n\vee 1)\cdot\mathbb{E}[\sigma_1^2],\cr
\mathbb{E}[\hat{\mu}_2(\sigma^2,\theta)] &= \frac{1}{n^2\tilde{p}}\mathbb{E}\biggl[ \sum_{j=1}^p \Bigl(\sum_{i=1}^n \sigma^2_j Y_{ij}^2\Bigr)^2 + \sum_{1\leq j\neq \ell \leq p} \Bigl( \sum_{i=1}^n \sigma_j\sigma_{\ell}Y_{ij} Y_{i\ell}\Bigr)^2 \biggr]\cr
&=\frac{1}{n^2\tilde{p}} \biggl[ np\, \mathbb{E}[\sigma_1^4]\, \mathbb{E}[Y_{11}^4] +  pn(n-1)\,\mathbb{E}[\sigma_1^4]  +  p(p-1)n \bigl(\mathbb{E}[\sigma_1^2]\bigr)^2\bigg]\cr
&= O(n^{-1}) +(\gamma_n\vee 1)\cdot \mathbb{E}[\sigma_1^4] +  \gamma_n(\gamma_n\vee 1)\cdot \bigl(\mathbb{E}[\sigma_1^2]\bigr)^2. 
\end{align*}
Note that $\sigma_1^2/\sigma^2\sim \mathrm{Gamma}(\theta,\theta,T_1,T_2)$. The density of $\mathrm{Gamma}(\theta,\theta,T_1,T_2)$ is equal to $x^{\theta-1}e^{-\theta x}\cdot (\int_{T_1}^{T_2}z^{\theta-1}e^{-\theta z}dz)^{-1}$. We immediately have
\begin{align*}
\mathbb{E}[\hat{\mu}_1(\sigma^2,\theta)] & =  (\gamma_n\vee 1) \sigma^2 \cdot \frac{\int_{T_1}^{T_2}x^{\theta}\exp(-\theta x)dx}{\int_{T_1}^{T_2}x^{\theta-1}\exp(-\theta x)dx}\\
\mathbb{E}[\hat{\mu}_2(\sigma^2,\theta)] & = O(\frac{1}{n}) +(\gamma_n\vee 1)\frac{\sigma^4 \int_{T_1}^{T_2}x^{\theta+1}\exp(-\theta x)dx}{\int_{T_1}^{T_2}x^{\theta-1}\exp(-\theta x)dx} +  \gamma_n(\gamma_n\vee 1)\frac{\sigma^4 \bigl[\int_{T_1}^{T_2}x^{\theta}\exp(-\theta x)dx\bigr]^2}{\bigl[ \int_{T_1}^{T_2}x^{\theta-1}\exp(-\theta x)dx\bigr]^2}.
\end{align*}  
Define $\Psi(\theta)=\Psi(\theta;T_1,T_2)\equiv (\int_{T_1}^{T_2} x^{\theta}e^{-\theta x}dx)/(\int_{T_1}^{T_2}x^{\theta-1}e^{-\theta x}dx)$. Let $\Phi(\theta)$ be the same as in the statement of this theorem. We plug the above equations into \eqref{thm-general-7(1)}-\eqref{thm-general-7(2)} to get
\begin{align} \label{thm-general-8}
\mu_1(\sigma^2,\theta) & = (\gamma_n\vee1)\sigma^2\cdot \Psi(\theta)+O_{\prec}(n^{-1}), \cr
\kappa(\theta) &= \frac{1}{(\gamma_n\vee 1)}\cdot \Phi(\theta) + \frac{\gamma_n}{(\gamma_n\vee 1)} + O_{\prec}(n^{-1}).  
\end{align}
Consider the mapping from $(\sigma^2, \theta)$ to $(\mu_1(\sigma^2,\theta), \; \kappa(\theta))$. The Jacobian matrix is
\[
J = (\gamma_n\vee 1) \begin{bmatrix}
\Psi(\theta) &  \sigma^2\cdot \Psi'(\theta)\\
0 & \frac{1}{(\gamma_n\vee 1)^2}\cdot \Phi'(\theta) 
\end{bmatrix} + O_{\prec}(n^{-1}). 
\]
First, since ${\cal J}_\theta$ is a bounded set, $\Psi(\theta)$, $\Psi'(\theta)$ and $\Phi'(\theta)$ are uniformly upper bounded by constants. Second,  we have $\Psi(\theta)>0$ in a fixed compact set ${\cal J}_\theta$. As a result, $\Psi(\theta)$ must be  uniformly lower bounded by a constant. Last, the assumption says that $\inf_{\theta\in {\cal J}_\theta}|\Phi'(\theta)|\geq \omega$, for a constant $\omega>0$. Combining these arguments with the formula of the inverse of a $2\times 2$ matrix, we have $\|J^{-1}\|\leq C$. It follows that
\begin{align} \label{thm-general-9}
& \bigl|\mu_1(\sigma^2,\theta)-\mu_1(\sigma_0^2,\theta_0)\bigr|^2+\bigl|\kappa(\theta)-\kappa(\theta_0)\bigr|^2\cr
\geq\;\; &  C\Bigl(|\sigma^2-\sigma_0^2|^2 + |\theta-\theta_0|^2\Bigr).  
\end{align}

We plug \eqref{thm-general-9} into \eqref{thm-general-6}, and then into \eqref{thm-general-5}, and then combine it with \eqref{thm-general-4}. It gives \eqref{thm-general-3}. \qed

\subsection{Proof of Lemma~\ref{lem:sensitivity-to-theta}}
Write 
\[
J_1(\theta)=(\int_{t_1}^{t_2}x^{\theta+1}exp(-\theta x)dx)(\int_{t_1}^{t_2}x^{\theta-1}exp(-\theta x)dx),\qquad J_2(\theta) = {(\int_{t_1}^{t_2}x^{\theta}exp(-\theta x)dx)^2}. 
\]
Then $\Psi(\theta)=J_1(\theta)/J_2(\theta)$ and 
\beq \label{lem-Psi-1}
\Psi'(\theta)=\frac{J_1'(\theta)J_2(\theta)-J_1(\theta)J_2'(\theta)}{J_2(\theta)^2}.
\eeq
By direct calculations, 
\[
\begin{split}
    J_1'(\theta)=&(\int_{t_1}^{t_2}\log(x) x^{\theta+1}exp(-\theta x)dx- \int_{t_1}^{t_2}x^{\theta+2}exp(-\theta x)dx)(\int_{t_1}^{t_2}x^{\theta-1}exp(-\theta x)dx)\\
    &+(\int_{t_1}^{t_2}\log(x) x^{\theta-1}exp(-\theta x)dx- \int_{t_1}^{t_2}x^{\theta}exp(-\theta x)dx)(\int_{t_1}^{t_2}x^{\theta+1}exp(-\theta x)dx),
\end{split}
\]
\[J_2'(\theta)=2(\int_{t_1}^{t_2}x^{\theta}exp(-\theta x)dx)(\int_{t_1}^{t_2}\log(x) x^{\theta}exp(-\theta x)dx- \int_{t_1}^{t_2}x^{\theta+1}exp(-\theta x)dx).
\]
Let $L(\alpha,\theta;t_1,t_2)$ denote $\int_{t_1}^{t_2}\log(x) x^{\alpha}exp(-\theta x)dx$ and $I(\alpha,\theta;t_1,t_2)$ denote $\int_{t_1}^{t_2} x^{\alpha}exp(-\theta x)dx$. When not causing any confusion, we write them as $L(\alpha)$ and $I(\alpha)$. Then
\[J_1(\theta)=I(\theta+1)\times I(\theta-1),\quad J_2(\theta)=I(\theta)^2\]
\[J_1'(\theta)=(L(\theta+1)-I(\theta+2))\times I(\theta-1)+(L(\theta-1)-I(\theta))\times I(\theta+1)\]
\[J_2'(\theta)=2(L(\theta)-I(\theta+1))\times I(\theta)\]
Plugging them into \eqref{lem-Psi-1}, we have 
\[
\Psi'(\theta)=\frac{I(\theta+1)I(\theta-1)}{I(\theta)^2}\Big(\Big(\frac{L(\theta+1)}{I(\theta+1)}+\frac{L(\theta-1)}{I(\theta-1)}-2\frac{L(\theta)}{I(\theta)}\Big)-\Big(\frac{I(\theta+2)}{I(\theta+1)}+\frac{I(\theta)}{I(\theta-1)}-2\frac{I(\theta+1)}{I(\theta)}\Big)\Big).
\]
Recall that we are interested in $\theta\in {\cal J}_\theta= [c,d]$. For $\alpha\in[c-1,d+2]$ and $\theta\in [c, d]$,
\[\int_{0}^{\infty}\log(x) x^{\alpha}exp(-\theta x)dx-L(\alpha,\theta;t_1,t_2)=\int_{0}^{t_1}\log(x) x^{\alpha}exp(-\theta x)dx+\int_{t_2}^{\infty}\log(x) x^{\alpha}exp(-\theta x)dx,\]
\[\Big|\int_{0}^{t_1}\log(x) x^{\alpha}exp(-\theta x)dx\Big|\leq \int_{0}^{t_1}(-\log(x)) x^{c-1}exp(-c x)dx\to 0,\quad \text{ as } t_1\to 0,\]
\[\Big|\int_{t_2}^{\infty}\log(x) x^{\alpha}exp(-\theta x)dx\Big|\leq \int_{t_2}^{\infty}\log(x) x^{d+2}exp(-c x)dx\to 0,\quad \text{ as } t_2\to \infty.
\]
This implies for $\alpha\in[c-1,d+2], \theta\in [c, d]$, as $(t_1,t_2)\to (0,\infty)$,  $L(\alpha,\theta;t_1,t_2)$ uniformly converges to $L_0(\alpha,\theta)= \int_{0}^{\infty}\log(x) x^{\alpha}exp(-\theta x)dx$. By a similar argument, we can show that $I(\alpha,\theta;t_1,t_2)$ uniformly converges to $I_0(\alpha,\theta)=\int_{0}^{\infty} x^{\alpha}exp(-\theta x)dx$. From the uniform convergence and the fact that $I_0(\alpha,\theta)$ is lower bounded by a common positive constant when $\alpha\in[c-1,d+2], \theta\in [c, d]$, we know that as $(t_1,t_2)\to (0,\infty)$ we have $\Psi'(\theta)$ uniformly converges to 
\[
\frac{I_0(\theta+1)I_0(\theta-1)}{I_0(\theta)^2}\Big(\Big(\frac{L_0(\theta+1)}{I_0(\theta+1)}+\frac{L_0(\theta-1)}{I_0(\theta-1)}-2\frac{L_0(\theta)}{I_0(\theta)}\Big)-\Big(\frac{I_0(\theta+2)}{I_0(\theta+1)}+\frac{I_0(\theta)}{I_0(\theta-1)}-2\frac{I_0(\theta+1)}{I_0(\theta)}\Big)\Big),
\]
for all $\theta \in [c, d]$. Here, $L_0(\alpha)$ and $I_0(\alpha)$ are short for $L_0(\alpha,\theta)$ and $I_0(\alpha,\theta)$. Let $Z\sim \text{Gamma}(\alpha,\theta)$ and let $\psi$ denote the digamma function. By properties of the Gamma distribution,  
\[
\frac{I_0(\alpha,\theta)}{I_0(\alpha-1,\theta)}=\mathbb{E}(Z)=\frac{\alpha}{\theta},\quad \frac{L_0(\alpha-1,\theta)}{I_0(\alpha-1,\theta)}=\mathbb{E}(\log(Z))=\psi(\alpha)-\log(\theta).
\]
Therefore, $\Psi'(\theta)$ uniformly converges to
\[\frac{\theta+1}{\theta}\Big(\Big(\psi(\theta+2)+\psi(\theta)-2\psi(\theta+1)\Big)-\Big(\frac{\theta+2}{\theta}+\frac{\theta}{\theta}-2\times \frac{\theta+1}{\theta}\Big)\Big)=\frac{\theta+1}{\theta}\Big(\frac{1}{\theta+1}-\frac{1}{\theta}\Big)=-\frac{1}{\theta^2}.\]
The first equation uses the recurrence relation of digamma function. By the uniform convergence, for any $\delta>0$ there exists $0<T_1^*<T_2^*<\infty$ such that $\sup_{\theta\in [c,d]} |\Psi'(\theta)-(-\frac{1}{\theta^2})|\leq \delta$. The claim follows by choosing $\delta=1/d^2-\omega$. 
\qed

\subsection{Proof of Theorem~\ref{thm:generalM-K}}
Let $d_j=\sigma_j^2+\mu_j$ for $1\leq k\leq K$ and $d_j=\sigma_j^2$ for $K+1\leq j\leq p$. 
Then, $d_1,d_2,\ldots,d_p$ are all the eigenvalues of $\bSigma$. Define
\beq \label{thm-genK-1}
\hat{G}(x) =-\frac{1}{x} + \frac{\gamma}{p}\sum_{j=1}^p\frac{1}{x+\sigma_j^{-2}}. 
\eeq
By Lemma 2.2 and Condition 3.6 of \cite{ding2020spiked}, this function $\hat{G}(x)$ has 2 critical points $0>\hat{x}_1>\hat{x}_2$;  furthermore, conditioning on $\bSigma$, the ESD converges to a limit whose support is $[\hat{G}(\hat{x}_2), \hat{G}(\hat{x}_1)]$. We apply Theorem 3.2 of \cite{ding2020spiked}. Using the first claim there, if $-1/d_k\geq \hat{x}_1+n^{1/3}$ for each $1\leq k\leq K$, then 
\[
|\hat{\lambda}_k - \hat{G}(-1/d_k)|\prec n^{-1/2}(-1/d_k - \hat{x}_1)^{1/2}, \qquad 1\leq k\leq K. 
\]
Using the second claim there, 
\[
|\hat{\lambda}_{K+1}- \hat{G}(\hat{x}_1)|\prec n^{-2/3}. 
\]
The above ``stochastic dominance'' arguments are conditioning on $\bSigma$. Under Model~\eqref{mod-D-truncated} for $\bSigma$, $\hat{G}(x)$ converges weakly to $G(x)$ defined in \eqref{define-right-edge}, and the critical points $(\hat{x}_1, \hat{x}_2)$ also converge to $(x_1^*, x_2^*)$, the critical points of $G(x)$, almost surely. 
Replacing $\hat{G}(\cdot)$ and $\hat{x}_1$ by $G(\cdot)$ and $x_1^*$ in the above inequalities has a negligible effect (e.g., see Example 3.9 of \cite{ding2020spiked}). 
It follows that 
\[
\max_{1\leq k\leq K}|\hat{\lambda}_k - G(-1/d_k)| \prec  n^{-1/2}, \qquad |\hat{\lambda}_{K+1}- G(x_1^*) | \prec n^{-2/3}. 
\]
Note that $d_k=\sigma_k^2+\mu_k\geq \mu_K+T_1$. The assumption of $-1/(T_1+\mu_K)\geq x_1^*+\tau$ guarantees that $G(-1/d_k)\geq G(-1/(T_1+\mu_K))\geq G(x_1^*+\tau)\geq G(x_1^*)+c$, where $c>0$ is a constant. Therefore,
\beq \label{thm-genK-2}
\min_{1\leq k\leq K}\{\hat{\lambda}_k \}-G(x_1^*)\; \geq\; c+O_{\prec} (n^{-1/2}),\qquad \hat{\lambda}_{K+1}- G(x_1^*) \; \prec\; n^{-2/3},
\eeq
where $O_{\prec}(b_n)$ means the absolute value is $\prec b_n$. 

The estimator $\hat{K}$ is obtained by thresholding the empirical eigenvalues at $\hat{T}_{\beta}$ as in \eqref{generalM-threshold}. Let $\hat{\lambda}_1^{*}=\hat{\lambda}_1^*(\sigma^2,\theta)$ be the largest empirical eigenvalue under the null model ($K=0$) with parameters $(\sigma^2,\theta)$. Applying Theorem 3.2 of \cite{ding2020spiked} again, for the same $x_1^*$ as above, 
\[
|\hat{\lambda}_1^{*}(\sigma^2,\theta)-G(x_1^*)|\prec n^{-2/3}. 
\]
In Theorem~\ref{thm:generalM-error}, we have shown $|\hat{\sigma}^2-\sigma^2|\prec n^{-1}$ and $|\hat{\theta}-\theta|\prec n^{-1}$. Now, let $\hat{x}_1^*$ be the largest critical point of $G(x)$ in \eqref{define-right-edge}, except that $(\sigma^2,\theta)$ is replaced by $(\hat{\sigma}^2,\hat{\theta})$. Then, we have $|G(\hat{x}_1^*)-G(x_1^*)|= O\bigl(\sqrt{|\hat{\sigma}^2-\sigma^2|^2+|\hat{\theta}-\theta|^2}\bigr)\prec n^{-1}$ and $|\hat{\lambda}_1^{*}(\hat{\sigma}^2,\hat{\theta})-G(\hat{x}_1^*)|\prec n^{-2/3}$. Combining these claims gives
\[
|\hat{\lambda}_1^{*}(\hat{\sigma}^2,\hat{\theta})-G(x_1^*)|\prec n^{-2/3}. 
\]
Note that $\hat{T}_\beta$ is the $(1-\beta)$-quantile of $\hat{\lambda}_1^*(\hat{\sigma}^2, \hat{\theta})$ (it means the quantile of $\hat{\lambda}_1^*(\sigma^2,\theta)$ evaluated at $(\sigma^2,\theta)=(\hat{\sigma}^2, \hat{\theta})$). 
The above inequality implies that there exists $\beta\to 0$ properly slow such that 
\beq \label{thm-genK-3}
n^{-2/3}\; \ll \;  \hat{T}_{\beta} -G(x_1^*)\; \ll \;  1.
\eeq
It follows from \eqref{thm-genK-2} and \eqref{thm-genK-3} that $\hat{K}=K$.\qed

\subsection{Proof of Lemma~\ref{lem:cdf-quantile}} \label{subsec:proof-lem-cdf-quantile} 
We only show the second inequality. The proof of the first inequality is similar and thus omitted. Note that $f(x)-g(x)$ is the derivative of $\bar{G}(x)-\bar{F}(x)$. Using integration by part, we have
\begin{align} \label{lem-cdf-quantile-1}
\mu_2(f)-\mu_2(g) = \int x^2[f(x)-g(x)]dx =2\int x[\bar{F}(x)-\bar{G}(x)]dx.  
\end{align}
We consider a change of variable from $x$ to $\alpha=\bar{F}(x)$. Note that $x=h_F(\alpha)$. It follows that
\begin{align*}
\int x[\bar{F}(x)-\bar{G}(x)]dx &= \int_0^1 h_F(\alpha)\bigl[\alpha - \bar{G}(h_F(\alpha))\bigr]h'_F(\alpha)d\alpha\cr
&= \int_0^1 h_F(\alpha)\bigl[\bar{G}(h_G(\alpha)) - \bar{G}(h_F(\alpha))\bigr]h'_F(\alpha)d\alpha. 
\end{align*}
By mean value theorem, there is $x^*$ between $h_F(\alpha)$ and $h_G(\alpha)$ such that $\bar{G}(h_G(\alpha)) - \bar{G}(h_F(\alpha)) = -g(x^*)[h_G(\alpha)-h_F(\alpha)]$. Recall that $\check{g}(x,y)=\max_{z\in [x,y]\cup [y,x]}g(z)$. It follows that $|\bar{G}(h_G(\alpha)) - \bar{G}(h_F(\alpha)|\leq \check{g}(h_F(\alpha), h_G(\alpha))\cdot |h_G(\alpha)-h_F(\alpha)|$. We plug it into the above equation to get
\begin{align*}
\Bigl| \int x[\bar{F}(x)-\bar{G}(x)]dx\Bigr|& \leq 
\int_0^1  |h_G(\alpha)) - h_F(\alpha)|\cdot \bigl| h_F(\alpha)\, \check{g}\bigl(h_F(\alpha), h_G(\alpha)\bigr)h'_F(\alpha)\bigr| d\alpha. 
\end{align*}
Since $h_F(\cdot)=\bar{F}^{-1}$, we have $h'_F(\alpha)= -1/f(h_F(\alpha))$. It follows that
\begin{align}  \label{lem-cdf-quantile-2}
\Bigl| \int x[\bar{F}(x)-\bar{G}(x)]dx\Bigr| & \leq \int_0^1 |h_G(\alpha)-h_F(\alpha)|\cdot \frac{h_F(\alpha)\cdot \check{g}(h_F(\alpha), h_G(\alpha))}{f(h_F(\alpha))}d\alpha\cr
&\leq \sqrt{\int_0^1|h_G(\alpha)-h_F(\alpha)|^2d\alpha}\; \sqrt{\int_0^1 \Bigl[ \frac{h_F(\alpha)\cdot \check{g}(h_F(\alpha), h_G(\alpha))}{f(h_F(\alpha))} \Bigr]^2d\alpha}\cr
&\leq \sqrt{\int_0^1|h_G(\alpha)-h_F(\alpha)|^2d\alpha}\cdot \sqrt{C_2}. 
\end{align}
Combining \eqref{lem-cdf-quantile-1}-\eqref{lem-cdf-quantile-2} gives the claim. \qed

\section{Robustness of BEMA on real data}

For the two real data sets in Section~\ref{sec:realdata}, we apply BEMA with different values of $\alpha$. The results are presented in the tables below. Both the point estimator and the confidence interval are very stable as long as $\alpha$ is in a reasonable range.

\begin{table}[!hbt]
\hspace{1cm}
\def~{\hphantom{0}}
\scalebox{0.9}{
\begin{tabular}{lllll}
\toprule
   &  BEMA (0.1)     & BEMA (0.2)    & BEMA (0.3) & BEMA (0.4)  \\ \midrule
$\hat{\theta}$        & 0.343    & 0.288       & 0.281      & 0.270 \\
$\hat{\sigma}^2$         & 0.869    & 0.926       & 0.949      & 1 \\
$\hat{K}$  $(\beta=0.1)$ & 1 & 1 & 1 & 1\\
\hline
90\% quantile     & 16.074      & 19.231       & 20.261   & 21.944    \\
10\% quantile         & 9.379    & 10.872       & 11.186      & 12.098 \\
confidence interval       & [1,4] & [1,4]   & [1,4]  & [1,2]  \\ \bottomrule
\end{tabular}}
\caption{Lung Cancer data. BEMA is applied with $\alpha\in\{0.1,0.2,0.3,0.4\}$ (denoted as BEMA ($\alpha$) in the table). The quantiles are from $\mathrm{Gamma}(\hat{\theta},\hat{\theta}/\hat{\sigma}^2)$, and they are used to construct the 80\% confidence interval.}
\end{table}

\begin{table}[!htb]
\hspace{1cm}
\def~{\hphantom{0}}
\scalebox{0.9}{
\begin{tabular}{lllll}
\toprule
   &  BEMA (0.1)     & BEMA (0.2)    & BEMA (0.3) & BEMA (0.4)  \\ \midrule
$\hat{\theta}$        & 4.256    & 4.239       & 4.198      & 4.261 \\
$\hat{\sigma}^2$        & 0.3779    & 0.3780       & 0.3782      & 0.3783 \\
$\hat{K}$  $(\beta=0.1)$ & 28 & 28 & 28 & 28\\
\hline
90\% quantile        & 6.895      & 6.899       & 6.909   & 6.903    \\
10\% quantile         & 6.822    & 6.829       & 6.838      & 6.831 \\
confidence interval       & [28,30] & [28,30]   & [28,29]  & [28,30]  \\ \bottomrule
\end{tabular}}
\caption{1000 Genomes data. BEMA is applied with $\alpha\in\{0.1,0.2,0.3,0.4\}$ (denoted as BEMA ($\alpha$) in the table). The quantiles are from $\mathrm{Gamma}(\hat{\theta},\hat{\theta}/\hat{\sigma}^2)$, and they are used to construct the 80\% confidence interval.}
\label{tb:misspecify}
\end{table}

\end{appendix}

\bibliographystyle{chicago}
\bibliography{factor}

\begin{thebibliography}{}

\bibitem[\protect\citeauthoryear{1000 Genomes Project~Consortium}{1000 Genomes
  Project~Consortium}{2015}]{10002015global}
1000 Genomes Project~Consortium, A. (2015).
\newblock A global reference for human genetic variation.
\newblock {\em Nature\/}~{\em 526\/}(7571), 68.

\bibitem[\protect\citeauthoryear{Bai and Ng}{Bai and
  Ng}{2002}]{bai2002determining}
Bai, J. and S.~Ng (2002).
\newblock Determining the number of factors in approximate factor models.
\newblock {\em Econometrica\/}~{\em 70\/}(1), 191--221.

\bibitem[\protect\citeauthoryear{Baik, Ben~Arous, and Peche}{Baik
  et~al.}{2005}]{BAP}
Baik, J., G.~Ben~Arous, and S.~Peche (2005).
\newblock Phase transition of the largest eigenvalue for nonnull complex sample
  covariance matrices.
\newblock {\em The Annals of Probability\/}~{\em 33\/}(5), 1643--1697.

\bibitem[\protect\citeauthoryear{Bao}{Bao}{2020}]{Bao-communication}
Bao, Z. (2020).
\newblock Personal communications.

\bibitem[\protect\citeauthoryear{Bloemendal, Knowles, Yau, and Yin}{Bloemendal
  et~al.}{2016}]{bloemendal2016principal}
Bloemendal, A., A.~Knowles, H.-T. Yau, and J.~Yin (2016).
\newblock On the principal components of sample covariance matrices.
\newblock {\em Probability Theory and Related Fields\/}~{\em 164\/}(1-2),
  459--552.

\bibitem[\protect\citeauthoryear{Braeken and Van~Assen}{Braeken and
  Van~Assen}{2017}]{braeken2017empirical}
Braeken, J. and M.~A. Van~Assen (2017).
\newblock An empirical kaiser criterion.
\newblock {\em Psychological Methods\/}~{\em 22\/}(3), 450.

\bibitem[\protect\citeauthoryear{Cai, Han, and Pan}{Cai
  et~al.}{2020}]{cai2017limiting}
Cai, T.~T., X.~Han, and G.~Pan (2020).
\newblock Limiting laws for divergent spiked eigenvalues and largest nonspiked
  eigenvalue of sample covariance matrices.
\newblock {\em Annals of Statistics\/}~{\em 48\/}(3), 1255--1280.

\bibitem[\protect\citeauthoryear{Ding}{Ding}{2020}]{ding2020spiked}
Ding, X. (2020).
\newblock Spiked sample covariance matrices with possibly multiple bulk
  components.
\newblock {\em Random Matrices: Theory and Applications\/}, 2150014.

\bibitem[\protect\citeauthoryear{Dobriban}{Dobriban}{2015}]{dobriban2015efficient}
Dobriban, E. (2015).
\newblock Efficient computation of limit spectra of sample covariance matrices.
\newblock {\em Random Matrices: Theory and Applications\/}~{\em 4\/}(04),
  1550019.

\bibitem[\protect\citeauthoryear{Dobriban and Owen}{Dobriban and
  Owen}{2019}]{dobriban2019deterministic}
Dobriban, E. and A.~B. Owen (2019).
\newblock Deterministic parallel analysis: an improved method for selecting
  factors and principal components.
\newblock {\em Journal of the Royal Statistical Society: Series B (Statistical
  Methodology)\/}~{\em 81\/}(1), 163--183.

\bibitem[\protect\citeauthoryear{Donoho, Gavish, and Johnstone}{Donoho
  et~al.}{2018}]{donoho2018optimal}
Donoho, D.~L., M.~Gavish, and I.~M. Johnstone (2018).
\newblock Optimal shrinkage of eigenvalues in the spiked covariance model.
\newblock {\em The Annals of Statistics\/}~{\em 46\/}(4), 1742.

\bibitem[\protect\citeauthoryear{Efron}{Efron}{2004}]{efron2004large}
Efron, B. (2004).
\newblock Large-scale simultaneous hypothesis testing: the choice of a null
  hypothesis.
\newblock {\em Journal of the American Statistical Association\/}~{\em
  99\/}(465), 96--104.

\bibitem[\protect\citeauthoryear{Fan, Guo, and Zheng}{Fan
  et~al.}{2020}]{fan2019estimating}
Fan, J., J.~Guo, and S.~Zheng (2020).
\newblock Estimating number of factors by adjusted eigenvalues thresholding.
\newblock {\em Journal of the American Statistical
  Association\/}~(just-accepted), 1--33.

\bibitem[\protect\citeauthoryear{Fan, Liao, and Mincheva}{Fan
  et~al.}{2013}]{fan2013large}
Fan, J., Y.~Liao, and M.~Mincheva (2013).
\newblock Large covariance estimation by thresholding principal orthogonal
  complements.
\newblock {\em Journal of the Royal Statistical Society: Series B (Statistical
  Methodology)\/}~{\em 75\/}(4), 603--680.

\bibitem[\protect\citeauthoryear{Gavish and Donoho}{Gavish and
  Donoho}{2014}]{gavish2014optimal}
Gavish, M. and D.~L. Donoho (2014).
\newblock The optimal hard threshold for singular values is $4/\sqrt{3}$.
\newblock {\em IEEE Transactions on Information Theory\/}~{\em 60\/}(8),
  5040--5053.

\bibitem[\protect\citeauthoryear{Gordon, Jensen, Hsiao, Gullans, Blumenstock,
  Ramaswamy, Richards, Sugarbaker, and Bueno}{Gordon
  et~al.}{2002}]{gordon2002translation}
Gordon, G.~J., R.~V. Jensen, L.-L. Hsiao, S.~R. Gullans, J.~E. Blumenstock,
  S.~Ramaswamy, W.~G. Richards, D.~J. Sugarbaker, and R.~Bueno (2002).
\newblock Translation of microarray data into clinically relevant cancer
  diagnostic tests using gene expression ratios in lung cancer and
  mesothelioma.
\newblock {\em Cancer Research\/}~{\em 62\/}(17), 4963--4967.

\bibitem[\protect\citeauthoryear{G{\"o}tze, Tikhomirov, et~al.}{G{\"o}tze
  et~al.}{2004}]{gotze2004rate}
G{\"o}tze, F., A.~Tikhomirov, et~al. (2004).
\newblock Rate of convergence in probability to the marchenko-pastur law.
\newblock {\em Bernoulli\/}~{\em 10\/}(3), 503--548.

\bibitem[\protect\citeauthoryear{Horn}{Horn}{1965}]{horn1965rationale}
Horn, J.~L. (1965).
\newblock A rationale and test for the number of factors in factor analysis.
\newblock {\em Psychometrika\/}~{\em 30\/}(2), 179--185.

\bibitem[\protect\citeauthoryear{Horn and Johnson}{Horn and
  Johnson}{2012}]{horn2012matrix}
Horn, R.~A. and C.~R. Johnson (2012).
\newblock {\em Matrix analysis}.
\newblock Cambridge university press.

\bibitem[\protect\citeauthoryear{Ji}{Ji}{2020}]{ji2020regularity}
Ji, H.~C. (2020).
\newblock Regularity properties of free multiplicative convolution on the
  positive line.
\newblock {\em International Mathematics Research Notices\/}.

\bibitem[\protect\citeauthoryear{Jin, Ke, and Wang}{Jin et~al.}{2017}]{JKW}
Jin, J., Z.~T. Ke, and W.~Wang (2017).
\newblock Phase transitions for high dimensional clustering and related
  problems.
\newblock {\em The Annals of Statistics\/}~{\em 45\/}(5), 2151--2189.

\bibitem[\protect\citeauthoryear{Jin and Wang}{Jin and
  Wang}{2016}]{jin2016influential}
Jin, J. and W.~Wang (2016).
\newblock Influential features {PCA} for high dimensional clustering.
\newblock {\em The Annals of Statistics\/}~{\em 44\/}(6), 2323--2359.

\bibitem[\protect\citeauthoryear{Johnstone}{Johnstone}{2001}]{Johnstone2001}
Johnstone, I.~M. (2001).
\newblock On the distribution of the largest eigenvalue in principal components
  analysis.
\newblock {\em The Annals of Statistics\/}~{\em 29\/}(2), 295--327.

\bibitem[\protect\citeauthoryear{Ke}{Ke}{2016}]{ke2016detecting}
Ke, Z.~T. (2016).
\newblock Detecting rare and weak spikes in large covariance matrices.
\newblock {\em arXiv preprint arXiv:1609.00883\/}.

\bibitem[\protect\citeauthoryear{Knowles and Yin}{Knowles and
  Yin}{2017}]{knowles2017anisotropic}
Knowles, A. and J.~Yin (2017).
\newblock Anisotropic local laws for random matrices.
\newblock {\em Probability Theory and Related Fields\/}~{\em 169\/}(1-2),
  257--352.

\bibitem[\protect\citeauthoryear{Kritchman and Nadler}{Kritchman and
  Nadler}{2009}]{kritchman2009non}
Kritchman, S. and B.~Nadler (2009).
\newblock Non-parametric detection of the number of signals: Hypothesis testing
  and random matrix theory.
\newblock {\em IEEE Transactions on Signal Processing\/}~{\em 57\/}(10),
  3930--3941.

\bibitem[\protect\citeauthoryear{Kwak, Lee, and Park}{Kwak
  et~al.}{2019}]{kwak2019extremal}
Kwak, J., J.~O. Lee, and J.~Park (2019).
\newblock Extremal eigenvalues of sample covariance matrices with general
  population.
\newblock {\em arXiv preprint arXiv:1908.07444\/}.

\bibitem[\protect\citeauthoryear{Marcenko and Pastur}{Marcenko and
  Pastur}{1967}]{marchenko1967distribution}
Marcenko, V.~A. and L.~A. Pastur (1967).
\newblock Distribution of eigenvalues for some sets of random matrices.
\newblock {\em Mathematics of the USSR-Sbornik\/}~{\em 1\/}(4), 457--483.

\bibitem[\protect\citeauthoryear{Onatski}{Onatski}{2009}]{onatski2009testing}
Onatski, A. (2009).
\newblock Testing hypotheses about the number of factors in large factor
  models.
\newblock {\em Econometrica\/}~{\em 77\/}(5), 1447--1479.

\bibitem[\protect\citeauthoryear{Onatski}{Onatski}{2010}]{onatski2010determining}
Onatski, A. (2010).
\newblock Determining the number of factors from empirical distribution of
  eigenvalues.
\newblock {\em The Review of Economics and Statistics\/}~{\em 92\/}(4),
  1004--1016.

\bibitem[\protect\citeauthoryear{Passemier and Yao}{Passemier and
  Yao}{2014}]{passemier2014estimation}
Passemier, D. and J.~Yao (2014).
\newblock Estimation of the number of spikes, possibly equal, in the
  high-dimensional case.
\newblock {\em Journal of Multivariate Analysis\/}~{\em 127}, 173--183.

\bibitem[\protect\citeauthoryear{Patterson, Price, and Reich}{Patterson
  et~al.}{2006}]{patterson2006population}
Patterson, N., A.~L. Price, and D.~Reich (2006).
\newblock Population structure and eigenanalysis.
\newblock {\em PLoS Genetics\/}~{\em 2\/}(12), e190.

\bibitem[\protect\citeauthoryear{Paul}{Paul}{2007}]{Paul2007}
Paul, D. (2007).
\newblock Asymptotics of sample eigenstructure for a large dimensional spiked
  covariance model.
\newblock {\em Statistica Sinica\/}~{\em 17\/}(4), 1617--1642.

\bibitem[\protect\citeauthoryear{Pillai and Yin}{Pillai and
  Yin}{2014}]{PillaiYin}
Pillai, N.~S. and J.~Yin (2014).
\newblock Universality of covariance matrices.
\newblock {\em The Annals of Applied Probability\/}~{\em 24\/}(3), 935--1001.

\bibitem[\protect\citeauthoryear{Shabalin and Nobel}{Shabalin and
  Nobel}{2013}]{shabalin2013reconstruction}
Shabalin, A.~A. and A.~B. Nobel (2013).
\newblock Reconstruction of a low-rank matrix in the presence of gaussian
  noise.
\newblock {\em Journal of Multivariate Analysis\/}~{\em 118}, 67--76.

\bibitem[\protect\citeauthoryear{Silverstein}{Silverstein}{2009}]{silverstein2009stieltjes}
Silverstein, J.~W. (2009).
\newblock The stieltjes transform and its role in eigenvalue behavior of large
  dimensional random matrices.
\newblock {\em Random Matrix Theory and Its Applications. Lect. Notes Ser.
  Inst. Math. Sci. Natl. Univ. Singap\/}~{\em 18}, 1--25.

\bibitem[\protect\citeauthoryear{Uhlig}{Uhlig}{1994}]{Uhlig94}
Uhlig, H. (1994).
\newblock On singular wishart and singular multivariate beta distributions.
\newblock {\em The Annals of Statistics\/}~{\em 22}, 395--405.

\bibitem[\protect\citeauthoryear{Wax and Kailath}{Wax and
  Kailath}{1985}]{wax1985detection}
Wax, M. and T.~Kailath (1985).
\newblock Detection of signals by information theoretic criteria.
\newblock {\em IEEE Transactions on Acoustics, Speech, and Signal
  Processing\/}~{\em 33\/}(2), 387--392.

\end{thebibliography}

\end{document}